\documentclass[11pt,a4paper]{article}
\pdfoutput=1
\usepackage{graphicx}
\usepackage[usenames,dvipsnames]{xcolor} % colour
\usepackage{amssymb,amsmath,mathtools,mathrsfs} % maths
\usepackage{epsfig,subfigure} % plots
\usepackage{booktabs,longtable} % tables
\usepackage{exscale,relsize} % scale 
\usepackage[normalem]{ulem} % editing
\usepackage{enumerate} 
\usepackage{jcappub}
\usepackage{caption}
\usepackage{hyperref}
\usepackage{float}
\usepackage[utf8]{inputenc}

\makeindex

\begin{document}
\hypersetup{pageanchor=false} 
\title{Cosmological perturbations in massive bigravity}

\author[a,b]{Macarena Lagos,}
\author[a]{Pedro G. Ferreira}
\affiliation[a]{Astrophysics, University of Oxford, DWB,\\
Keble road, Oxford OX1 3RH, UK}
\affiliation[b]{Theoretical physics, Blackett Laboratory, Imperial College London,\\
Prince Consort Road, London SW7 2BZ, UK}

\emailAdd{m.lagos13@imperial.ac.uk}
\emailAdd{p.ferreira1@physics.ox.ac.uk}

\abstract{We present a comprehensive analysis of classical scalar, vector and tensor cosmological perturbations in ghost-free massive bigravity. In particular, we find the full evolution equations and analytical solutions in a wide range of regimes. We show that there are viable cosmological backgrounds but, as has been found in the literature, these models generally have exponential instabilities in linear perturbation theory. However, it is possible to find stable scalar cosmological perturbations for a very particular choice of parameters. For this stable subclass of models we find that vector and tensor perturbations have growing solutions. We argue that special initial conditions are needed for tensor modes in order to have a viable model.}

%\arxivnumber{1234.5678}


\maketitle

%--------------------------------------------------------------------------------------------------------------------------------------------------
\section{Introduction}

%Einstein's general relativity (GR), through the Big Bang theory and cosmic inflation, has been the standard and most successful model for theoretically describing our Universe. However, this model is not exempt of problems as, for instance, it assumes the existence of a yet unknown inflationary field, along with some dark matter and dark energy. Regarding this last issue, if GR was correct, some dark energy density would be needed to explain the observed accelerated expansion of the universe *cite*, and it would correspond to around $68\%$ of the total energy density of the universe. However, it is still not known what kind of particle would be responsible for dark energy and, furthermore, the observed value for dark energy is in disagreement with what we could expect from quantum field theory arguments *cite*. These issues have motivated the idea that GR might not be the appropriate theory to describe the Universe at large scales, and that therefore, some modifications might be needed. As a consequence, nowadays many alternative gravitational theories have been proposed *cite*.

Massive bigravity, as proposed by Hassan and Rosen in \cite{Hassan:2011zd}, is an alternative to general relativity, and an extension of the dRGT (de Rham, Gabadadze and Tolley) massive gravity  \cite{deRham:2010kj}. One of the main attractions of this model is that it can predict viable cosmological homogeneous and isotropic solutions with late time self-acceleration {\it without} including a cosmological constant. Furthermore, if one assumes the presence of a large vacuum energy in this model, it has been argued that an appropriate value for the graviton's mass may lead to screening of long wavelength modes, reconciling the value of the {\it measured} cosmological constant with quantum field theory \cite{Hinterbichler:2011tt}. As such, massive bigravity seems to be an appealing candidate for a theory of the universe.

Massive bigravity has five more degrees of freedom (dof) than general relativity (GR) -- due to an extra massive graviton propagating -- which could be a source of concern. Only recently has GR been shown to be well-behaved, i.e. that the initial value problem is sufficiently well posed that the theory can be considered classically predictive \cite{Lehner:2001wq}. With an extra five degrees of freedom, it is conceivable that massive bigravity will not be as obliging. A possible hint of there being any problem would be the presence of classical instabilities and a natural first step would be to study linear cosmological perturbations. A first analysis of such perturbations has been undertaken in \cite{Comelli:2012db,Konnig:2014dna,Berg:2012kn,Solomon:2014dua}, where unstable solutions on sub-horizon scales were found for some parameters of the theory in homogeneous and isotropic backgrounds\footnote{As of now, these type of backgrounds have been the only ones considered on cosmological studies of massive gravity.}. A subsequent analysis in \cite{Konnig:2014xva} identified a particular class of parameters that lead to stable solutions and, as such, might be used to construct a viable cosmology. In this paper, we undertake an independent analysis of the evolution and stability of linear cosmological perturbations using the gauge fixing method proposed in \cite{Lagos:2013aua}. We confirm previous results for scalar perturbations but also analyse vector and tensor perturbations finding a number of interesting instabilities. Our results confirm the obvious: that it is a phenomenologically rich theory which needs to be studied in great detail if it is to be cosmologically considered on par with GR. 

The outline of this paper is as follows. In section \ref{SecMassiveGravity} we review the massive bigravity model. In section \ref{SectionCosmologicalPerturbations} we review the standard Friedmann-Robertson-Walker (FRW) cosmological background in the presence of a perfect fluid, and we find the equations of motion for first order cosmological perturbations. Here, we use the formalism developed in \cite{Lagos:2013aua} to fix the gauge, simplify the problem, and to identify the physical degrees of freedom. In section \ref{SecScalarPert}, we study the evolution of the two physical scalar degrees of freedom, in section \ref{SecVectorPert} we study vector perturbations, and in section \ref{SecTensorPert} we study tensor perturbations. In section \ref{Discussion} we summarise our findings and discuss the prospects of massive bigravity as a viable theory of gravity and cosmology. Throughout this paper we will be using Planck units.

%--------------------------------------------------------------------------------------------------------------------------------------------------
\section{Bimetric Massive gravity}
\label{SecMassiveGravity}

A linear theory of a massive spin-2 field in Minkowski space was proposed by Fierz and Pauli in 1939 \cite{1939RSPSA.173..211F}. It consists of a covariant quadratic action, known as the Fierz-Pauli action, describing a free massive spin-2 particle which propagates five degrees of freedom -- namely, modes with helicity $\pm 2$, $\pm 1$ and $0$. This action is the only possible instability-free quadratic action for a massive spin-2 particle \cite{VanNieuwenhuizen:1973fi}. In the presence of matter, the Fierz-Pauli action has the so-called van Dam, Veltman, Zakharov (vDVZ) discontinuity \cite{vanDam:1970vg,Zakharov:1970cc}, which arises when taking the massless limit; the helicity-0 mode couples to the trace of the stress-energy tensor, and therefore still propagates in the massless limit, where one would expect to only propagate the helicity $\pm 2$ modes.

Non-linear massive gravitational theories (which reduce to the Fierz-Pauli action at the linear level) were studied extensively following the non-linear proposal by Vainshtein in 1972 \cite{Vainshtein:1972sx}. It was then argued that non-linearities could cure the vDVZ discontinuity as these interactions would become comparable to the linear terms even for very weak fields, for small values of $m$. Such non-linear interactions would give rise to a screening of the helicity-0 mode at observable scales, rendering the theory compatible with observational tests of gravity \cite{Vainshtein:1972sx,Deffayet:2001uk}. Vainshtein's model was flawed as it contained an instability, the so-called Boulware-Deser ghost \cite{Boulware:1973my}, i.e.~an extra scalar degree of freedom whose kinetic term had the wrong sign.

In 2010 major progress was made when a particular family of ghost-free interaction potentials was constructed by de Rham, Gabadadze and Tolley in \cite{deRham:2010kj} and confirmed to be ghost-free by Hassan and Rosen in \cite{Hassan:2011hr} (see also \cite{Hassan:2011ea}). dRGT massive gravity \cite{deRham:2010gu, deRham:2010ik, Hassan:2011vm}, as it is known, contains the space-time metric $g_{\mu\nu}$ as well as a fixed non-dynamical second metric $f_{\mu\nu}$. A bimetric ghost-free extension of the dRGT massive gravity was proposed by Hassan and Rosen in \cite{Hassan:2011zd} (see also \cite{Hassan:2011tf}), where the new metric $f_{\mu\nu}$ is also dynamical. Concern on these types of theories may arise as for dRGT massive gravity some issues may been found (see \cite{Burrage:2011cr,Izumi:2013poa,Deser:2014hga,Deser:2013qza,Deser:2012qx,Deser:2013eua,deRham:2014zqa} for related discussions). However, none of these problems have been seen in the bimetric model yet. For a more detailed review on massive gravity and its origins, see \cite{Hinterbichler:2011tt, deRham:2014zqa}. 

In this paper, we will focus on the massive bigravity model proposed in \cite{Hassan:2011zd}:
\begin{equation}\label{MGaction}
S=\; \frac{M_g^2}{2}\int d^4x \sqrt{-g}R(g) +\frac{M_f^2}{2}\int d^4x \sqrt{-f}R(f) - m^2M_{g}^2\int d^4x \sqrt{-g}\sum_{n=0}^4\beta_n e_n\left(\sqrt{g^{-1}f}\right)+S_m.
\end{equation}
In this action there are two dynamical metric fields $g_{\mu\nu}$ and $f_{\mu\nu}$, with their associated Ricci scalars $R(g)$ and $R(f)$, respectively, along with a coupling to matter, $S_m$. In addition, this action contains interactions between both metrics that preserve general covariance, and are expressed in terms of the functions $e_n \left(\sqrt{g^{-1}f}\right)$, which correspond to the elementary symmetric polynomials of the eigenvalues $\lambda_n$ of the matrix $\sqrt{g^{-1}f}$, which satisfies $\sqrt{g^{-1}f}\sqrt{g^{-1}f}=g^{\mu\lambda}f_{\lambda\nu}$. Note that there is an ambiguity in $\sqrt{g^{-1}f}$, as different matrices may result in $g^{\mu\lambda}f_{\lambda\nu}$ when squared. Finally, $\beta_n$ are free dimensionless coefficients while $M_g$, $M_f$, and $m$ are mass scales.
For simplicity, we will be considering the case where matter is minimally coupled to $g_{\mu\nu}$ only, and therefore $g_{\mu\nu}$ will be describing the space-time evolution. 

As shown in \cite{Hassan:2011vm}, the equations of motion for $g_{\mu\nu}$ and $f_{\mu\nu}$ are:
\begin{align}
& \; R(g)_{\mu\nu}-\frac{1}{2}g_{\mu\nu}R(g) +\frac{m^2}{2}\sum_{n=0}^3(-1)^n\beta_n\left[g_{\mu\lambda}Y^\lambda_{(n)\nu}\left(\sqrt{g^{-1}f}\right)+ g_{\nu\lambda}Y^\lambda_{(n)\mu}\left(\sqrt{g^{-1}f}\right)\right]=\frac{T_{\mu\nu}}{M_g^2},\label{Eqg}\\
 & \; R(f)_{\mu\nu}-\frac{1}{2}f_{\mu\nu}R(f) + \frac{m^2}{2M_*^2}\sum_{n=0}^3(-1)^n\beta_{4-n}\left[f_{\mu\lambda}Y^\lambda_{(n)\nu}\left(\sqrt{f^{-1}g}\right)+f_{\nu\lambda}Y^\lambda_{(n)\mu}\left(\sqrt{f^{-1}g}\right)\right]=0,\label{Eqf}
\end{align}
where $M_*^2\equiv M_f^2/M_g^2$, and $T^{\mu\nu}$ is the matter stress-energy tensor, and where we have used the following relation for the interaction terms:
\begin{equation}\label{InteractionRelation}
\sqrt{-g}\sum_{n=0}^4 \beta_n e_n\left(\sqrt{g^{-1}f}\right)=\sqrt{-g}\sum_{n=0}^4 \beta_n \frac{e_{4-n}\left(\sqrt{f^{-1}g}\right)}{\det\left(\sqrt{g^{-1}f}\right)}=\sqrt{-f}\sum_{n=0}^4 \beta_{4-n} e_n\left(\sqrt{f^{-1}g}\right),
\end{equation}
where the matrix $\sqrt{f^{-1}g}$ is the inverse of $\sqrt{g^{-1}f}$. Note that to satisfy this relation we need to have $\sqrt{-g}\det(\sqrt{g^{-1}f})=\sqrt{-f}$. Otherwise, we would have a minus sign in the RHS of eq.~(\ref{InteractionRelation}), and therefore a minus sign in the interaction terms of eq.~(\ref{Eqf}). In addition, matrices $Y^\lambda_{(n)\mu}(\mathbb{X})$ are defined as:
\begin{align}
Y_{(0)}=&\mathbb{I},\nonumber\\
Y_{(1)}=&\mathbb{X}-\mathbb{I}[\mathbb{X}],\nonumber\\
Y_{(2)}=&\mathbb{X}^2-\mathbb{X}[\mathbb{X}]+\frac{1}{2}\mathbb{I}\left([\mathbb{X}]^2-[\mathbb{X}^2]\right),\nonumber\\
Y_{(3)}=&\mathbb{X}^3-\mathbb{X}^2[\mathbb{X}]+\frac{1}{2}\mathbb{X}\left([\mathbb{X}]^2-[\mathbb{X}^2]\right) -\frac{1}{6}\mathbb{I}\left([\mathbb{X}]^3-3[\mathbb{X}][\mathbb{X}^2]+2[\mathbb{X}^3]\right),
\end{align}
where $\mathbb{I}$ is the identity matrix and $[\mathbb{X}]$ stands for the trace of the matrix $\mathbb{X}$. We must also add a matter equation which, in this case, will correspond to the local conservation of energy-momentum:
\begin{equation}
\nabla^{\mu}_gT_{\mu\nu}=0,
\end{equation}
where $\nabla^{\mu}_g$ is the covariant derivative with respect to the metric $g_{\mu\nu}$.

%--------------------------------------------------------------------------------------------------------------------------------------------------
\section{Cosmological perturbations}\label{SectionCosmologicalPerturbations}

In this section we first review previous results on solutions for homogeneous and isotropic universes in massive bigravity. We then consider general linear cosmological perturbations using the standard classification of scalar, vector and tensor \cite{Mukhanov:1990me}. 
\subsection{Background}

For simplicity we will assume that both metrics share the same characteristics: homogeneous, isotropic and flat:
\begin{align}
 ds_f^2 &= Y(\tau)^2[-X(\tau)^2d\tau^2+\delta_{ij}dx^i dx^j], \label{sq}\\
 ds_g^2 &= a(\tau)^2[-d\tau^2+\delta_{ij}dx^i dx^j], \label{sg}
\end{align}
where $\tau$ is the conformal time, $a(\tau)$ is the scale factor of the space-time metric, and $X(\tau)$ with $Y(\tau)$ describe the evolution of the metric $f_{\mu\nu}$.  

In addition, we will assume the type of matter coupled to gravity to be a perfect fluid: 
\begin{equation}\label{BackStress}
T^{\mu}{}_{\nu}=(p_0+\rho_0)u^{\mu}_0u_{0\nu}+p_0\delta^{\mu}{}_{\nu},
\end{equation}
 where $p_0=p_0(\tau)$ is the pressure of the fluid, $\rho_0=\rho_0(\tau)$ its rest energy density and $u^{\mu}_0=(1/a,0,0,0)$ its isotropic 4-velocity.

If we replace eq.~(\ref{sq})-(\ref{BackStress}) into eq.~(\ref{Eqg})-(\ref{Eqf}), we find the following equations of motion:
\begin{align}
\mathcal{H}^2& = \frac{a^2}{3}\left[\frac{\rho_0 }{M_g^2} +m^2\left(\beta_0 + 3\beta_1N+3\beta_2N^2+\beta_3N^3   \right) \right],\label{EqFried}\\
\mathcal{H}'&=\frac{a^2}{2}\left[-\frac{p_0}{M_g^2}-\frac{\mathcal{H}^2}{a^2} +m^2\left(\beta_0+\beta_1N\left[2+X\right]+\beta_2N^2\left[1+2X\right] +\beta_3N^3X \right)   \right],\label{EqDerH}    \\
h^2& =\frac{a^2}{3}\left(\frac{X^2}{N}\right)\nu^2\left(\beta_1+3\beta_2N+3\beta_3N^2+\beta_4N^3\right),  \label{Eqh2}\\
h'&=\frac{a^2}{2}\left[ \frac{2}{a^2}h_x h-\frac{h^2}{a^2} +\left(\frac{X}{N}\right)\nu^2\left(\beta_1+\beta_2N[2+X]+\beta_3N^2[1+2X]+\beta_4N^3X\right) \right]\label{EqDerh}
\end{align}
where it is implicit that all variables depend only on $\tau$, all primes represent conformal time derivatives, and we have defined $\mathcal{H}=a'/a$, $h=Y'/Y$, $h_x=X'/X$, $\nu=m/M_*$, and $N=Y/a$. Note that the parameter $M_*$ is redundant, as we can rescale the metric $f_{\mu\nu}$ to make $M_*$ take any value we want and redefine $\beta$s such that the action remains invariant. For simplicity, from now on we will use $M_*=1$.

It is important to note that in order to obtain the previous equations, we had to make a choice for the matrix $\sqrt{g^{-1}f}$. For simplicity, we have chosen the diagonal form: $\sqrt{g^{-1}f}=\mbox{diag}\left(NX,N,N,N\right)$. As we will explain later, some solutions allow $X$ to change sign, and therefore this matrix can change sign at some point. Then, in order to satisfy $\sqrt{-g}\det(\sqrt{g^{-1}f})=\sqrt{-f}$, and therefore eq.~(\ref{InteractionRelation}), we need to make the unconventional (multivalued) choice of $\sqrt{-g}$ and $\sqrt{-f}$ {\it without} absolute values, allowing them to change signs. As explained in \cite{Gratia:2013gka,Gratia:2013uza} if $\sqrt{g^{-1}f}$ can change sign we can find continuous solutions through singularities in $f_{\mu\nu}$.

We also have the matter equation of motion:
\begin{equation}
\rho_0'=-3\mathcal{H}(\rho_0+p_0),
\end{equation}
which has the standard form, as matter has been minimally coupled to the metric $g_{\mu\nu}$. In addition, we have Bianchi constraints 
for both metrics, resulting from the Bianchi identities and the local conservation of the matter stress-energy tensor. However, due to the diffeomorphism invariance, they are both equivalent, so we have only one relevant Bianchi constraint, given in this case by:
\begin{equation}
\left(X\mathcal{H}-h\right)\left(\beta_1+2\beta_2N+\beta_3N^2\right)=0.
\end{equation}
We can easily identify two cases for the solutions: 
\begin{description}
\item[\underline{$\left(\beta_1+2\beta_2N+\beta_3N^2\right)=0$:}] This case leads to a constant $N=\bar{N}$, such that 
\begin{equation}\label{Bianchi1}
\beta_1+2\beta_2\bar{N}+\beta_3\bar{N}^2=0.
\end{equation}
As a consequence, $\mathcal{H}=h$ and the Friedmann equation becomes:
\begin{equation}
\mathcal{H}^2 = \frac{a^2}{3}\left[\frac{\rho_0 }{M_g^2} +\Lambda \right]; \; \Lambda= m^2\left(\beta_0 + 3\beta_1\bar{N}+3\beta_2\bar{N}^2+\beta_3\bar{N}^3   \right),
\end{equation}
which corresponds to general relativity with a cosmological constant. This case is not particularly interesting at the background level as it does not bring new features. Furthermore, as pointed out in \cite{vonStrauss:2011mq}, when studying first order perturbations, the interaction terms between $g_{\mu\nu}$ and $f_{\mu\nu}$ vanish when imposing the constraint (\ref{Bianchi1}), and the model results in just two copies of general relativity. %It might be possible to obtain deviations from GR for higher order perturbations.

\item[\underline{$\left(X\mathcal{H}-h\right)=0$:}] This constraint can be replaced into eq.~(\ref{Eqh2}), and then compared to eq.~(\ref{EqFried}), to find the following consistency equation:
\begin{equation}\label{Density}
\tilde{\rho} \equiv \frac{\rho_*}{m^2}=\frac{\beta_1}{N}+3\beta_2-\beta_0+3N(\beta_3-\beta_1)+N^2(\beta_4-3\beta_2)-N^3\beta_3; \quad \rho_*=\rho_0/M_g^2, 
\end{equation}
which relates $N$ and the density $\rho_0$.

For a standard equation of state $p_0=w\rho_0$ (with $w$ constant), according to eq.~(\ref{Density}), at late times ($\tilde{\rho} \ll 1$), $N$ will approach a constant value, and both metrics enter an accelerated de-Sitter phase. However, at early times ($\tilde{\rho} \gg 1$), two types of behaviours can be identified: one where $N\ll 1$ (and $\beta_1\not=0$) and another where $N\gg 1$. The branch characterised by $N\ll 1$ at early times will be called {\it expanding branch}, as in this case both metrics expand in time. While the branch characterised by $N\gg 1$ will be called {\it bouncing branch}, as in this case $g_{\mu\nu}$ expands but $f_{\mu\nu}$ bounces.

The expanding branch is usually identified as the physical one as, in this case, the contribution of the graviton mass to the Friedmann equation will always be small (for appropriate choices of parameters), as expected. However, in the bouncing branch, the contribution of the graviton mass may be comparable to the matter energy density $\rho_0$ at early times. Furthermore, in the bouncing branch, if $w>0$ at early times, then $X<0$ at early times, and tend to $X=1$ at late times. This means that $X$ crosses a zero point, where $f_{00}=0$, and therefore $f^{-1}_{\mu\nu}$ diverges. At this point also $\det(\sqrt{g^{-1}f})=0$. As explained in \cite{Gratia:2013gka,Gratia:2013uza}, this divergence stays hidden from the matter sector as $g_{\mu\nu}$ does not experience any divergence, and the corresponding vielbein fields are continuous through this point. We confirm this at the level of the background, where no divergence is present in the set of equations of motion eq.~(\ref{EqFried})-(\ref{EqDerh}), nor in their solutions\footnote{One might worry about eq.~(\ref{EqDerh}), as the first term in the RHS contains $h_x$, which diverges when $X=0$. However, the full relevant quantity in that equation is $h_xh$, which is finite. This can be seen from eq.~(\ref{Eqh2}), where we observe that $h\propto X$, cancelling the $X$ in the denominator of $h_x$ and rendering the relevant term finite.}. In addition, in the next sections we find non-divergent solutions for linear perturbations through this point. Therefore, our results suggest that this divergence might have a mathematical origin instead of a physical one\footnote{The Ricci scalar associated to the metric $f_{\mu\nu}$ diverges, while the one for $g_{\mu\nu}$ does not. Given that the latter one represents the space-time metric, it will determine the relevant physical properties of space-time. Furthermore, the Ricci scalar of $f_{\mu\nu}$ will always appear multiplied by the determinant of $f_{\mu\nu}$, rendering it finite.}. Then, even though solutions in the bouncing branch are exotic, they will be analysed in this paper at the level of perturbations. However, it is clear to us that further research is needed to understand completely the nature of this branch. 

\end{description}

Throughout this paper we will focus on the second branch of solutions satisfying $X\mathcal{H}=h$, as this one brings relevant modifications to general relativity. Background solutions and viable cosmologies in this branch have been studied in detail in \cite{Konnig:2013gxa,Akrami:2013pna,Akrami:2012vf,Comelli:2011zm,Volkov:2011an, vonStrauss:2011mq}. Given these results, the next logical step is the study of cosmological perturbations in this background. We will use the standard tensor-vector-scalar decomposition, and find the relevant equations of motion for these three types of perturbations.

\subsection{Scalar perturbations}

Let us consider linear scalar perturbations \cite{Konnig:2014xva,Comelli:2012db,Konnig:2014dna,Berg:2012kn,Solomon:2014dua}. We use the following Ansatz for the perturbed metrics:
\begin{align}
 ds_f^2 &= Y^2[-X^2(1+2\phi_1)d\tau^2+2B_{1,i}X dx^id\tau  +[(1-2\psi_1)\delta_{ij}+2E_{1,ij}]dx^i dx^j], \label{spertq}\\
 ds_g^2 &= a^2[-(1+2\phi_2)d\tau^2+2 B_{2,i}dx^id\tau  +[(1-2\psi_2)\delta_{ij}+2 E_{2,ij}]dx^i dx^j], \label{spertg}
\end{align}

\noindent where $ds_f^2$ and $ds_g^2$ are the line elements for the metrics $f_{\mu\nu}$ and $g_{\mu\nu}$ respectively. We read from here that we have four scalar perturbation fields for each metric: $\phi_1$, $B_1$, $E_1$, $\psi_1$ for $f_{\mu\nu}$ and $\phi_2$, $B_2$, $E_2$, $\psi_2$ for $g_{\mu\nu}$.

For matter, we have a perfect fluid with an equation of state $p=w\rho$, and therefore the perturbed stress-energy tensor coupled to these scalar perturbations can be written as:
\begin{align}\label{pset}
 \delta T^0{}_0=& -(\rho_0+p_0)(3\psi_2-E_{2,ii}-\chi_{,ii}),\nonumber\\
 \delta T^i{}_0=&-(\rho_0+p_0)\chi'_{,i},\nonumber\\
 \delta T^0{}_i=&(\rho_0+p_0)(B_{2,i}+\chi'_{,i}),\nonumber\\
 \delta T^i{}_j=&w(\rho_0+p_0)(3\psi_2-E_{2,ll}-\chi_{,ll})\delta^i{}_j.
\end{align}
Note that we describe matter perturbations with only one scalar field $\chi$, in a non-conventional but useful way proposed in \cite{Mukhanov:1990me}. Consequently, we have nine scalar fields describing first order perturbations in this theory. As we will see later, from these nine fields there will be only two propagating physical degrees of freedom: one coming from matter perturbations and another one from the helicity-0 mode of the massive graviton. All the other seven scalar fields are simply auxiliary fields, i.e.~they appear without time derivatives and therefore they are not physical dynamical fields. 
This unconventional description for perfect fluid perturbations is useful in order to apply the tools developed in \cite{Lagos:2013aua} to eliminate ambiguities related to the gauge-symmetry present in the theory. 

The action given in eq.~(\ref{MGaction}) is invariant under diffeomorphisms, and the nine perturbation scalar fields in the model transform under this symmetry as: 
\begin{align}\label{transf}
 & \tilde{\phi}_2=\phi_2-\mathcal{H}\xi^0-\xi^{0'},\; \tilde{\psi}_2=\psi_2+\mathcal{H}\xi^0,\;\tilde{B}_2=B_2+\xi^0-\xi',\nonumber\\
& \tilde{E}_2=E_2-\xi,\; \tilde{\phi}_1=\phi_1-\left[h+h_x\right]\xi^0-\xi^{0'}, \nonumber \\
& \tilde{\psi}_1=\psi_1+h\xi^0,\;\tilde{B}_1=B_1-\frac{\xi'}{X}+\xi^0X,\; \tilde{E}_1=E_1-\xi, \nonumber \\
& \tilde{\chi}=\chi+\xi,
\end{align} 
where $\xi$ and $\xi^0$ are the two scalar gauge parameters. As these fields are gauge-dependent, anything you calculate from them will depend on your gauge-choice. This ambiguity is usually eliminated by defining a new set of independent gauge-invariant scalar fields. In this paper we will approach this problem by fixing the gauge in a convenient way, as in \cite{Lagos:2013aua}. First, we look at the Noether identities associated to the gauge symmetry:
\begin{align}
   &\mathcal{E}_{\phi_1}'-\mathcal{E}_{\phi_1}\left[ h+h_x\right]+ \mathcal{E}_{\psi_1}h+\mathcal{E}_{B_1}X   +\mathcal{E}_{\phi_2}' + \left(\mathcal{E}_{\psi_2} -\mathcal{E}_{\phi_2} \right)\mathcal{H}+  \mathcal{E}_{B_2} = 0, \nonumber\\
 &\mathcal{E}_{\xi}- \mathcal{E}_{E_1}+\left(\frac{\mathcal{E}_{B_1}}{X}\right)' - \mathcal{E}_{E_2}+ \mathcal{E}_{B_2}' =0,
\end{align}
where we have denoted $\mathcal{E}_x$ as the equation of motion for the field $x$. From here we can recognise those fields with redundant equations of motions, and therefore the ones that are good candidates to be fixed with the gauge-freedom. The appropriate candidates are the following:
\begin{equation}
 (\psi_1, \psi_2) + (E_1, E_2, \chi),
\end{equation}
which means that we can use our two gauge parameters to fix one field of the first parenthesis and one of the second parenthesis. Specifically, we will choose the gauge such that $\psi_1=\chi=0$ . The advantages of fixing the gauge, and particularly in this way, is that: (1) we easily simplify the problem by reducing the number of fields by two, (2) we eliminate the redundant equations of motion, and all the remaining ones form the independent set of relevant equations, (3) all the remaining dynamical fields are still gauge-invariant, in the sense that the following gauge-invariant variables:
\begin{align}
&\zeta \equiv \psi_2-\frac{1}{3(\rho_0+p_0)}\delta \rho = \frac{1}{3}(E_{2,ii}+\chi_{,ii})\\
&\zeta_1 \equiv \frac{1}{3}(E_{1,ii}+\chi_{,ii})
%\Psi\equiv & \psi_2-\frac{\mathcal{H}}{h}\psi_1,\nonumber
\end{align}
become $E_2$, $E_1$ in our gauge-choice, and as we will see later, these two fields are the only physical ones. 

After fixing the gauge, let us consider the equation of motions for the seven remaining fields in Fourier space: 
\begin{align}
&2  \mathcal{H} \left(3\psi_2'+ k^2 E_2'\right)+\left( (1+w)\rho_*(3\psi_2+k^2E_2)+ m^2  NZ (3\psi_2+k^2(E_2-E_1))\right)a^2\nonumber  \\
& +2 \left(\psi_2 k^2+\mathcal{H}(3\phi_2\mathcal{H}-k^2 B_2)\right)=0 ,\label{G00}\\
&2(X+1)\psi_2'+2\mathcal{H}(X+1)\phi_2-m^2ZN(XB_1-B_2)+(1+w)\rho_*(1+X)B_2=0 ,   \label{G0i}\\
& 2 (k^2E_2^{''}+3\psi_2^{''}) + 2\mathcal{H}(3\phi_2'+6\psi_2'+2 k^2 E_2') -2k^2B_2' + 3 Za^2 m^2 N (\phi_1+\phi_2)X\nonumber\\
& + a^2\left( -3(1+w)\rho_*(2\phi_2+w(3\psi_2+k^2 E_2))  + 2 N m^2 (-3\phi_2 Z+(3\psi_2+k^2(E_2-E_1)) \tilde{Z}) \right) \nonumber\\
&+2(9 \mathcal{H}^2- k^2)\phi_2+2k^2(\psi_2-2\mathcal{H} B_2)=0 , \label{Gii} \\
& E_2^{''}-B_2'+2\mathcal{H} E_2'+(E_2 - E_1)a^2 m^2 N \tilde{Z}-\phi_2-2 \mathcal{H} B_2+\psi_2=0 \label{Gij}, 
\end{align}
and also
\begin{align}
& 2 N  h k^2 E_1' -a^2 \nu^2 Z (k^2 E_2-k^2 E_1+3\psi_2) X^2-2 N  h k^2 B_1 X +6 \phi_1 h^2 N=0, \label{F00} \\
& 2h\phi_1N(X+1)+\nu^2 X a^2 Z(XB_1-B_2)=0,   \label{F0i} \\
& N X E_1^{''}- N (-2 X h +X')E_1'- X^2 \left( B_1' N + N\phi_1 X + 2NB_1 h+ \nu^2 a^2 \tilde{Z}(E_2-E_1)\right)=0,   \label{Fij}
\end{align}
where we have defined $Z=\beta_1+2\beta_2N+\beta_3N^2$, $\tilde{Z}=\beta_1+\beta_2N(1+X)+\beta_3N^2X$. We have omitted the explicit dependence of variables, but it should be clear that the perturbation fields are now in Fourier space and depend on the conformal time $\tau$ and the wavenumber $k$.

From the equations (\ref{G00}), (\ref{G0i}), (\ref{F00}), and (\ref{F0i}) we can see that $B_1$, $B_2$, $\phi_1$ and $\phi_2$ appear as auxiliary variables, as they do not have any time derivatives and therefore they can be easily worked out in terms of $\psi_2$, $E_1$ and $E_2$ (see Appendix \ref{AppScalarsAux}). After replacing these four fields in the remaining three equations, we notice from eq.~(\ref{Gii}) that $\psi_2$ becomes an auxiliary variable as all its time derivatives cancelled. Therefore, we can now work out $\psi_2$ in terms of $E_1$ and $E_2$. If we do this, we end up with two equations for the only two physical scalar degrees of freedom:
\begin{equation}\label{EqE1E2}
E_a^{''}+c_{ab}E_b'+d_{ab}E_b=0,
\end{equation}
where the indices $a$ and $b$ can take the values $(1,2)$, and the coefficients $c_{ab}$ and $d_{ab}$ depend on the background functions and the wavenumber $k$. More specifically, these coefficients depend only on $k$, $\mathcal{H}$, $N$ and $a$, which are the four relevant quantities. The explicit expressions for these equations are given in the Appendix \ref{AppScalarEq}.

\subsection{Vector perturbations}

Let us consider vector perturbations for both metrics:
\begin{align}
 ds_f^2&=Y^2[-X^2 d\tau^2 - 2S_{1i}Xdx^id\tau+( \delta_{ij}+F_{1i,j}+F_{1j,i})dx^i dx^j],\label{vpertq}\\
ds_g^2&=a^2[-d\tau^2-2S_{2i}dx^id\tau+(\delta_{ij}+F_{2i,j}+F_{2j,i})dx^i dx^j].\label{vpertg}
\end{align}
\noindent From here we can see that the vector perturbations are $S_{1i}$ and $F_{1i}$ for the metric $f_{\mu\nu}$, and $S_{2i}$ and $F_{2i}$ for $g_{\mu\nu}$. These vector fields satisfy:
\begin{equation}\label{veccond}
 S_i{}^{,i}=F_i{}^{,i}=0,
\end{equation}
\noindent which means that they are purely transverse vectors with no scalar contributions. Here we lower and raise three-space indices by using the Kronecker delta, $\delta_{ij}$, and its inverse, $\delta^{ij}$.
The perturbed stress-energy tensor for a perfect fluid coupled to vector perturbations is:
\begin{align}\label{pvet}
\delta T^0{}_0 &= 0,\nonumber\\
\delta T^i{}_0 &= -(\rho_0+p_0)\chi^{iT'},\nonumber\\
\delta T^0{}_i &= (\rho_0+p_0)(\chi^{iT'}-S_{2i}),\nonumber\\
\delta T^i{}_j &= 0,
\end{align}
\noindent where $v^{iT}\equiv \chi^{iT'}$ represents the vorticity of the fluid and satisfies $v^{iT}{}_{,i}=0$.
Hence we have five vector perturbation fields: two for each metric and one for matter. In GR we only have one propagating degree of freedom but it is cosmologically irrelevant as it decays with the expansion of the universe. However, in massive gravity we will have 3 degrees of freedom: one from matter and two polarisations from the massive graviton.

In analogy to scalar perturbations, we analyse the gauge symmetry present in the massive bigravity action to fix a gauge. In this case, vector fields transform as:
\begin{equation}
\tilde{F}_{2i}=F_{2i}-  \xi^{T}_i, \; \tilde{S}_{2i}=S_{2i}+\xi^{T'}_i, \; \tilde{v}^{T}_i= v^{T}_i+\xi^{T'}_i,\;  \tilde{F}_{1i}=F_{1i}-  \xi^{T}_i,\; \tilde{S}_{1i}=S_{1i}+\xi^{T'}_i, 
\end{equation}  
\noindent where $\xi^{iT}$ is an infinitesimal arbitrary gauge field, also satisfying $\xi^{iT}{}_{,i}=0$. Consequently, the Noether identity associated to this gauge parameter is:
\begin{equation}
\mathcal{E}_{F_{2i}}+\mathcal{E}_{F_{1i}}+\mathcal{E}_{S_{2i}}'+\mathcal{E}_{S_{1i}}'+\mathcal{E}_{v_{i}^T}'=0,
\end{equation}
and we can use the gauge freedom to fix either $F_{1i}$ or $F_{2i}$. With the gauge choice $\tilde{F}_{1i}=0$, the relevant equations of motion are:
\begin{align}
 & \left( (k^2N +2m^2a^2 Z) X + k^2N\right)S_{1i}-2m^2a^2S_{2i}Z =0, \label{EqS1}  \\
 & -2a^2\rho_* (1+X)(1+w)v^{T}_i+k^2  (1+X)F_{2i}'+2S_{2i}\rho_*(1+X)(1+w)a^2 + S_{2i}k^2(1+X) \label{EqS2}\nonumber\\
& +2Zm^2Na^2(S_{2i}-X S_{1i})=0,  \\  
 &   F_{2i}''+2\mathcal{H}F_{2i}'+S_{2i}' + m^2Na^2\tilde{Z}F_{2i}+2\mathcal{H}S_{2i}= 0, \\
 & v^{T'}_i-\mathcal{H}(3w-1)v^{T}_i-S_{2i}'+(3w-1)\mathcal{H}S_{2i}= 0. 
\end{align}  

We see that $S_{1i}$ and $S_{2i}$ appear as auxiliary variables in (\ref{EqS1}) and (\ref{EqS2}). Therefore they can be worked out in terms of the remaining fields. When doing that we obtain only two relevant equations for $F_{2i}$ and the vorticity field $v^{T}_i$. 

The full equations for the vector field $F_{2i}$ and the vorticity field $v^{T}_i$ are the following:
\begin{align}\label{EqvT}
&v^{T'}_i+\frac{1}{D_v}\left[-2a^2 \rho_* Z(N k^2+2 m^2 a^2 Z)(1+w)X'-\mathcal{H} \left( \frac{}{} -4k^2a^2\tilde{Z}\rho_* N(1+X)(1+w)  \right. \right. \nonumber\\
& +2a^2m^2\left(-4\rho_* X^2(1+w) a^2+ (3w-1) (N^2+X)k^2 \right)Z^2 \nonumber \\
& \left. \left.+ (1+X)k^2N\left((3w-1)k^2-2\rho_* (1+X)(1+w)a^2\right) Z\right)\right]v^{T}_i \nonumber \\
& -\frac{k^2}{D_v}\left[  -X'Z(Nk^2+2m^2a^2Z) + \mathcal{H}\left( 2a^2XZ^2 (2X-1+3w) m^2+ (X+3w)(1+X)k^2NZ \right.\right.\nonumber \\
& \left. \left . + 2Nk^2\tilde{Z}(1+X)\right)\right]F_{2i}' -\frac{\tilde{Z}(k^2(1+X)N+2Xm^2a^2Z)}{2ZN}F_{2i}=0,
\end{align}
\begin{align}\label{EqF2}
&F_{2i}''+\frac{1}{D_v}\left[-X'k^2Z(k^2 N +2a^2m^2Z)+\mathcal{H} \left(4a^2m^2\left(k^2N^2+X(2\rho_*(1+w)a^2+k^2X)\right)Z^2\right. \right. \nonumber  \\
&\left. \left. + (1+X)\left(4\rho_*(1+w)a^2+(1+X)k^2\right)k^2NZ + 2Nk^4\tilde{Z}(1+X)\right)\right]F_{2i}' \nonumber\\
& -\frac{2a^2(1+w)\rho_*}{D_v}\left[  -X'Z(Nk^2+2m^2a^2Z) + \mathcal{H}\left( 4a^2Xm^2Z^2(X-1) + Nk^2(X-1)(1+X)Z\right. \right.\nonumber  \\
& \left. \left. + 2Nk^2\tilde{Z}(1+X) \right) \right]v^{T}_i + \frac{\tilde{Z}(2m^2a^2ZN^2+k^2(1+X)N+2Xm^2a^2Z)}{2NZ}F_{2i}=0,
\end{align}
where $D_v$ is given by: 
\begin{equation}
D_v=Z[4\rho_*m^2 ZX(1+w)a^4+2k^2(m^2N^2Z+\rho_*(1+X)(1+w)N+m^2ZX)a^2+k^4N(1+X)].
\end{equation}

Since $v^{T}_i$ and $F_{2i}$ satisfy $v^{T}_i k^i=F_{2i}k^i=0$, and the equation for $v^{T}_i$ is of first order, these set of equations actually propagate three degrees of freedom, as expected. 
\subsection{Tensor perturbations}
Let us consider tensor perturbations for both metrics:
\begin{align}
 ds_f^2&=Y^2[-X^{2}d\tau^2+(\delta_{ij}+h_{1ij})dx^i dx^j],\label{tpertq}\\
ds_g^2&=a^2[-d\tau^2+(\delta_{ij}+h_{2ij})dx^i dx^j],\label{tpertg}
\end{align}
such that
\begin{equation}\label{hcond}
 h_{bi}{}^i=0 , \quad h_{bij}{}^{,i}=0\; ; \; b=(1,2).
\end{equation}
\noindent From here we can see that the tensor perturbations are $h_{1ij}$ for the metric $f_{\mu\nu}$, and $h_{2ij}$ for $g_{\mu\nu}$. These perturbations satisfy (\ref{hcond}). Here, we use the metric $\delta_{ij}$ and its inverse $\delta^{ij}$ to lower and raise spatial indices. Since, in the perfect fluid model, there are no tensor matter perturbations, the perturbed stress-energy tensor to be considered here coupled to tensor perturbations $h_{bij}$ is zero. 

Because of (\ref{hcond}), each $h_{bij}$ has two degrees of freedoms, or polarisations, which are usually indicated as $p=+,\times$. More precisely, 
\begin{equation}\label{polexp}
h_{bij}(\vec{x},\tau)=\int \frac{d^3 k}{(2\pi)^{3/2}}h_{bij}(k,\tau)e^{i\vec{k}\cdot\vec{x}}, \quad h_{bij}(k,\tau)=h_{b+}(k,\tau) e^+_{ij}(k)+ h_{b\times}(k,\tau) e^{\times}_{ij}(k),
\end{equation}
\noindent where $e^+_{ij}$ and $e^{\times}_{ij}$ are the polarisation tensors, which have the following properties:
\begin{align}
e^p_{ij}=e^p_{ji}, \quad k^i&e^p_{ij}  =0, \quad e^p_{ii}=0,\nonumber\\
 e^p_{ij}(k)=e^{p*}_{ij}(-k), &\quad  e^{p*}_{ij}(k)e^{p'}_{ij}(k)=2\delta_{pp'}.  
\end{align}
\noindent Notice also that $h_{bij}$ are gauge-invariant and therefore they represent physical degrees of freedom. For simplicity, we choose a specific direction $\vec{k}=k\hat{z}$ so tensor perturbations lie in the $xy$ plane. As a result, tensor metric perturbations can be written as:
\begin{align}
ds^2_f &=Y^2\left[-X^{2}d\tau^2+[(1+h_{1+})dx^2+(1-h_{1+})dy^2+dz^2+2h_{1\times} dxdy]\right],\\
ds^2_g &=a^2\left[-d\tau^2+[(1+h_{2+})dx^2+(1-h_{2+})dy^2+dz^2+2h_{2\times} dxdy]\right],
\end{align}
where these tensor perturbations now depend only on $\tau$ and $z$. If we replace this Ansatz in the equations of motion (\ref{Eqg}) and (\ref{Eqf}) we find:
\begin{align}
& h_{2p}''+2\mathcal{H}h_{2p}'+h_{2p}k^2+m^2a^2N\tilde{Z}(h_{2p}-h_{1p})=0,\label{EqTensorh2}\\
&h_{1p}''-(h_x-2h)h_{1p}'+X^2k^2h_{1p}+\frac{Xm^2a^2\tilde{Z}}{N}(h_{1p}-h_{2p})=0.\label{EqTensorh1}
\end{align}

Summarising, in this section we described the possible background cosmological solutions in the massive bigravity theory, and found the relevant equations for first order cosmological perturbations. Note that for scalar and vector perturbations all the coefficients in their equations of motion are continuous and finite in the expanding and bouncing branches. However, if we recall that $h_x=X'/X$, for the tensor perturbations we see in eq.~(\ref{EqTensorh1}) that the coefficient of $h_{1p}'$ diverges when $X=0$ in the bouncing branch. Nevertheless, this coefficient is not a problem given that $h'_{1p}=0$ when $X=0$, in such a way that the complete second term in eq.~(\ref{EqTensorh1}) stays finite, regardless of the initial conditions. We can see this analytically near the bounce time, $\tau_b$, where $X(\tau_b)=0$. For the large $k$ limit, eq.~(\ref{EqTensorh1}) is approximated by:
\begin{equation}
h_{1p}''-\frac{h_{1p}'}{\left(\tau-\tau_b\right)}+x_0^2k^2\left(\tau-\tau_b\right)^2h_{1p}=0,
\end{equation}
where we have used that $h= 0$ and $X= x_0\left(\tau-\tau_b\right)$ near $\tau_b$. The solution to this equation is $h_{1p}\propto e^{\pm i kx_0(\tau-\tau_b)^2/2}$, and its derivative is $h_{1p}'\propto (\tau-\tau_b) e^{\pm i kx_0(\tau-\tau_b)^2/2}$, which goes to zero as fast as $X$ when $\tau\rightarrow\tau_b$. Similarly, for the small $k$ limit, eq.~(\ref{EqTensorh1}) is approximated by:
\begin{equation}
h_{1p}''-\frac{h_{1p}'}{\left(\tau-\tau_b\right)}=0, 
\end{equation}
where we have ignored the interaction term with $h_{2p}$, as this one is proportional to $(\tau-\tau_b)$, and 
is then negligible. The solution to this equation is $h_{1p}\propto (\tau-\tau_b)^2$, whose derivative also goes to zero as fast as $X$ when $\tau\rightarrow\tau_b$. 

%In the next three sections we find approximated analytical solutions for the equations of motion found in this section, as well as conditions on the parameters of the theory in order to get well-behaved solutions. Note that a condition on the parameters of the theory has already been mentioned in \cite{Fasiello:2013woa}, where a Higuchi bound was found for perturbations in the cosmological background given by eq.~(\ref{sq})-(\ref{sg}). However, the analysis of this bound is beyond the scope of this paper, so here we give an independent analysis on the parameters by analysing the equations of motion only. 

%--------------------------------------------------------------------------------------------------------------------------------------------------
\section{Scalar perturbations}\label{SecScalarPert}
In order to study the evolution of the two physical scalar fields, we need to analyse the form of the coefficients given in eq.~(\ref{EqE1E2}). Since it is not possible to find exact analytical solutions to these equations, we focus on a number of different relevant regimes and use suitable approximations in order to have a better understanding of the evolution of perturbations.

\subsection{Expanding branch}
As mentioned before, the expanding branch is characterised by $N\ll 1$ at early times and a de-Sitter phase at late times.

\subsubsection{Early times}
Let us assume that the early times are dominated by radiation. At this stage, we have $\tilde{\rho} \gg 1$, and $N\ll 1$, therefore, we can expand the solutions in powers of $N$. For example, at first order we have that eq.~(\ref{Density}) becomes:
\begin{equation}
\tilde{\rho}=\frac{\beta_1}{N}+\mathcal{O}(N^0); \; \beta_1>0
\end{equation}
Note that at early times this equation is solely characterised in terms of $\beta_1$, and therefore all models in this branch will behave in the same way at early times, regardless of the specific values for the other $\beta$s. We can then find
approximate equations of motion for super-horizon and sub-horizon scales when considering only the leading order terms in $1/N$ in eq.~(\ref{EqE1E2}).

\begin{description}

\item[1. Super-Horizon scales ($x=k\mathcal{H}^{-1} \ll 1$):]
the evolution equations reduce to
\begin{align}
& E_2''+ 2\mathcal{H}E_2'-\frac{x^2}{15}\mathcal{H}N^2 E_1'+3N^2\mathcal{H}^2(E_2-E_1)=0,\nonumber\\
& E_1''+ 10\mathcal{H}E_1'-\frac{5}{3}\mathcal{H}x^2 E_2'+15\mathcal{H}^2(E_1-E_2)=0,
\end{align}
and when considering only lowest orders in $x^2$ and $N$, the solutions are:
\begin{align}
&E_2=c_1+\frac{c_2}{\tau},\nonumber\\
&E_1=c_1+\frac{15}{7}\frac{c_2}{\tau}+c_\pm\tau^{n_\pm} \quad n_{\pm}=\frac{1}{2}(-11\pm \sqrt{21})<0,
\end{align} 
where $c_1$, $c_2$ and $c_\pm$ are some integration constants. As we can observe, in this regime both functions are decaying to the same constant $c_1$.

\item[2. Sub-Horizon scales ($x=k\mathcal{H}^{-1}\gg 1$):]
the evolution equations reduce to
\begin{align}
& E_2''+\frac{12\mathcal{H}}{x^2}E_2'-\frac{27}{2}\frac{N^2\mathcal{H}}{x^4}E_1'+\frac{x^2\mathcal{H}^2}{3}E_2-\frac{45}{2}\frac{N^2\mathcal{H}^2}{x^2}E_1=0,\label{EqE2SubHEarly}\\
& E_1''+6\mathcal{H}(E_1'-E_2')-\frac{5}{3}x^2\mathcal{H}^2 E_1 + 2x^2\mathcal{H}^2E_2=0, \label{EqE1SubHEarly}
\end{align}
and when considering only the highest orders in $x^2$ the solutions are:
\begin{align}
&E_2\propto e^{\pm ik\tau/\sqrt{3}}, \nonumber\\
&E_1=\frac{1}{(k\tau)^3}c_{\pm}e^{\pm \frac{\sqrt{15}}{3}k\tau } + E_2, \label{SolScalarFiniteRad}
\end{align}
where $c_\pm$ are some integration constants. We can see that $E_2$ is oscillating, while $E_1$ has an exponential instability.
\end{description}

\begin{figure}[H]
\begin{center}
\begin{subfigure}  
\centering
\includegraphics[scale=0.41]{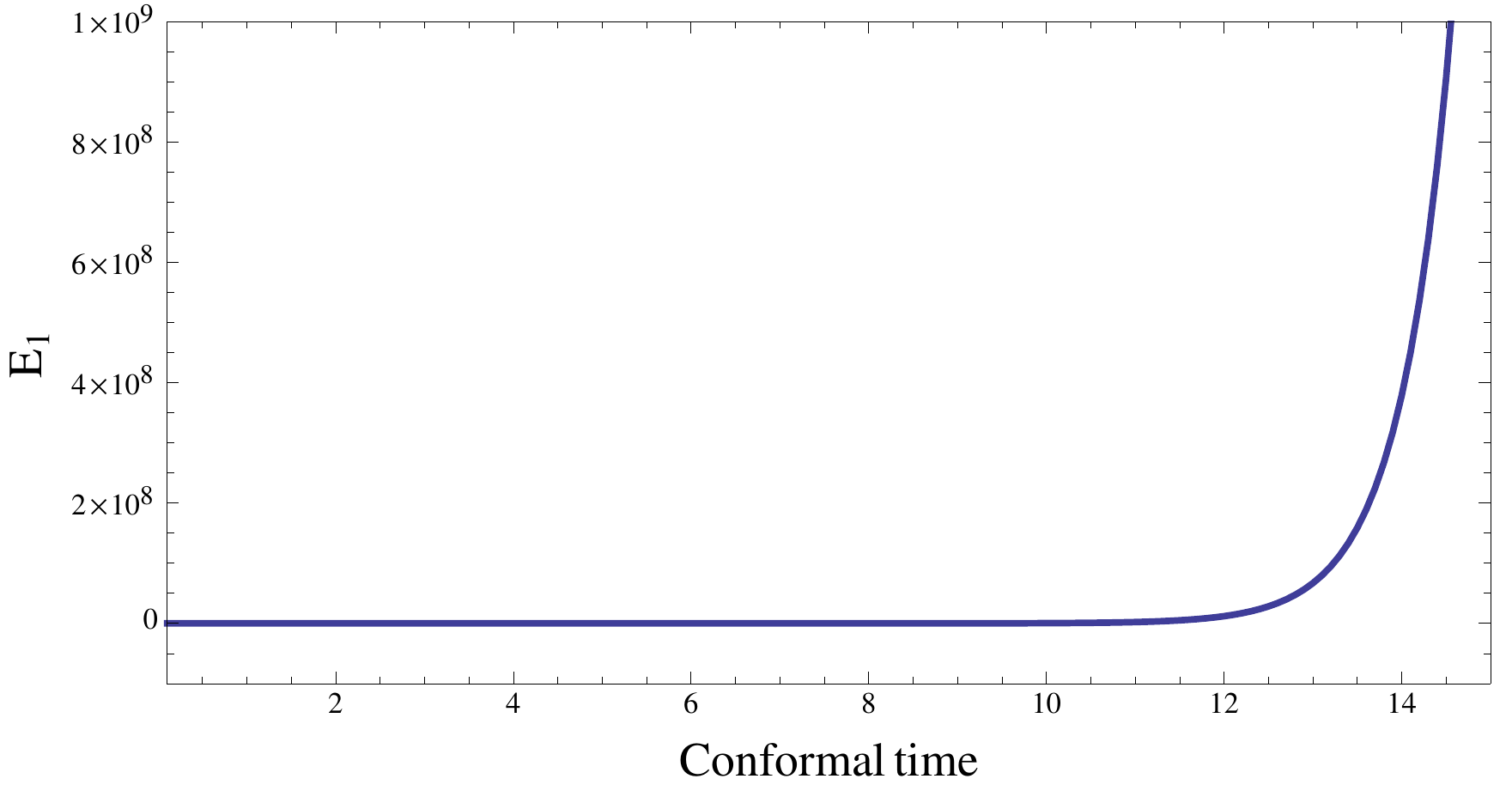}
 \end{subfigure}
\; \begin{subfigure}    
\centering
\includegraphics[scale=0.41]{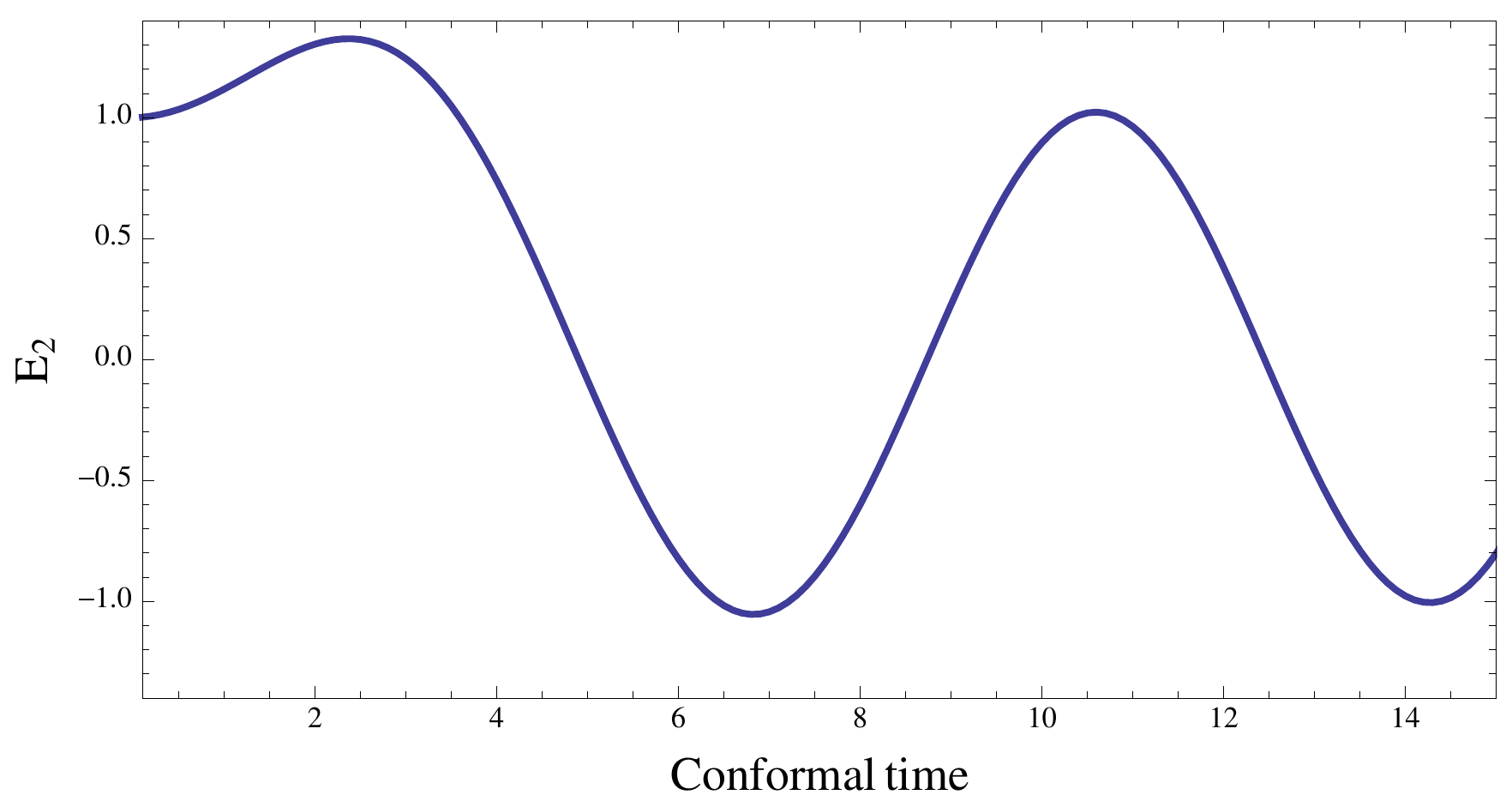}
 \end{subfigure}
\end{center}
\caption{Evolution of $E_1$ and $E_2$ as a function of $\tau$ for a given sub-horizon scales at early times during the radiation-dominated era. We have set $\beta_1m^2=10^{-2}$ and the other $\beta$s vanishing.}
\label{FigScalarFiniteRad} 
\end{figure}

We confirm the general behaviour previously described with numerical plots given in Fig.~\ref{FigScalarFiniteRad}, obtained solving the full equations of motion. In this figure we show the evolution of $E_1$ and $E_2$ as a function of conformal time (with arbitrary units) at early times during the radiation-dominated era for a given sub-horizon scale- we have set $m^2\beta_1=10^{-2}$, with the other $\beta$s vanishing, and arbitrary initial conditions of order 1 for both fields. For this plot and all the following numerical plots in this paper we will set $M_g=1$. As we expected, $E_2$ oscillates while $E_1$ grows exponentially fast, increasing its value in many orders of magnitude, and eventually breaking the validity of linear perturbations. Note that large scales will not be affected by the exponential growth as much as small scales, as the former ones enter the horizon later, and therefore, experience the exponential expansion for a shorter period. Note also that in eq.~(\ref{EqE1SubHEarly}), the exponential solution for $E_1$ is due to the minus sign in the coefficient $E_1$, which when calculated for a general $w$, will be negative for $w>-1/2$. Therefore, during the matter-dominated era, there will also be an exponential growth in $E_1$. 

\begin{figure}[H]
\begin{center}
\begin{subfigure}  
\centering
\includegraphics[scale=0.41]{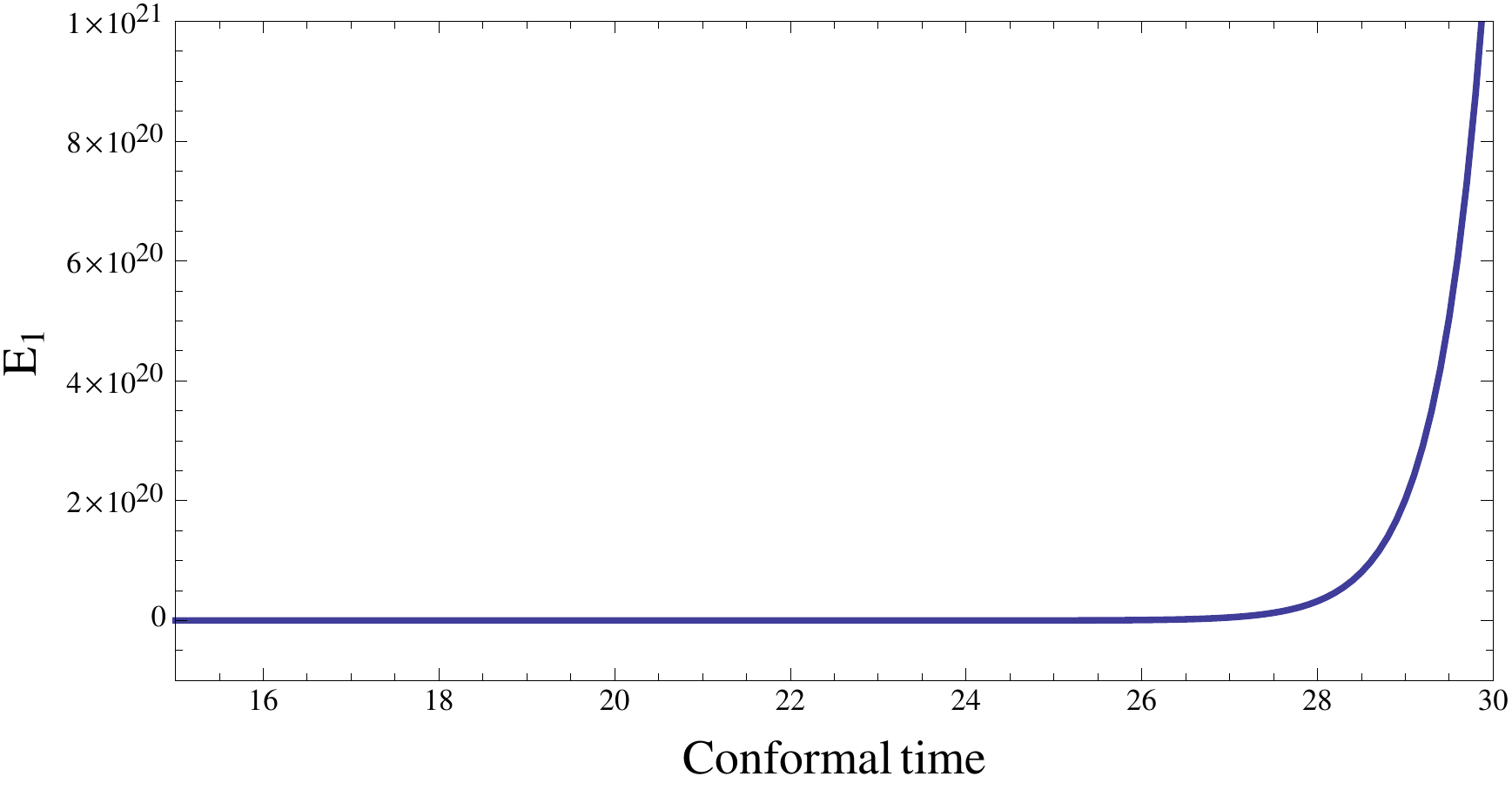}
 \end{subfigure}
\; \begin{subfigure}    
\centering
\includegraphics[scale=0.41]{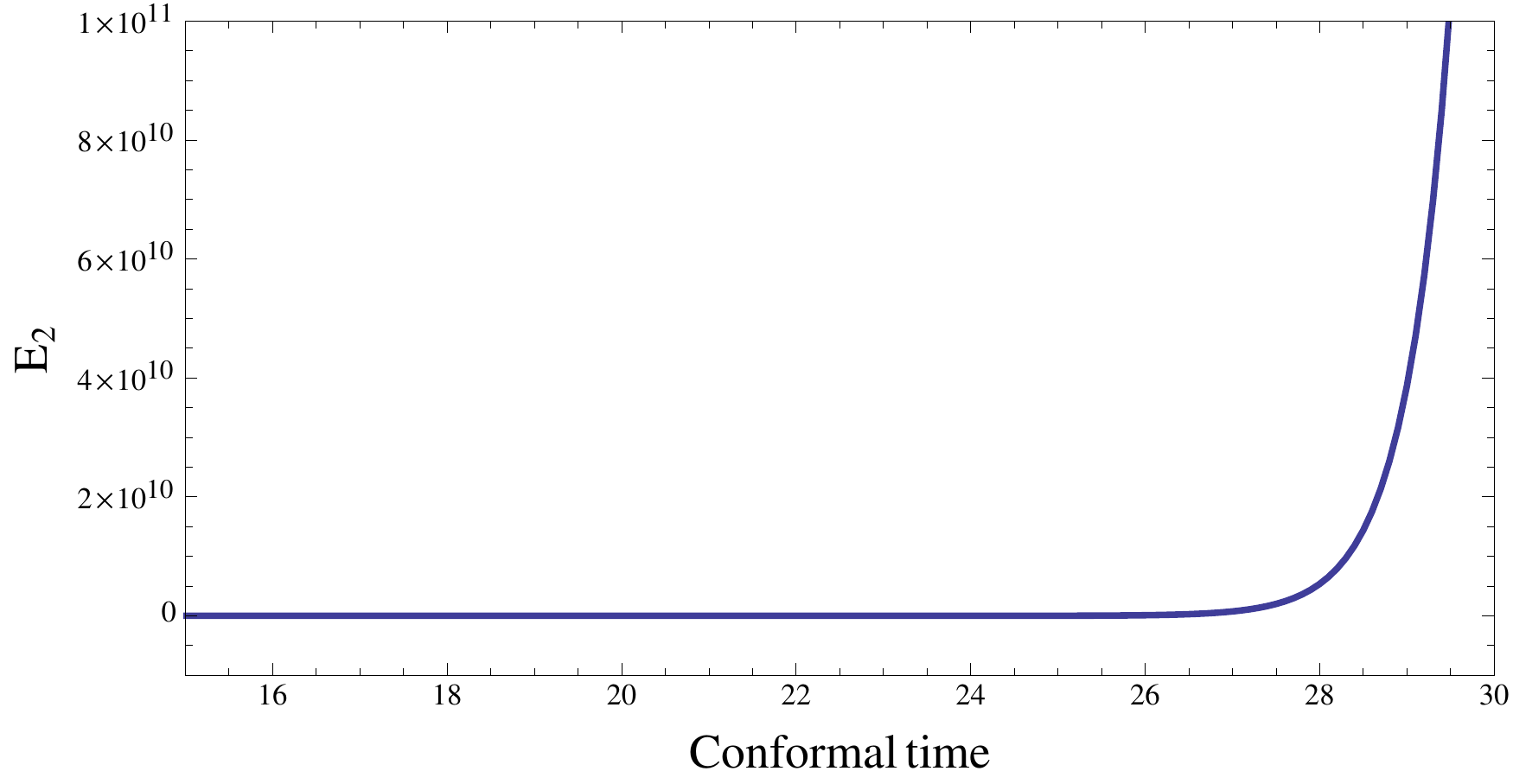}
 \end{subfigure}
\end{center}
\caption{Evolution of $E_1$ and $E_2$ as a function of $\tau$ in the radiation-dominated era for a given scale sub-horizon $k$. At later times, both perturbation fields are growing exponentially fast, becoming several orders of magnitude larger than their early time value.}
\label{FigMidTimes}
\end{figure}

Furthermore, we can see that in eq.~(\ref{EqE2SubHEarly}), we ignored the terms with $E_1$ and $E_1'$ to find the analytical solutions in eq.~(\ref{SolScalarFiniteRad}). However, as time goes on, $E_1$ will grow many orders of magnitude and it will not be possible to discard the coupling between the two fields; $E_1$ will feed back into the equation for $E_2$, making this latter field grow as well. Roughly, we expect that to happen when the terms for $E_1$ are larger than those of $E_2$ in eq.~(\ref{EqE2SubHEarly}), i.e.~when $x^7 e^{-\sqrt{15}x/3}\ll N^2$. Fig.~\ref{FigMidTimes} is a continuation of Fig.~\ref{FigScalarFiniteRad}, as it shows the evolution at later times, where we can see the unstable behaviour in $E_2$.

We have studied the behaviour of scalar perturbations at early times during the radiation-dominated era, showing that generically, there is an exponential instability at early times in both scalar perturbations. During the matter era, the same instability appears. This exponential growth breaks the validity of first order perturbations and therefore we cannot trust the results. This instability could correspond to an actual physical problem of the model, or could be cured by higher order perturbations. A further analysis is needed to understand the nature of this instability and what it tells us, more generally, about the theory.

\subsubsection{Late times}

At late times, the background will approach a de-Sitter phase, where $N\rightarrow \bar{N}$, $X\rightarrow 1$, $\tilde{Z}\rightarrow Z=\bar{Z}$, and $\mathcal{H}\rightarrow aH_0$, with $\bar{N}$, $\bar{Z}$ and $H_0$ constants. Notice that the exact value of $\bar{N}$ depends on the parameters $\beta$, and also
%As it was pointed out in \cite{Comelli:2012db}, during de-Sitter only one scalar degree of freedom propagates, so we expect to have one field freezing out.
\begin{align}
&\bar{Z}=\beta_1+2\beta_2\bar{N}+\beta_3\bar{N}^2,\nonumber\\
&H_0^2=\frac{1}{3}\frac{m^2}{\bar{N}}\left( \beta_1+3\beta_2\bar{N}+3\beta_3\bar{N}^2+\beta_4\bar{N}^3 \right),\nonumber\\
& a=\frac{1}{-H_0\tau},
\end{align} 
where, in these coordinates, the infinite future is characterised by $\tau\rightarrow 0$.

We now study the evolution for super-horizon and sub-horizon scales in this de-Sitter phase, assuming $w=0$.  
\begin{description}
\item[1. Super-horizon scales:]
the evolution equations are now
\begin{align}
& E_2''+\left(\frac{2\bar{N}^2+1}{\bar{N}^2+1}\right)\mathcal{H}E_2'-\left(\frac{\bar{N}^2}{\bar{N}^2+1}\right)\mathcal{H}E_1'+ q\bar{N} \mathcal{H}^2(E_2-E_1)=0,\nonumber\\
& E_1''+\left(\frac{\bar{N}^2+2}{\bar{N}^2+1}\right)\mathcal{H}E_1'-\left(\frac{1}{\bar{N}^2+1}\right)\mathcal{H}E_2'+\left(\frac{q}{\bar{N}}\right) \mathcal{H}^2 (E_1 -E_2)=0,
\end{align}
where $q\equiv m^2\bar{Z}/H_0^2$. These equations are solved by: 
\begin{align}
&E_1=c_0+c_1\tau^2+c_{\pm}\tau^{n_\pm},\\
&E_2=c_0+c_1\tau^2-\bar{N}^2c_{\pm}\tau^{n_\pm},
\end{align}
where $c_0$, $c_1$ and $c_\pm$ are some integration constants, and $n_\pm$ is such that $Re(n_\pm)>0$. Therefore, both functions decay to the same constant.

\item[2. Sub-horizon scales:] the evolution equations are now
\begin{align}
& E_2''+\mathcal{H}E_2'-\frac{9}{4}\frac{q[q(\bar{N}^2+1)-2\bar{N}]}{x^4} \mathcal{H} E_1'+\frac{1}{2} q \bar{N}\mathcal{H}^2(E_2-E_1)=0,\nonumber\\
& E_1''+6\mathcal{H}E_1'-5 \mathcal{H}E_2'+x^2\mathcal{H}^2 (E_1 -E_2)=0,
\end{align}
and when considering only the highest orders in $x^2$, the solutions are hypergeometric functions with power laws decaying to the same constant.
\end{description}

\begin{figure}[H]
\begin{center}
\begin{subfigure}  
\centering
\includegraphics[scale=0.427]{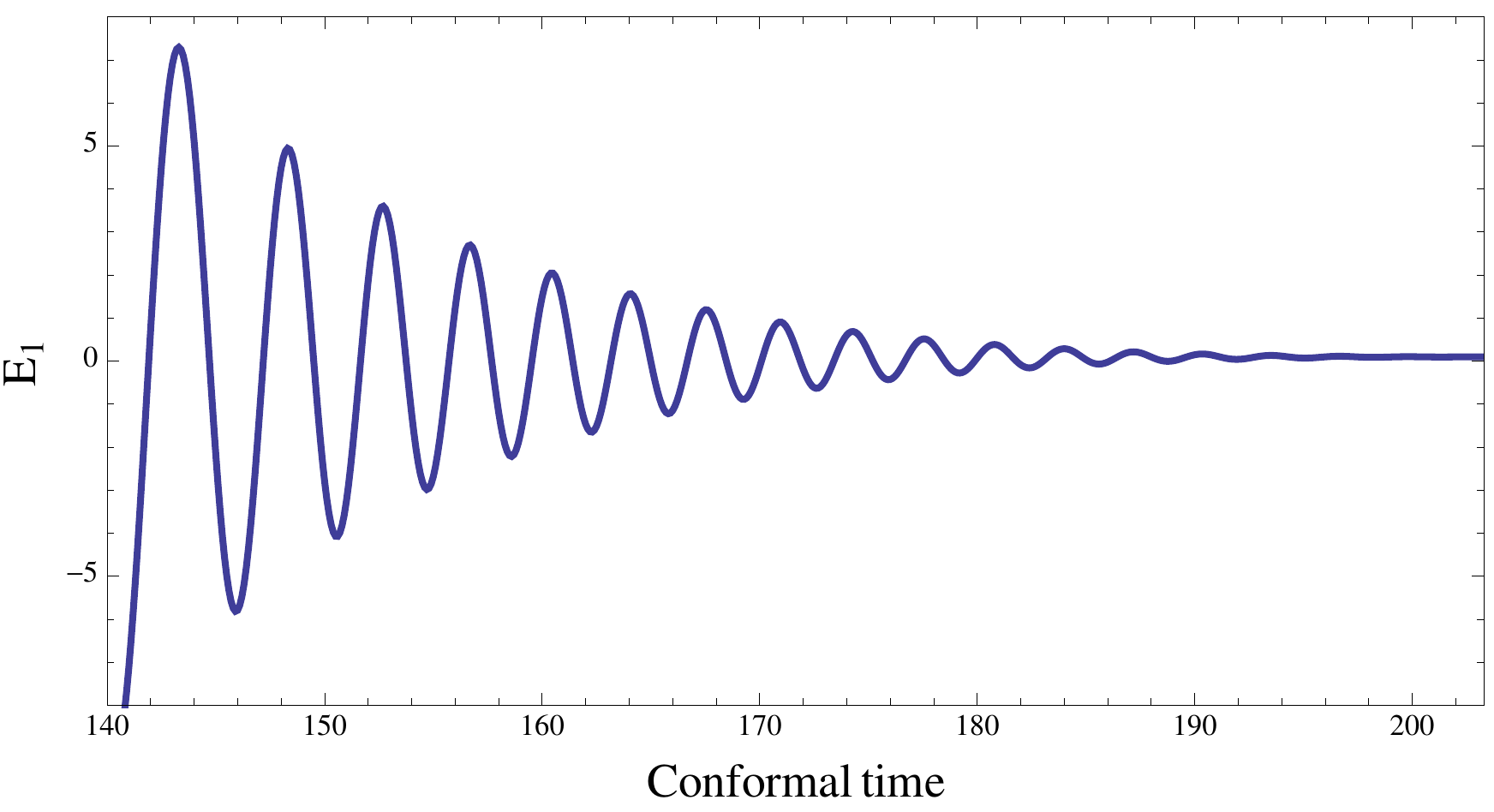}
 \end{subfigure}
 \begin{subfigure}    
\centering
\includegraphics[scale=0.43]{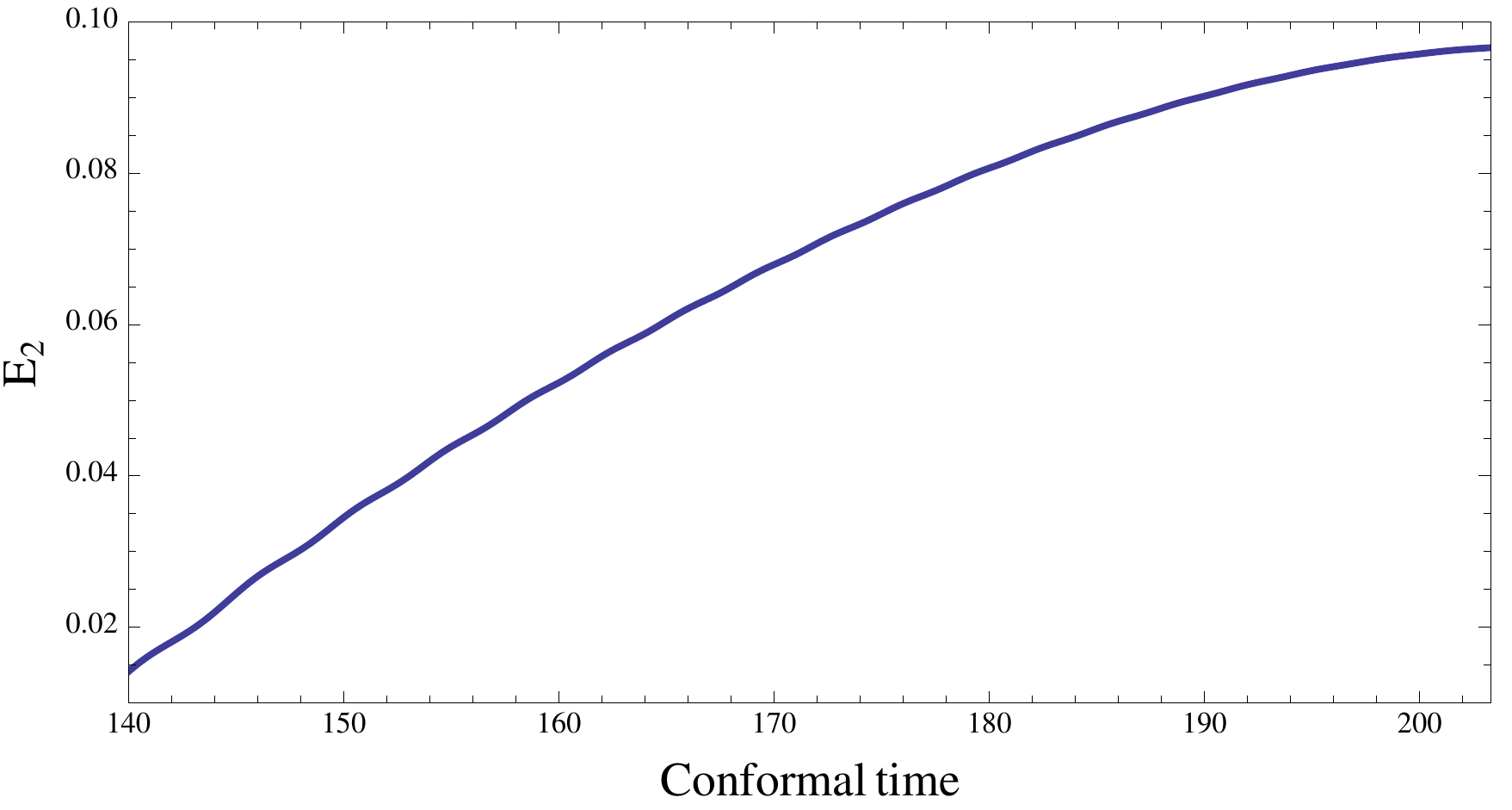}
 \end{subfigure}
\end{center}
\caption{Evolution of scalar perturbations as a function of the conformal time during the de-Sitter phase at late times in the matter-dominated era. }
\label{FigScalarDS}
\end{figure}

Figure \ref{FigScalarDS} shows numerical results on the evolution of both scalar perturbations in the de-Sitter phase in the matter-dominated era for a given sub-horizon scale. As in previous plots, we considered $m^2\beta_1=10^{-2}$ and the other $\beta$s vanishing, and arbitrary initial conditions of the same order for both fields. Here both fields are oscillating and approaching the same constant value.
% Since during early times both scalars fields grew exponentially fast, in these plots we have rescaled the late-time values of both fields as they originally were many orders of magnitude larger. 

\subsection{Bouncing branch}\label{SecScalarPert2}

In this subsection we will show some approximate analytical solutions for the two physical scalar fields in the bouncing branch. First of all, note that the differences between the background evolutions in the expanding and bouncing branches occur only at early times, as at late times in both cases the metrics will enter a de-Sitter phase. Consequently, in the bouncing branch the evolution of perturbations at late times is the same as in the expanding branch. For this reason, in this subsection we focus on early times only. It is relevant in this case to show the evolution during the radiation-dominated era and the matter-dominated era, as fields do not evolve in the same way in both stages.

In this branch we can have different background solutions depending on the parameter values. We will distinguish the following cases: (a) $\beta_3\not=0$; (b) $\beta_3=0$ and $(\beta_4-3\beta_2)\not=0$; (c) $\beta_3=0$ and $(\beta_4-3\beta_2)=0$; (d) $\beta_3=\beta_2=0$. All the viable solutions with other combinations of null parameters are contained in these cases. As stated in \cite{Konnig:2014xva}, only case (d) is physically possible, as all the other cases have an exponential instability for sub-horizon scales at early times, similar to the one found in the expanding branch. For this reason, from now on we study perturbations for case (d) only. For more details about the other cases see Appendix \ref{AppInstability}. 

For case (d), notice that at early times $N\gg 1$ and then eq.~(\ref{Density}) approximates to:
\begin{equation}\label{BounceCondB4}
\tilde{\rho}=N^2\beta_4,
\end{equation}
and therefore we need to impose $\beta_4>0$. Conditions on the remaining parameters $\beta_0$ and $\beta_1$ are also present, as at late times the Friedmann equation (\ref{EqFried}) becomes:
\begin{equation}\label{BounceCondB0}
\mathcal{H}^2 = \frac{a^2}{3}m^2\left(\beta_0 + 3\beta_1\bar{N} \right),
\end{equation}
where $\bar{N}$ is the late time value of the function $N$. Consequently, we also need to impose $\beta_0 + 3\beta_1\bar{N} >0$. In general, we could satisfy this condition when both $\beta$s are positive or when one of them is negative (for some appropriate values). However, as we will see later, cases with $\beta_1<0$ bring instabilities in the solutions for scalars, vectors and tensor perturbations during the radiation-dominated era. Therefore, from now on we will assume $\beta_1>0$.

\subsubsection{Early times radiation-dominated era}

At early times $N\gg 1$, and therefore we consider only leading order terms in $N$ in the equations of motion and we assume $w=1/3$. We again study the evolution in super-horizon and sub-horizon scales, focusing on case (d), where $\beta_3=\beta_2=0$. 

\begin{description}
\item[Super-horizon scales:] for super-horizon scales the equations become
\begin{align}
&E_2''+ 2\mathcal{H}E_2'+\frac{9}{2x^2}\frac{\mathcal{H}}{N}\frac{\beta_1}{\beta_4}E_1'-\frac{1}{3}x^2\mathcal{H}^2E_2-\frac{m^2\beta_1a^2N}{2}E_1=0,\\
&E_1''+6\frac{\beta_1}{\beta_4}\frac{\mathcal{H}}{N}\left(E_1'-\frac{x^2}{6}E_2'\right)+\frac{x^2\mathcal{H}^2}{3}E_1-2m^2\beta_1a^2N\frac{x^2}{3}E_2=0
\end{align}
and when keeping only the lowest orders of $x^2$ and the highest orders of $N$, the solutions are $E_2=c_1+c_2/\tau$ and $E_1=c_3+c_4 \mbox{erf}(p\tau)$, where $c_1$, $c_2$, $c_3$ and $c_4$ are some integration constants and $p^2=3\beta_1/(\beta_4N^2\tau^2)=const$. Therefore, both functions decay to a constant in this regime. 

Notice that if $\beta_1$ were negative, the solution for $E_1$ would be $E_1=c_3+c_4\mbox{erf}(i|p|\tau)$, which would grow exponentially fast, breaking the linear perturbation approximation.

\item[Sub-horizon scales:]
the evolutions equations are now
\begin{align}
&E_2''+\frac{12}{x^2}\mathcal{H}E_2'+\frac{27}{x^4}\frac{\mathcal{H}}{N}\frac{\beta_1}{\beta_4}E_1'+\frac{1}{3}x^2\mathcal{H}^2E_2-\frac{3m^2\beta_1a^2N}{x^2}E_1=0,\\
&E_1''+6\frac{\beta_1}{\beta_4}\frac{\mathcal{H}}{N}\left(E_1'-E_2'\right)+\frac{1}{3}x^2\mathcal{H}^2E_1-4m^2\beta_1a^2N E_2=0,
\end{align}
and when keeping only the terms of order $x^2$, the solutions are $E_i\propto e^{\pm i k\tau/\sqrt{3}}$. Unlike in the expanding branch, in this case scalar perturbations are well behaved.
\end{description}

\begin{figure}[H]
\begin{center}
\begin{subfigure}  
\centering
\includegraphics[scale=0.423]{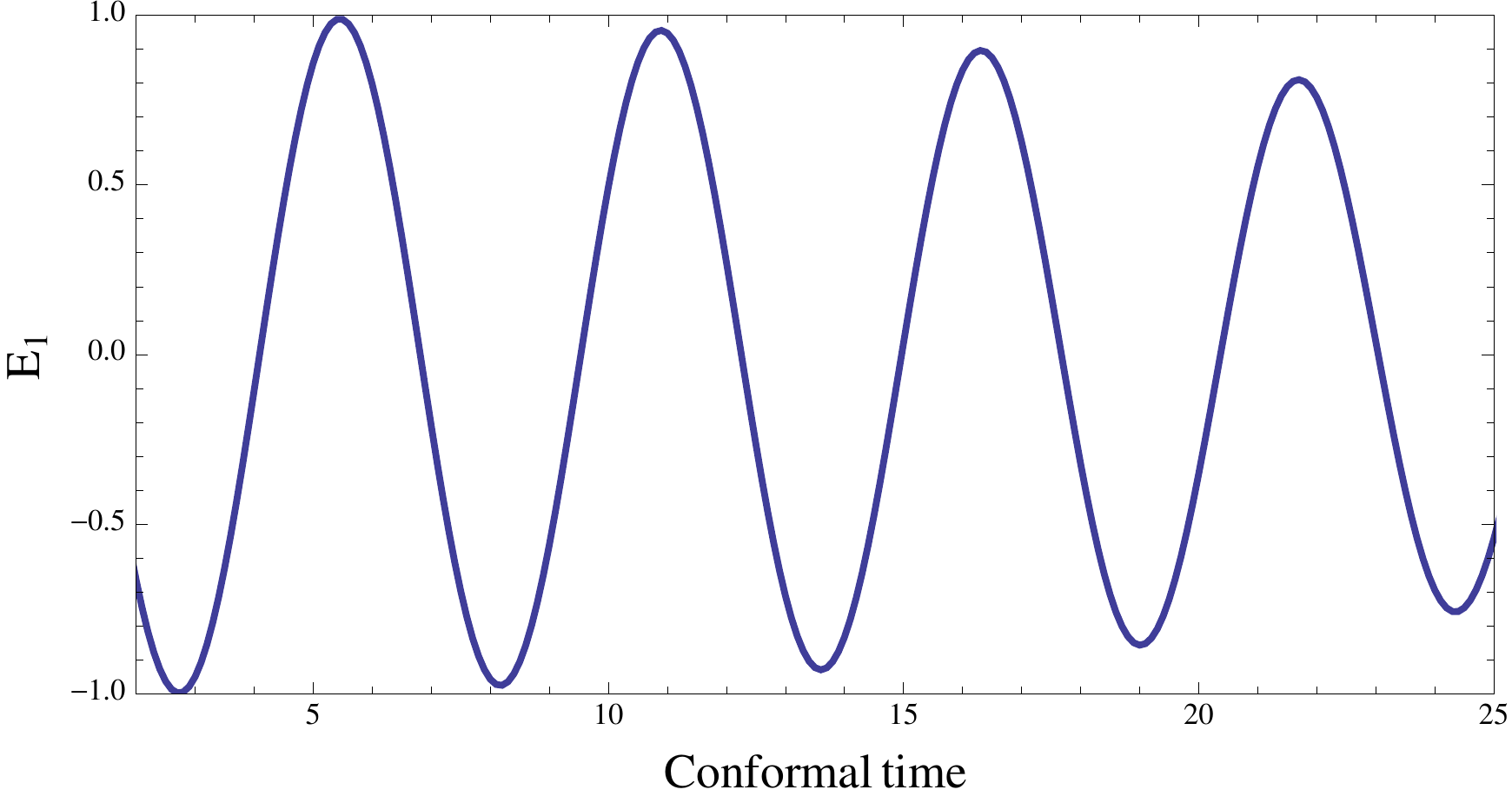}
 \end{subfigure}
\; \begin{subfigure}    
\centering
\includegraphics[scale=0.423]{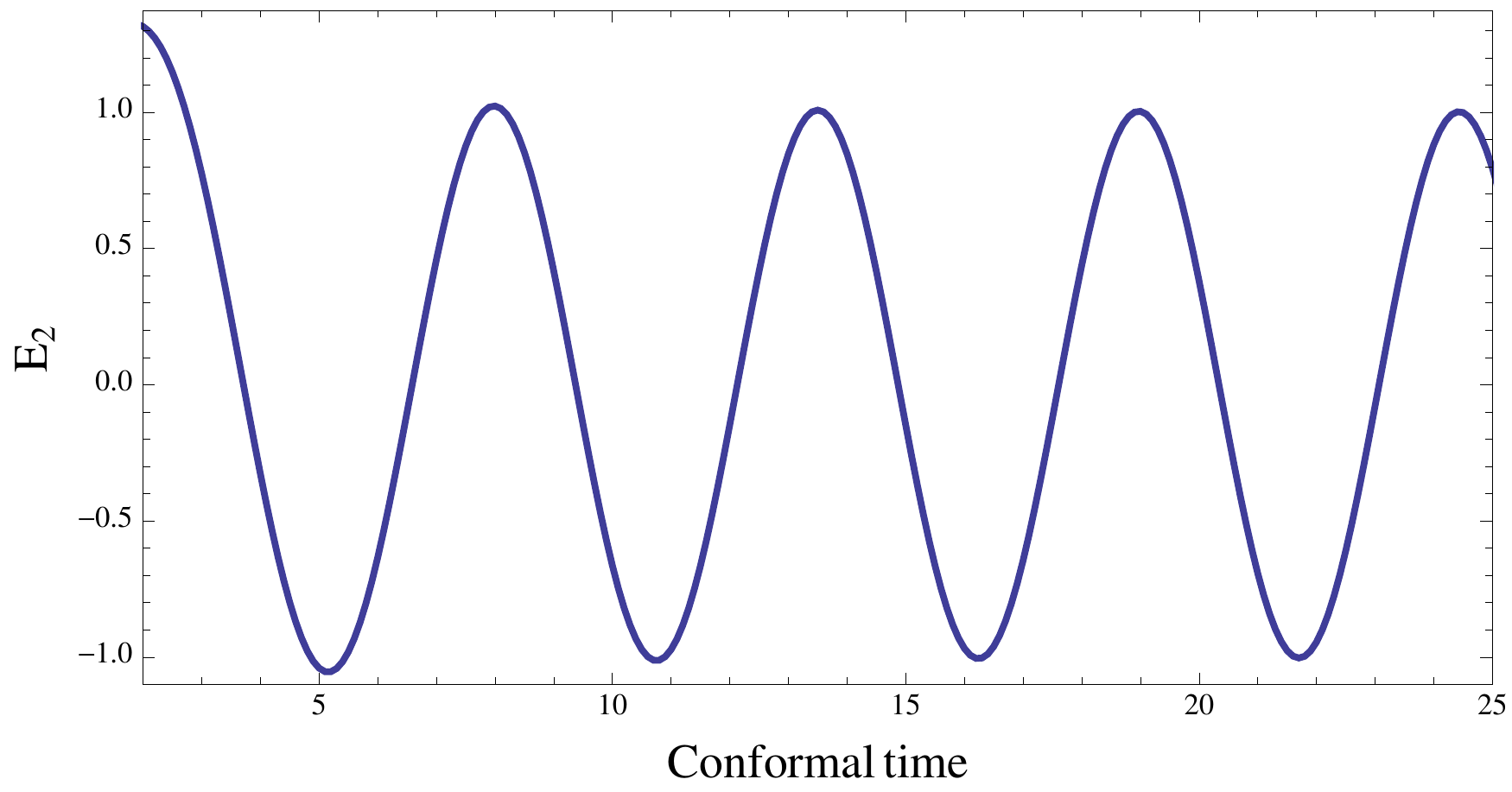}
 \end{subfigure}
\end{center}
\caption{Evolution of scalar perturbations as a function of $\tau$, during early times in the radiation-dominated for a given sub-horizon scales.}
\label{FigScalarInfiniteRad}
\end{figure}

Fig.~\ref{FigScalarInfiniteRad} shows numerical results for the evolution of scalar perturbations as a function of the conformal time (in arbitrary units), for a given sub-horizon scale during the radiation-dominated era at early times, confirming our previous analytical results. In this particular case we set $m^2\beta_1=m^2\beta_4=10^{-2}$, and arbitrary initial conditions of order 1 for both fields.

\subsubsection{Early times matter-dominated era}

As above, let us consider only leading terms in $N$ but now assume $w=0$.
\begin{description}
\item[Super-horizon scales:]
the evolution equations are
\begin{align}
& E_2''+2\mathcal{H}E_2'-2\frac{\mathcal{H}}{N}\frac{\beta_1}{\beta_4}E_1'-\frac{1}{3}x^2\mathcal{H}^2E_2-m^2\beta_1a^2N E_1=0,\\
& E_1''+\frac{5}{2}\mathcal{H}E_1'-\mathcal{H}\frac{x^2}{3}E_2'+\frac{5}{6}x^2\mathcal{H}^2E_1-\frac{1}{3}x^2\mathcal{H}^2E_2=0,
\end{align}
and when keeping only terms with the lowest orders in $x^2$ (and highest powers in $N$) we get: $E_i=c_{1i}+c_{2i}/\tau^{n_i}$, where $c_{1i}$ and $c_{2i}$ are some integration constants, and $n_1=4$ and $n_2=3$.

\item[Sub-horizon scales]
the evolution equations now reduce to
\begin{align}
& E_2''+\mathcal{H}E_2'+\frac{27}{2x^4}\frac{\mathcal{H}}{N}\frac{\beta_1}{\beta_4}E_1'-\frac{3}{2}\mathcal{H}^2E_2-\frac{m^2\beta_1a^2N}{2}E_1=0,\\
& E_1''+\frac{3}{2}\mathcal{H}E_1'-\frac{1}{2}\mathcal{H}E_2'+\frac{1}{2}\mathcal{H}^2x^2E_1-\mathcal{H}^2E_2=0,
\end{align}
and when considering only the highest orders in $x^2$ (and highest powers in $N$) the solutions are $E_1\propto e^{\pm i k\tau/\sqrt{2}}$ and $E_2=c_1/\tau^3+c_2\tau^2$, where $c_1$ and $c_2$ are some integration constants. Here we can see that $E_2$ grows as a power law in time, which will affect $E_1$ at later times, where this one will also start to grow as a power law. 
\end{description}

\begin{figure}[H]
\begin{center}
\begin{subfigure}  
\centering
\includegraphics[scale=0.422]{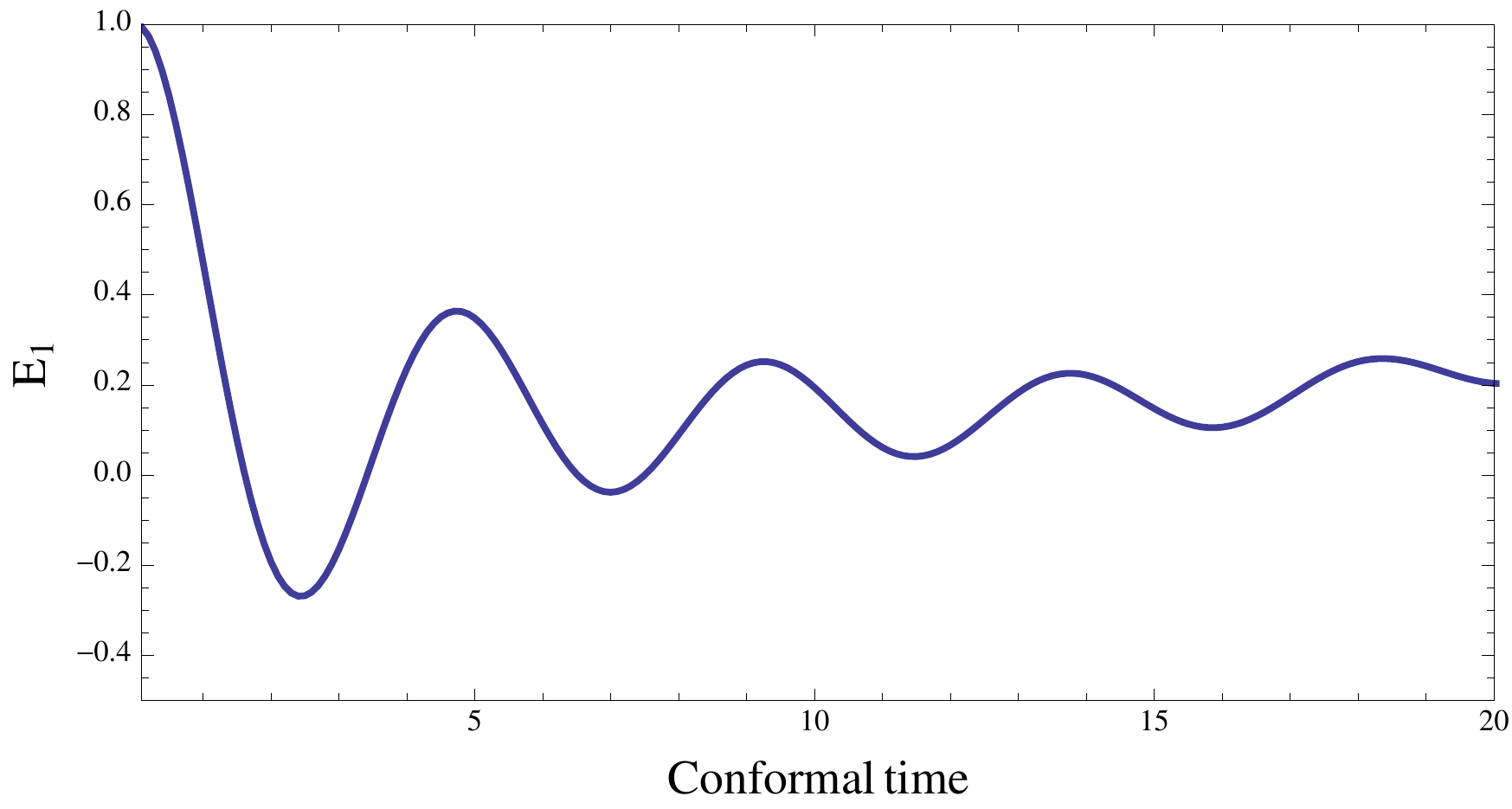}
 \end{subfigure}
\; \begin{subfigure}    
\centering
\includegraphics[scale=0.422]{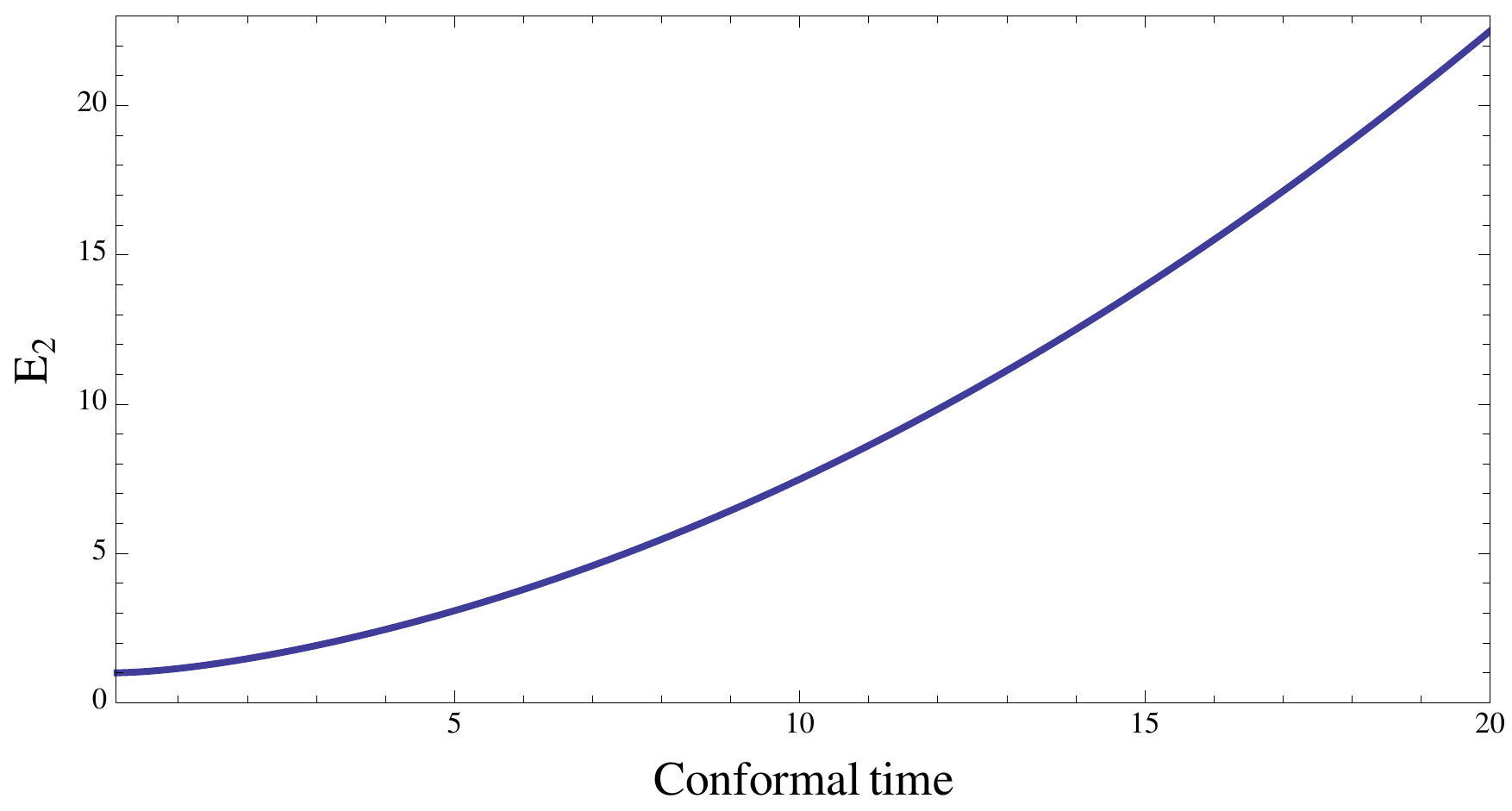}
 \end{subfigure}
\end{center}
\caption{Evolution of scalar perturbations as a function of $\tau$, during early times in the matter-dominated era for a given sub-horizon scale.}
\label{FigScalarInfiniteMat}
\end{figure}

Fig.~\ref{FigScalarInfiniteMat} shows numerical solutions for both scalar fields for a given sub-horizon scale during early times in the matter-dominated era. In this case we set $m^2\beta_1=m^2\beta_4=10^{-2}$, and arbitrary initial conditions of order one for both fields. As found in the analytical solutions, $E_1$ oscillates while $E_2$ grows as a power law. 

\begin{figure}[H]
\begin{center}
\begin{subfigure}  
\centering
\includegraphics[scale=0.423]{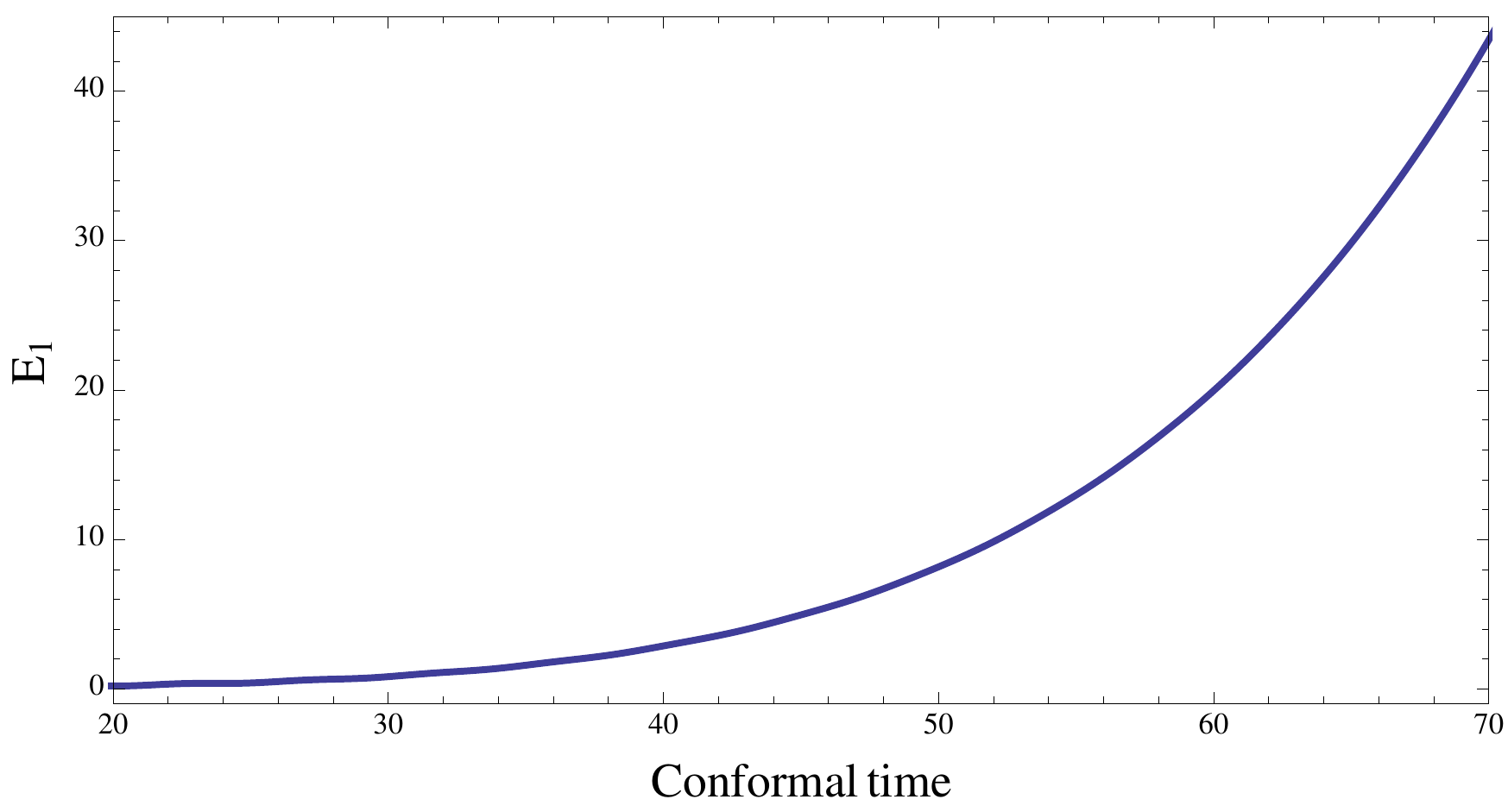}
 \end{subfigure}
\; \begin{subfigure}    
\centering
\includegraphics[scale=0.423]{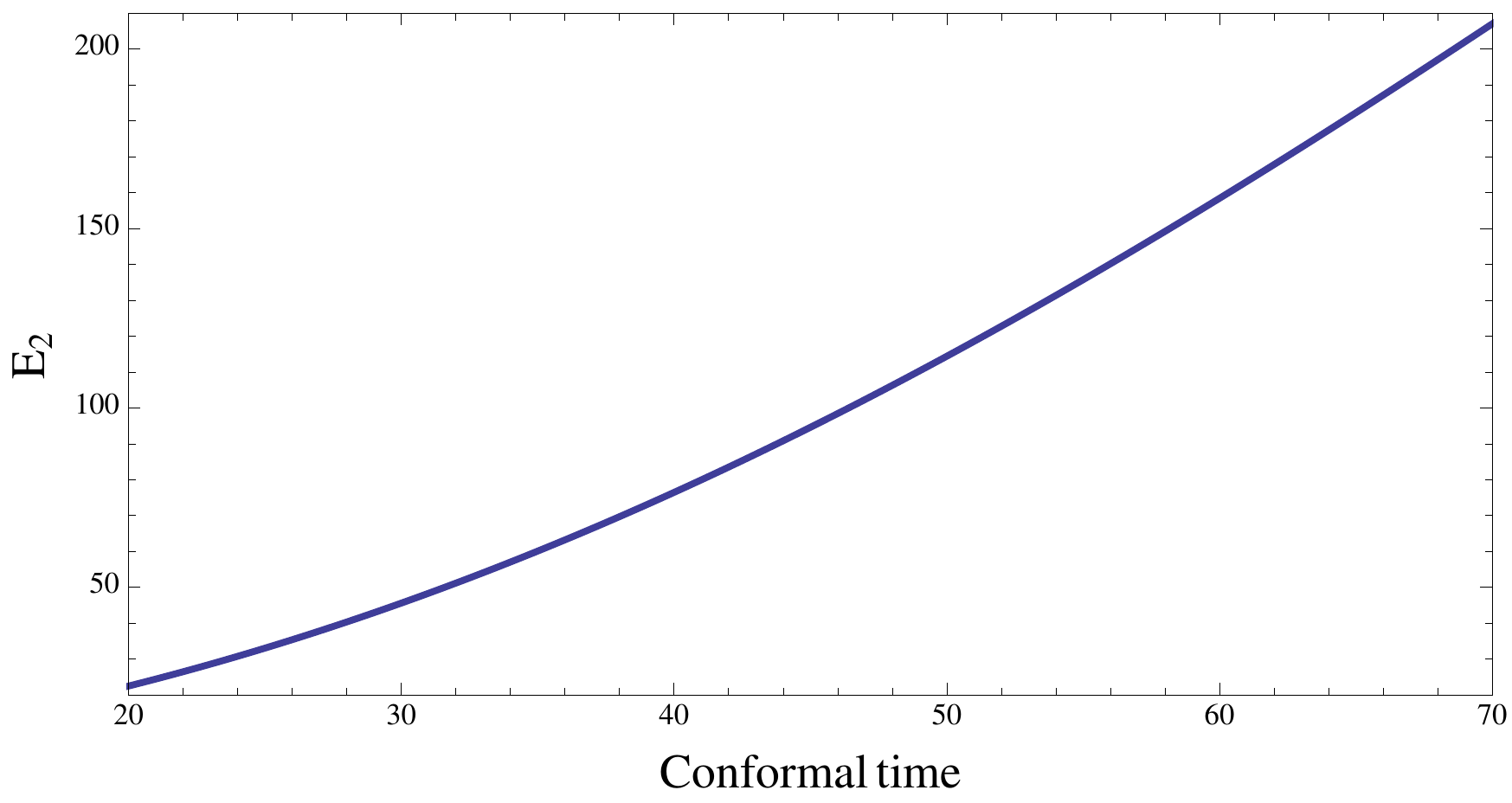}
 \end{subfigure}
\end{center}
\caption{Evolution of scalar perturbations during early times in the matter-dominated for a given sub-horizon scale. }
\label{FigScalarInfiniteMat2}
\end{figure}

Analogous to the results for the expanding branch, in this case the quadratic growth in $E_2$ will affect $E_1$ at later times, making the latter field grow as a power law as well, as we observe in Fig.~\ref{FigScalarInfiniteMat2} (this figure is a continuation of Fig.~\ref{FigScalarInfiniteMat}). 

In addition, we can study the evolution of the gauge-invariant form for the density contrast $\delta_{GIk}=\delta \rho_{GI}/\rho_0$,
\begin{equation}
\delta_{GIk}=[\delta\rho+\rho_0'(B_2-E_2')]/\rho_0,
\end{equation}
where $\delta \rho$ is given by the $\delta T^0 {}_0$ in eq.~(\ref{pset}). After fixing the gauge, and eliminating the auxiliary variables, $\delta_{GIk}$ can be expressed entirely in terms of $E_i$ and $E_i'$ (see Appendix \ref{AppDensityContrast}). In Fig.~\ref{DensityContrastPlot} we see numerical results for the evolution of $d \ln \delta_{GIk}/ d\ln a$ as a function of the conformal time (in arbitrary units) for a given sub-horizon scale during the matter-dominated era. In this case we have also set $m^2\beta_1=m^2\beta_4=10^{-4}$. We observe that at early times $\delta_{GIk}$ grows nearly proportional to the scale factor $a$, and then it starts decaying faster as we enter into the de-Sitter phase, analogously to GR. A more detailed study on the comparison of this model with observations was done in \cite{Konnig:2014xva}.

\begin{figure}[H]
\centering
\includegraphics[scale=0.45]{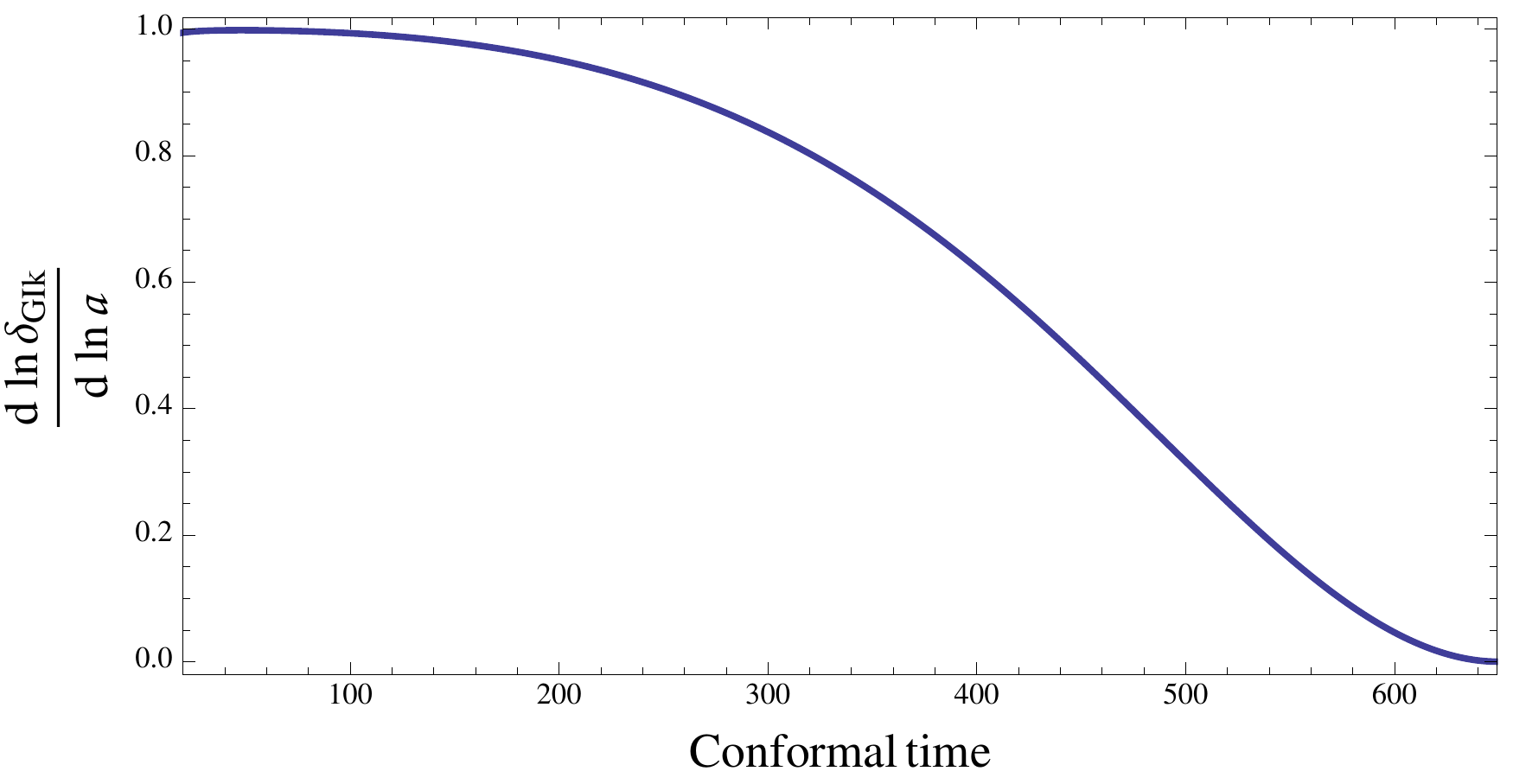}
\caption{Evolution of density constrast as a function of the conformal time $\tau$ in the matter-dominated era for a given sub-horizon scale.}
\label{DensityContrastPlot}
\end{figure}

It is important to remark that even though classical scalar fields do not evidence exponential instabilities in this branch, they do not satisfy the Higuchi bound (see Appendix \ref{AppHiguchi} for details), and therefore one scalar field propagates as a ghost, i.e.~with a negative kinetic term. Consequently, instabilities might appear when studying higher order perturbations, and negative norm states would appear when quantising the linear theory massive gravity (see \cite{Higuchi:1986py}).

%--------------------------------------------------------
\section{Vector perturbations}
\label{SecVectorPert}
Analogously to the previous section, we now study the evolution of vector perturbations in different regimes, by making relevant approximations to the full equations of motion given by eq.~(\ref{EqvT})-(\ref{EqF2}).
\subsection{Expanding branch}
Recall that the expanding branch is characterised by $N\ll 1$ at early times and a de-Sitter phase at late times. 
\subsubsection{Early times radiation-dominated era}
Considering $w=1/3$ and leading order terms in $1/N$, the equations for vector perturbations become:
\begin{align}
&F_{2i}''+ \frac{2(4x^4+33x^2+40)\mathcal{H}}{(8+x^2)(x^2+5)}F_{2i}' -\frac{16(3x^2+20)\mathcal{H}}{(8+x^2)(x^2+5)}v^T_i+3(x^2+5)\mathcal{H}^2F_{2i}=0, \\
& v^{T'}_i+\frac{8(8x^2+50)\mathcal{H}}{(8+x^2)(x^2+5)}v^T_i-\frac{2(4x^2+25)\mathcal{H}x^2}{(8+x^2)(x^2+5)}F_{2i}'-3(x^2+5)\mathcal{H}^2F_{2i}=0,
\end{align}
where $x=k/\mathcal{H}$.
\begin{description}
\item[Super-horizon scales:]
the evolution equations reduce to
\begin{align}
& F''_{2i}+2\mathcal{H}F'_{2i}-8\mathcal{H}v^T_i+15\mathcal{H}^2F_{2i}=0,\\
&v^{T'}_i+10\mathcal{H}v^T_i-\frac{5}{4}x^2\mathcal{H}F'_{2i}-15\mathcal{H}^2F_{2i}=0,
\end{align}
and, ignoring terms of order $x^{2}$, the solutions are $F_{2i}=c_1/\tau+c_{\pm}\tau^{n_\pm}$ and $v^T_i=c_2/\tau^2+b_\pm \tau^{n_\pm}$, where $n\pm<0$, and where $c_1$, $c_2$, $c_\pm$ and $b_\pm$ are some integration constants related to each other. Therefore, both vector perturbations decay to zero in this regime.
\item[Sub-horizon scales:]
the evolution equations reduce to
\begin{align}
& F''_{2i}+8\mathcal{H}F'_{2i}-\frac{48}{x^2}\mathcal{H}v^T_i+3x^2\mathcal{H}^2F_{2i}=0,\\
&v^{T'}_i+\frac{64}{x^2}\mathcal{H}v^T_i-8\mathcal{H}F'_{2i}-3x^2\mathcal{H}^2F_{2i}=0,
\end{align}
and when ignoring terms of order $x^{-2}$, the solutions are
\begin{align}
&F_{2i}\propto e^{\pm ik\sqrt{3}\tau}/\tau^4,\\
&v^T_i=c_1-c_{\pm}e^{\pm ik\sqrt{3}\tau}/\tau^4,
\end{align} 
where $c_1$ and $c_\pm$ are come integration constants related to those of $F_{2i}$. Therefore, in this regime both functions decay as $a^4$.
\end{description}

\begin{figure}[H]
\begin{center}
\begin{subfigure}  
\centering
\includegraphics[scale=0.431]{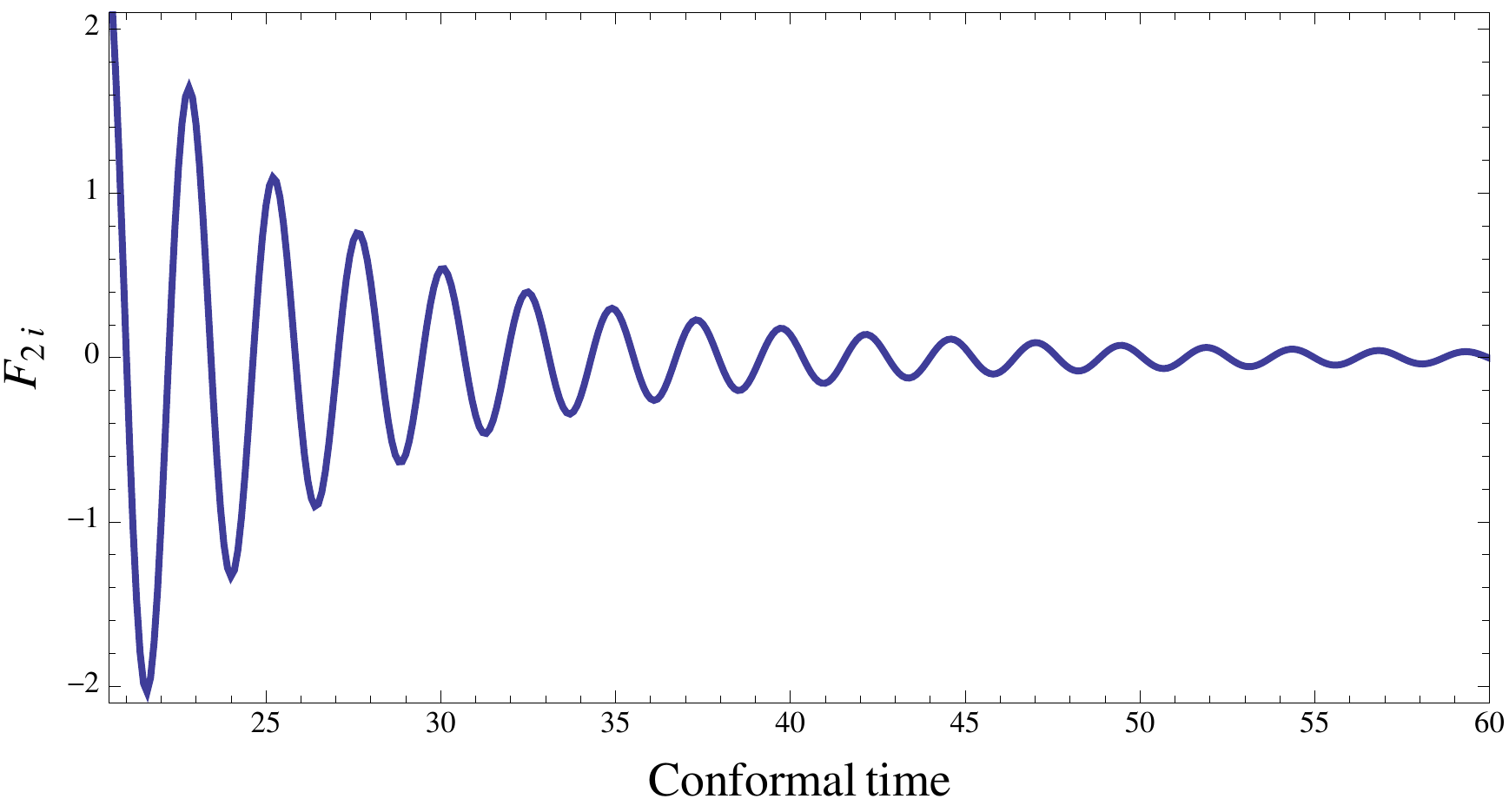}
 \end{subfigure}
 \begin{subfigure}    
\centering
\includegraphics[scale=0.432]{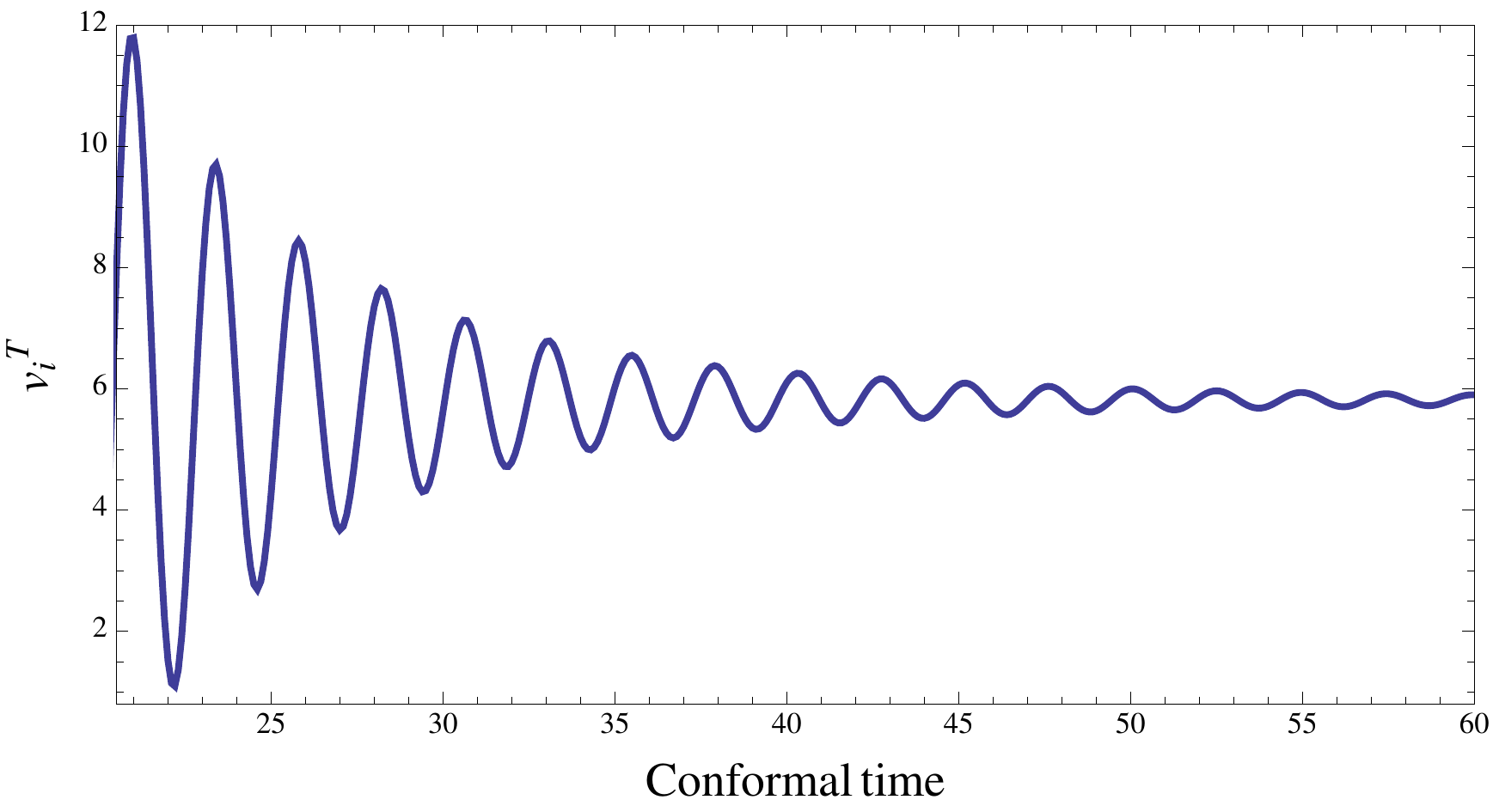}
 \end{subfigure}
\end{center}
\caption{Evolution of vector perturbations as a function of $\tau$, during early times in the radiation-dominated era for a given sub-horizon scale.}
\label{PlotVecFinRad}
\end{figure}

Figure \ref{PlotVecFinRad} shows numerical results for the evolution of vector perturbations as a function of $\tau$, during early times in the radiation-dominated era for a given sub-horizon scale; we have set $m^2\beta_1=10^{-2}$ while all other $\beta$s are vanishing, and we have chosen arbitrary initial conditions of the same order for both fields. We can clearly see that both fields decay in the same way, but while $F_{2i}$ is oscillating around 0, $v^T_i$ oscillates around a constant value. We find similar behaviour the matter-dominated era. 

\subsubsection{Late times}
We now assume $w=0$ and a de-Sitter space-time where $N$ takes the constant value $\bar{N}$, and $a\propto 1/\tau$, with $\tau\rightarrow 0$ being the infinite future.

\begin{description}
\item[Super-horizon scales:]
the evolution equations are 
\begin{align}
&F_{2i}''+ 2\mathcal{H}F_{2i}' +a^2x_2F_{2i}=0, \; x_2=m^2\bar{Z}(\bar{N}^2+1)/\bar{N},\\
& v^{T'}_i+\mathcal{H}v^T_i-\frac{1}{\bar{N}^2+1}\mathcal{H}F_{2i}'-a^2x_1F_{2i}=0, \; x_1=m^2\bar{Z}/\bar{N},
\end{align}
and are solved by $F_{2i}\propto \tau^{n_\pm}$ and $v^T_i=c_1\tau+c_\pm \tau^{n\pm-1}$; $Re(n_\pm)>1$, while $c_1$ and $c_\pm$ are integration constants related to those of $F_{2i}$.
\item[Sub-horizon scales:] the evolution equations reduce to
\begin{align}
&F_{2i}''+ 4\mathcal{H}F_{2i}' +x^2\mathcal{H}^2F_{2i}=0, \\
& v^{T'}_i+\mathcal{H}v^T_i-3\mathcal{H}F_{2i}'-x^2\mathcal{H}^2F_{2i}=0,
\end{align}
and are solved by $F_{2i}\propto \tau^2 e^{\pm i k\tau}$ and $v^{T}_i=c_1\tau+c_{\pm}\tau^2e^{\pm i k\tau}$, where $c_1$ and $c_\pm$ are integration constants related to those of $F_{2i}$. In this case, both perturbations are decaying.

\end{description}

\begin{figure}[H]
\begin{center}
\begin{subfigure} 
\centering
\includegraphics[scale=0.427]{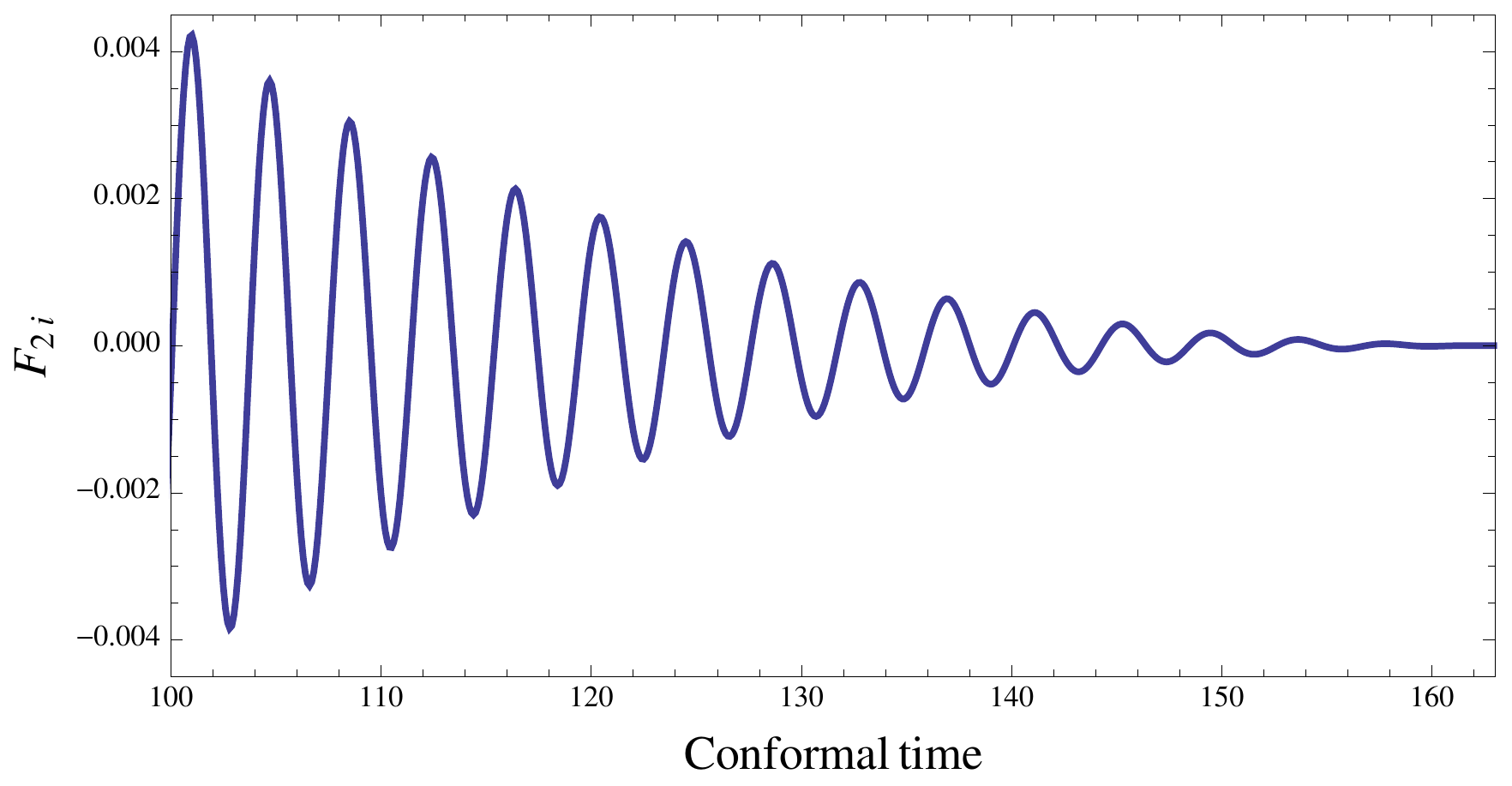}
 \end{subfigure}
\begin{subfigure}    
\centering
\includegraphics[scale=0.427]{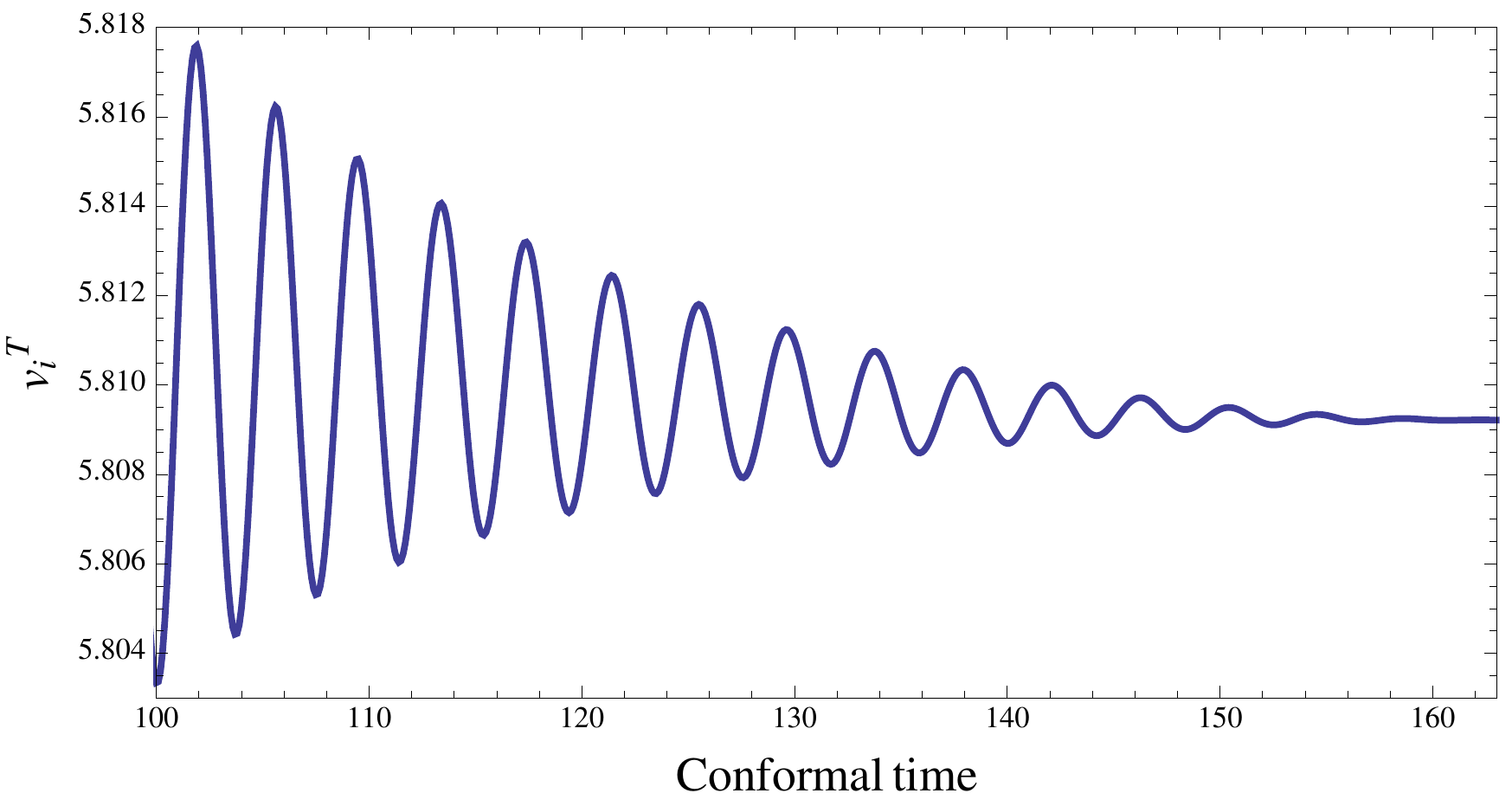}
 \end{subfigure}
\end{center}
\caption{Evolution of vector perturbations during late times in the de-Sitter phase for a sub-horizon scale.}
\label{PlotVecFinDS}
\end{figure}

Figure \ref{PlotVecFinDS} shows numerical solutions for the evolution of vector perturbations as a function of $\tau$, during late times for a given sub-horizon scale. Again, we have set $m^2\beta_1=10^{-2}$ and all other $\beta$s vanishing, and arbitrary initial conditions of the same order for both fields. Both fields oscillate and decay in the same way, but while $F_{2i}$ is oscillating around 0, $v^T_i$ oscillates around a decaying function. 

\subsection{Bouncing branch}\label{SubSecBounceVector}
As we have mentioned before, the bouncing branch is characterised by $N\gg 1$ at early times and a de-Sitter phase at late times. Next, we study the evolution of vector perturbations at early times in the same way we previously did for scalar perturbations.

\subsubsection{Early times radiation-dominated era}
We will start by assuming $w=1/3$. When considering only leading terms in $N$, the equations of motion become:
\begin{align}
& F_{2i}''+\frac{20\mathcal{H}}{(x^2+10)}F_{2i}'+\frac{16\mathcal{H}}{x^2+10}v^T_i+\frac{3\beta_1}{2\beta_4}\frac{\mathcal{H}^2}{N}(2+x^2)F_{2i}=0,\\
&v_i^{T'}+\frac{8}{3}\frac{\mathcal{H}}{N}\frac{(9\beta_1^2-4\beta_0\beta_4)}{\beta_1\beta_4(x^2+10)}v^T_i-\frac{x^2}{3}\frac{\mathcal{H}}{N}\frac{(9\beta_1^2-4\beta_0\beta_4)}{\beta_1\beta_4(x^2+10)} F_{2i}'-\frac{3\beta_1}{2\beta_4}\frac{x^2\mathcal{H}^2}{N}F_{2i}=0.
\end{align}
We now study these equations for sub-horizon and super-horizon scales.
\begin{description}
\item[Super-horizon scales:] the evolution equations reduce to
\begin{align}
& F_{2i}''+2\mathcal{H}F_{2i}'+\frac{8}{5}\mathcal{H} v^T_i+\frac{3\beta_1}{\beta_4}\frac{\mathcal{H}^2}{N}F_{2i}=0,\\
&v_i^{T'}+\frac{4}{15}\frac{\mathcal{H}}{N}\frac{(9\beta_1^2-4\beta_0\beta_4)}{\beta_1\beta_4}v^T_i-\frac{x^2}{30}\frac{\mathcal{H}}{N}\frac{(9\beta_1^2-4\beta_0\beta_4)}{\beta_1\beta_4} F_{2i}'-\frac{3\beta_1}{2\beta_4}\frac{x^2\mathcal{H}^2}{N}F_{2i}=0.
\end{align}
Ignoring terms of order $x^2$ and lowest order terms of $N$, the solutions are $F_{2i}=c_1+c_2/\tau$ and $v^T_i\propto e^{-p^2\tau^2}$, where $c_1$ and $c_2$ are some integration constants, and $p^2=2(9\beta_1^2-4\beta_0\beta_4)/(15\beta_1\beta_4N\tau^2)=const$. Notice that here we have assumed that $(9\beta_1^2-4\beta_0\beta_4)/(\beta_1\beta_4)>0$, since otherwise $v^{iT}$ would grow exponentially fast, creating an instability in the solutions.

\item[Sub-horizon scales:] the evolution equations reduce to
\begin{align}
& F_{2i}''+\frac{20}{x^2}\mathcal{H}F_{2i}'+\frac{16\mathcal{H}}{x^2}v^T_i+\frac{3\beta_1}{2\beta_4}\frac{\mathcal{H}^2x^2}{N}F_{2i}=0,\\
&v_i^{T'}+\frac{8}{3}\frac{\mathcal{H}}{N}\frac{(9\beta_1^2-4\beta_0\beta_4)}{\beta_1\beta_4x^2}v^T_i-\frac{1}{3}\frac{\mathcal{H}}{N}\frac{(9\beta_1^2-4\beta_0\beta_4)}{\beta_1\beta_4} F_{2i}'-\frac{3\beta_1}{2\beta_4}\frac{x^2\mathcal{H}^2}{N}F_{2i}=0.\label{EqvTBounceRad}
\end{align}
Considering only highest order terms in $x^{2}$, the solutions are $F_{2i}\propto e^{\pm iK\tau^2/2}/\sqrt{\tau}$ and $v^T_i\propto e^{\pm iK\tau^2/2}\sqrt{\tau}$, where $K^2=\frac{3\beta_1}{2\beta_4}\frac{k^2}{N\tau^2}$. We then see that, contrary to GR, $F_{2i}$ decays but the vorticity field $v^T_i$ grows. This modification happens as the dominant term in eq.~(\ref{EqvTBounceRad}) corresponds to the interaction term with $F_{2i}$ instead of the term with $v^{T}_i$. % and furthermore, the factor $1/N\propto \tau^2$ on this term determines the positive power on time on the solution for $v^T_i$.

Notice that if $\beta_1$ were negative, solutions for $F_{2i}$ and $v^{iT}$ would be combinations of Bessel I and K functions, which would grow exponentially fast, creating an instability in the solutions.

\end{description}

\begin{figure}[H]
\begin{center}
\begin{subfigure}  
\centering
\includegraphics[scale=0.427]{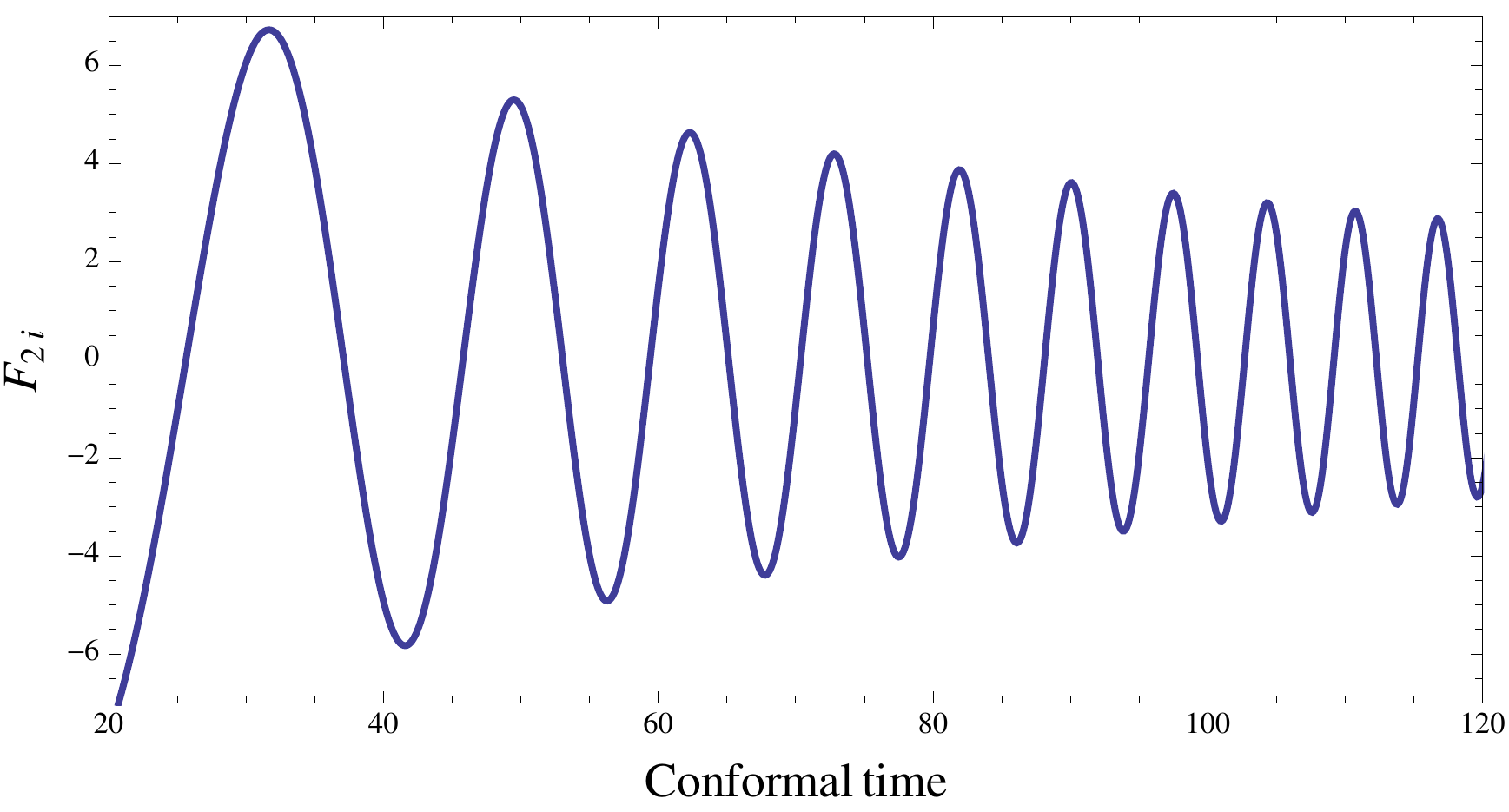}
 \end{subfigure}
\begin{subfigure}    
\centering
\includegraphics[scale=0.427]{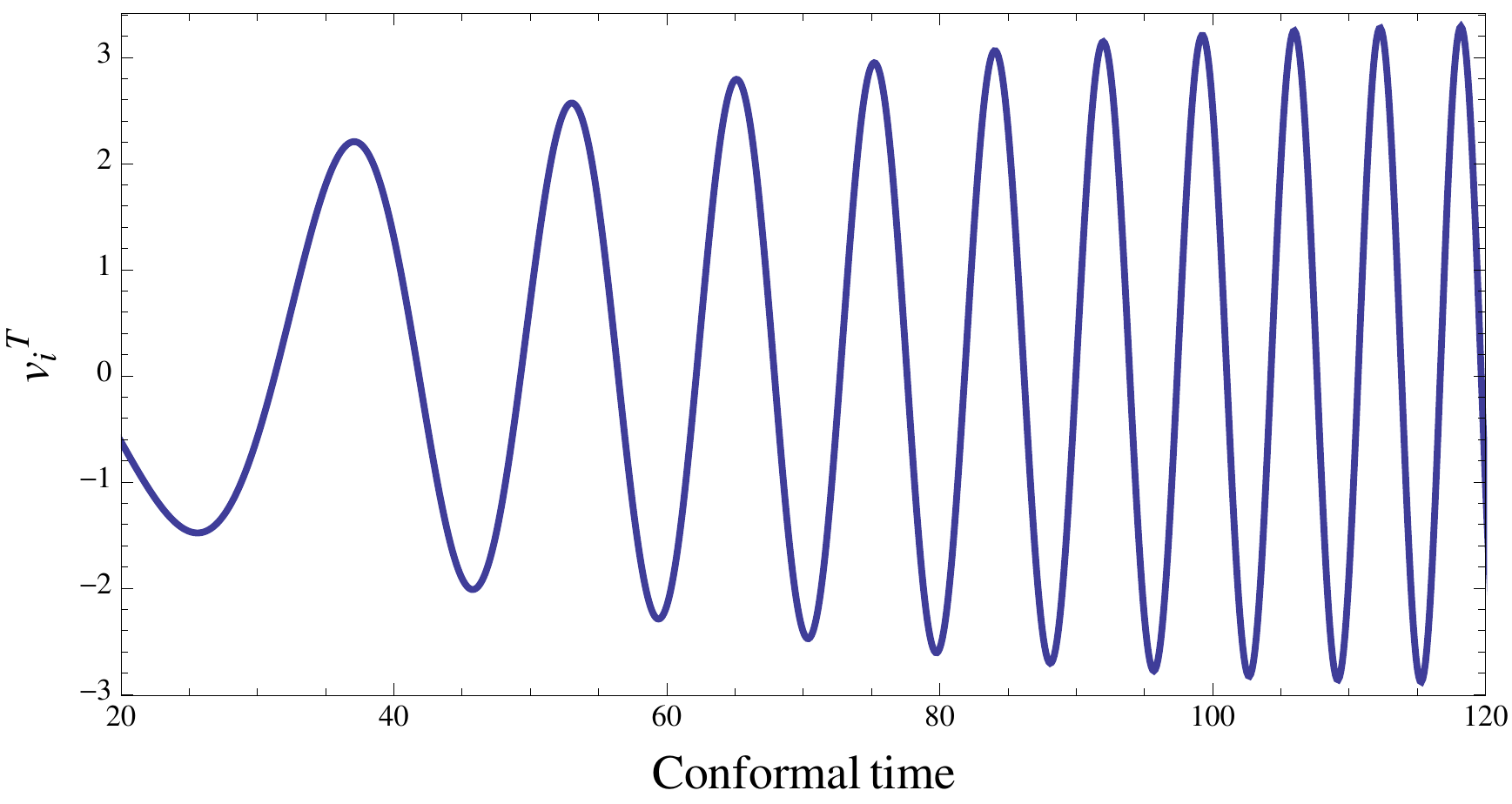}
 \end{subfigure}
\end{center}
\caption{Evolution of vector perturbations as a function of $\tau$, during early times in the radiation-dominated era for a sub-horizon scale.}
\label{PlotVecInfinRad}
\end{figure}

Figure \ref{PlotVecInfinRad} shows numerical results for the evolution of vector perturbations as a function of $\tau$, during early times for a given sub-horizon scale in the radiation-dominated era. In this case we have set $m^2\beta_1=m^2\beta_4=10^{-2}$, and arbitrary initial conditions of the same order for both fields. As expected due to the analytical solutions, $F_{2i}$ decays in time while $v^T_i$ grows.

\subsubsection{Early times matter-dominated era}
Let us now assume that $w=0$, and consider only leading order terms in $N$ in the equations of motion to find
\begin{align}
& F_{2i}''+\mathcal{H}\frac{(5x^2+24)}{2(x^2+6)}F_{2i}'-\frac{3\mathcal{H}}{x^2+6}v^T_i+\frac{1}{4}x^2\mathcal{H}^2 F_{2i}=0,\\
&v_i^{T'}+\mathcal{H}\frac{(x^2+15)}{(x^2+6)}v^T_i-\frac{3}{2}\mathcal{H}\frac{x^2}{(x^2+6)} F_{2i}'-\frac{1}{4}x^2\mathcal{H}^2F_{2i}=0.
\end{align}
\begin{description}
\item[Super-horizon scales:] 
the evolution equations become
\begin{align}
& F_{2i}''+2\mathcal{H}F_{2i}'-\frac{1}{2}\mathcal{H}v^T_i+\frac{1}{4}x^2\mathcal{H}^2 F_{2i}=0,\\
&v_i^{T'}+\frac{15}{6}\mathcal{H}v^T_i-\frac{1}{4}\mathcal{H}x^2 F_{2i}'-\frac{1}{4}x^2\mathcal{H}^2F_{2i}=0,
\end{align}
and, when ignoring terms of order $x^2$, the solutions are $F_{2i}=c_1/\tau^4+c_2/\tau^3+c_3$ and $v^T_i\propto 1/\tau^5$, where $c_1$, $c_2$ and $c_3$ are some integration constants. Therefore, both functions decay in time.
\item[Sub-horizon scales] the evolution equations now reduce to
\begin{align}
& F_{2i}''+\frac{5}{2}\mathcal{H}F_{2i}'-\frac{3\mathcal{H}}{x^2}v^T_i+\frac{1}{4}x^2\mathcal{H}^2 F_{2i}=0,\\
&v_i^{T'}+\mathcal{H}v^T_i-\frac{3}{2}\mathcal{H} F_{2i}'-\frac{1}{4}x^2\mathcal{H}^2F_{2i}=0,
\end{align}
and, when ignoring terms of order $x^{-2}$, the solutions are $F_{2i}\propto e^{\pm ik\tau/2}/\tau^{3/2}$ and $v^T_i=c_1/\tau^2+c_\pm  e^{\pm ik\tau/2}/\tau^{3/2}$, where $c_1$ and $c_\pm$ are integration constants.
\end{description}

\begin{figure}[H]
\begin{center}
\begin{subfigure}  
\centering
\includegraphics[scale=0.43]{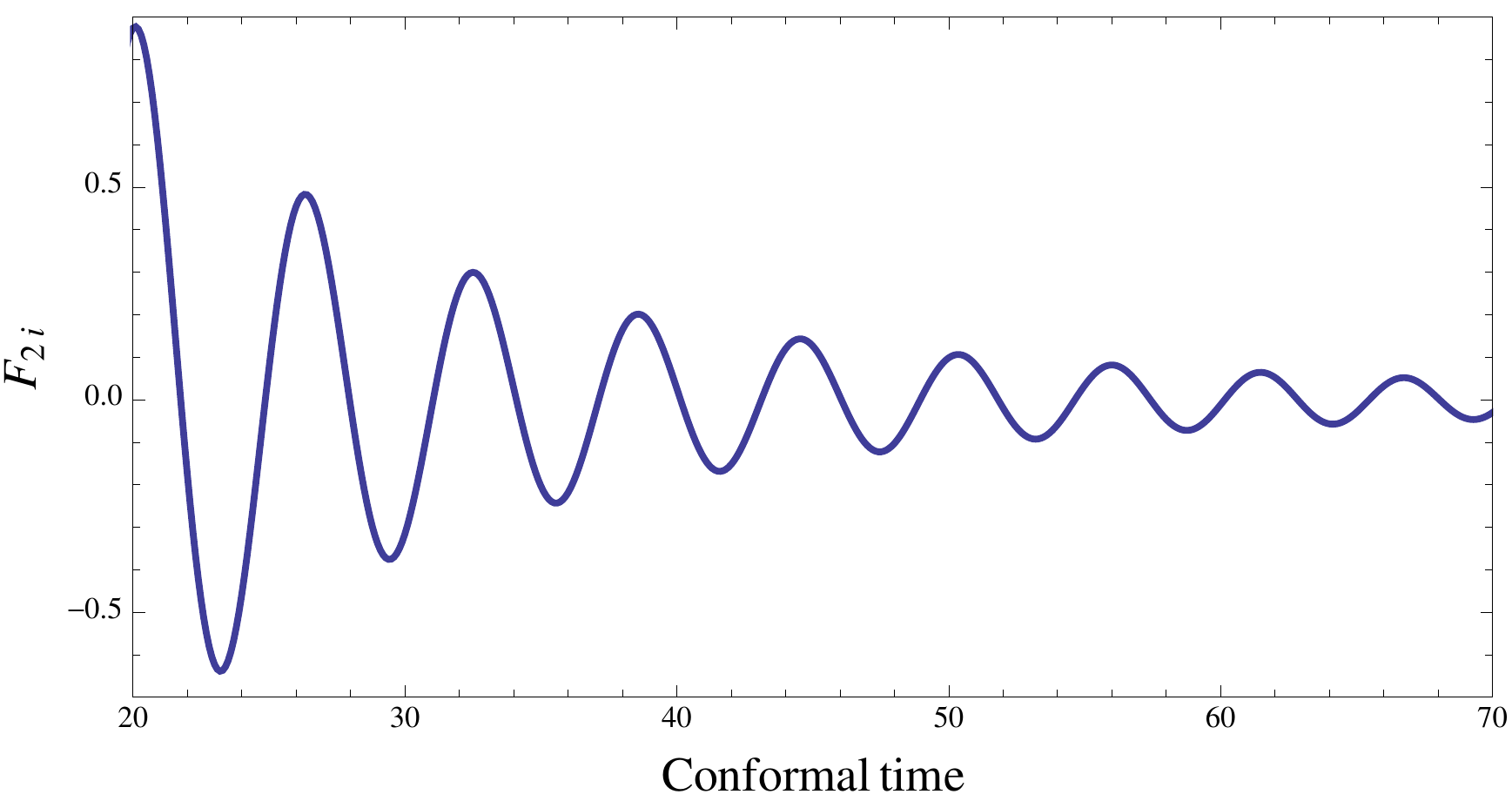}
 \end{subfigure}
\begin{subfigure}    
\centering
\includegraphics[scale=0.43]{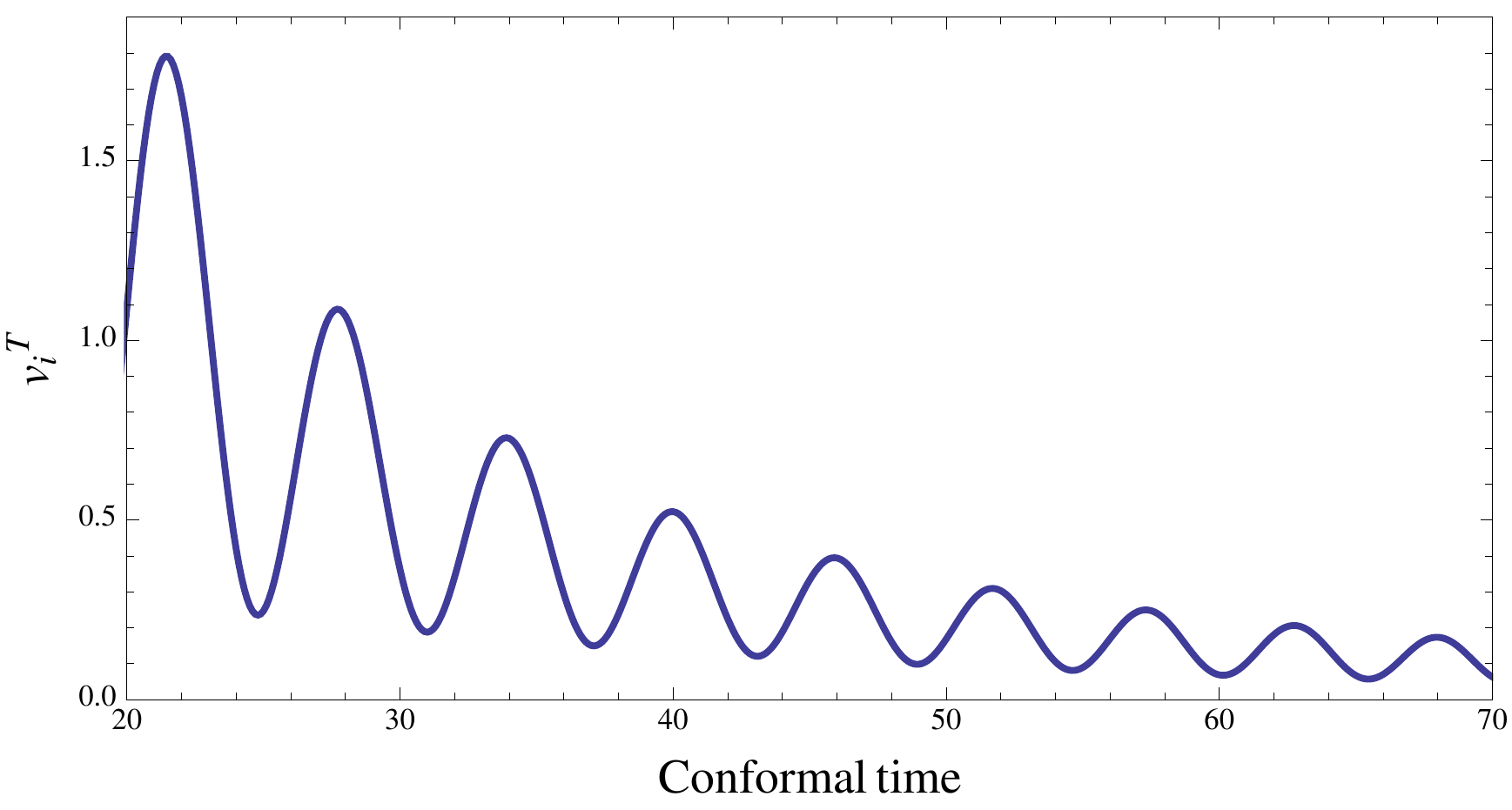}
 \end{subfigure}
\end{center}
\caption{Evolution of vector perturbations as a function of $\tau$, during early times in matter-dominated era for a given sub-horizon scale.}
\label{PlotVecInfinMat}
\end{figure}

Figure \ref{PlotVecInfinMat} shows numerical results for the evolution of vector perturbations as a function of $\tau$, during early times for a given sub-horizon scale in the matter-dominated era. In this case we have set $m^2\beta_1=m^2\beta_4=10^{-2}$, and arbitrary initial conditions of the same order for both fields. With these plots we confirm our analytical results.

%--------------------------------------------------------------------------------------------------------------------------------------------------
\section{Tensor perturbations}\label{SecTensorPert}

In this section we find approximate analytical solutions for the tensor modes in the relevant regimes for both branches. As mentioned previously, in the bouncing branch, we restrict our study of the tensor modes for the case $\beta_3=\beta_2=0$.

\subsection{Expanding branch}
As before, we study the solutions of tensor perturbations at early and late times.
\subsubsection{Early times}
\begin{figure}[H]
\begin{center}
\begin{subfigure}  
\centering
\includegraphics[scale=0.421]{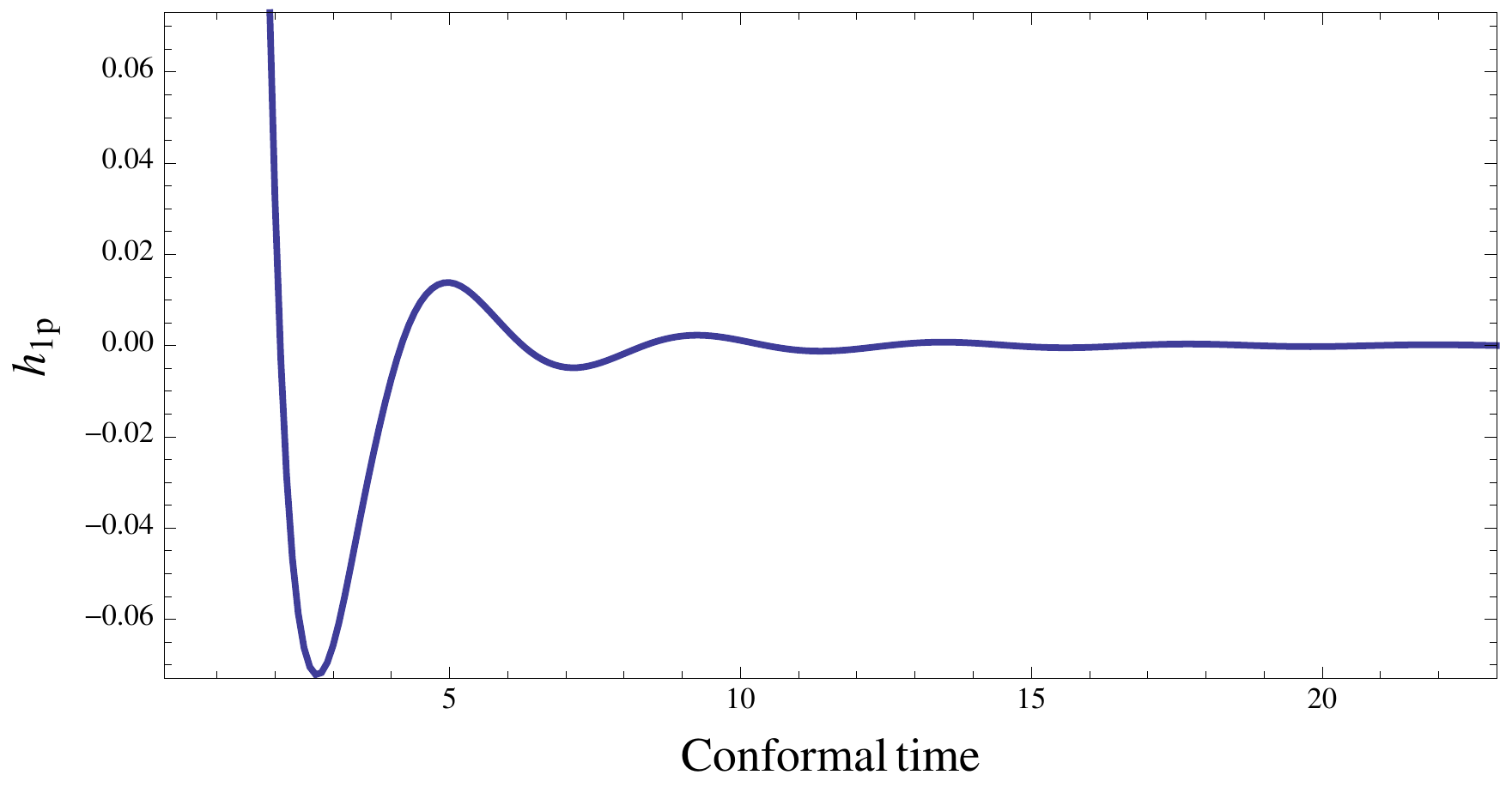}
 \end{subfigure}
\; \begin{subfigure}    
\centering
\includegraphics[scale=0.421]{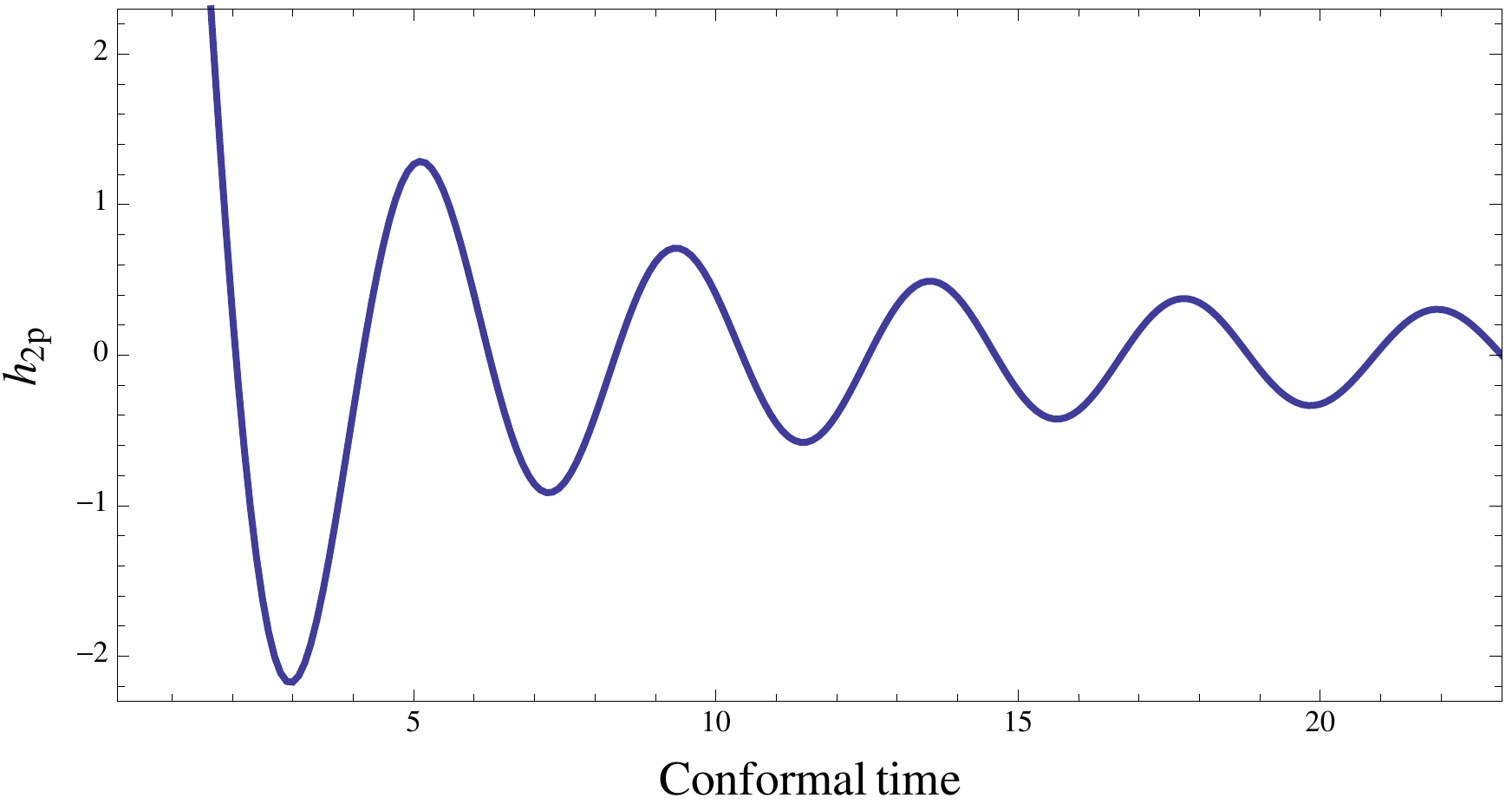}
 \end{subfigure}
\end{center}
\caption{Evolution of tensor perturbations as a function of the conformal time during early times in the radiation-dominated era for a given sub-horizon scale.}
\label{FigTensorFiniteRad}
\end{figure}

Let us consider only leading order terms in $1/N$, as $N\ll 1$ at early times in this branch. In this approximation eq.~(\ref{EqTensorh2})-(\ref{EqTensorh1}) become:
\begin{align}
& h_{2p}''+2\mathcal{H}h_{2p}'+x^2\mathcal{H}^2h_{2p}+m^2a^2N\beta_1(h_{2p}-h_{1p})=0,\\
&h_{1p}''+2(4+3w)\mathcal{H}h_{1p}'+(4+3w)^2x^2\mathcal{H}^2h_{1p}+ 3(4+3w)\mathcal{H}^2(h_{1p}-h_{2p})=0.
\end{align}
\begin{description}
\item[Super-horizon scales:] the equations simplify to the form
\begin{align}
& h_{2p}''+2\mathcal{H}h_{2p}'=0,\\
&h_{1p}''+10\mathcal{H}h_{1p}'+15\mathcal{H}^2(h_{1p}-h_{2p})=0,
\end{align}
and are solved by $h_{2p}=c_1+c_2/\tau$ and $h_{1p}=c_3+c_4/\tau +c_\pm \tau^{n_\pm}$, with $n_\pm=-(9\pm \sqrt{21})/2<0$, where $c_1$, $c_2$, $c_3$, $c_4$ and $c_\pm$ are integrations constants, related to each other. Therefore, both solutions decay to a constant.
\item[Sub-horizon scales:] the evolution equations become
\begin{align}
& h_{2p}''+2\mathcal{H}h_{2p}'+x^2\mathcal{H}^2h_{2p}+\mathcal{O}(N^{3/2})(h_{2p}-h_{1p})=0,\\
&h_{1p}''+10\mathcal{H}h_{1p}'+25x^2\mathcal{H}^2h_{1p}=0,
\end{align}
with solutions $h_{2p}\propto e^{\pm ik\tau}/\tau$ and $h_{1p}\propto e^{\pm i5k \tau}/\tau^5$.
\end{description}
Unlike scalar perturbations, tensor perturbations in the expanding branch are not unstable- they oscillate and decay. We find the same behaviour in the matter-dominated era. Fig.~\ref{FigTensorFiniteRad} shows numerical results for the evolution of both tensor perturbations as a function of $\tau$ (in arbitrary units), at early times during the radiation-dominated era for a given sub-horizon scale. In this particular case we set $m^2\beta_1=10^{-2}$, and all other $\beta$s vanishing, and arbitrary initial conditions of the same order for both fields. As expected due to the analytical solutions, we observe that $h_{1p}$ decays faster than $h_{2p}$.

\subsubsection{Late times}
Now, let us study the behaviour in the de-Sitter phase, in the matter-dominated era. In this phase $N$ takes the constant value $\bar{N}$, and $a\propto 1/\tau$, with $\tau\rightarrow 0$ being the infinite future. The equations of motion become:
\begin{align}
& h_{2p}''+2\mathcal{H}h_{2p}'+k^2h_{2p}+x_2\mathcal{H}^2(h_{2p}-h_{1p})=0, \label{Solh2Late}\\
&h_{1p}''+2\mathcal{H}h_{1p}'+k^2h_{1p}+x_1\mathcal{H}^2(h_{1p}-h_{2p})=0, \label{Solh1Late}
\end{align} 
where $x_2=m^2\bar{N}\tilde{\bar{Z}}/H_0^2$ and $x_1=m^2\tilde{\bar{Z}}/(H_0^2\bar{N})=x_2/\bar{N}^2$. 
\begin{description}
\item[Super-horizon scales:] the evolutions equations simplify to
\begin{align}
& h_{2p}''+2\mathcal{H}h_{2p}'+x_2\mathcal{H}^2(h_{2p}-h_{1p})=0, \\
&h_{1p}''+2\mathcal{H}h_{1p}'+x_1\mathcal{H}^2(h_{1p}-h_{2p})=0, 
\end{align}
and are solved by $h_{1p}=c_1+c_2\tau^3+c_{\pm}\tau^{n_\pm}$ and $h_{2p}=c_1+c_2\tau^3-\frac{x_2}{x_1}c_{\pm}\tau^{n_\pm}$, where $c_1$, $c_2$ and $c\pm$ are integration constants and $n_\pm=\frac{1}{2}(3\pm \sqrt{9-4x_1-4x_2})$. Since $Re(n_\pm)>0$, both solutions decay in time to a constant.
\item[Sub-horizon scales:] the evolutions equations now become
\begin{align}
& h_{2p}''+2\mathcal{H}h_{2p}'+k^2h_{2p}=0,\\
&h_{1p}''+2\mathcal{H}h_{1p}'+k^2 h_{1p}=0,
\end{align}
and are solved by $h_{bp}\propto e^{\pm ik\tau}\tau$, which are decaying, as in this regime $\tau\rightarrow 0$ in the infinite future.

\end{description}

Note that since $h_{1p}$ decays considerably faster than $h_{2p}$ during early times, $h_{2p}$ could start in the de-Sitter phase being some orders of magnitude larger that $h_{1p}$ (which will happen if the initial conditions at early times for both fields were of the same order of magnitude). In this case, there is an intermediate phase in the full solutions of eq.~(\ref{Solh2Late})-(\ref{Solh1Late}), when the $k^2h_{1p}\sim x_1\mathcal{H}^2h_{2p}$. In this phase $h_{2p}$ could affect the evolution of $h_{1p}$, as $h_{1p}$ will start growing,``reaching" the magnitude of $h_{2p}$, until $k^2\ll x_1\mathcal{H}^2$, when the scale is super-horizon, and both fields will approach the same constant.

\begin{figure}[H]
\begin{center}
\begin{subfigure}  
\centering
\includegraphics[scale=0.427]{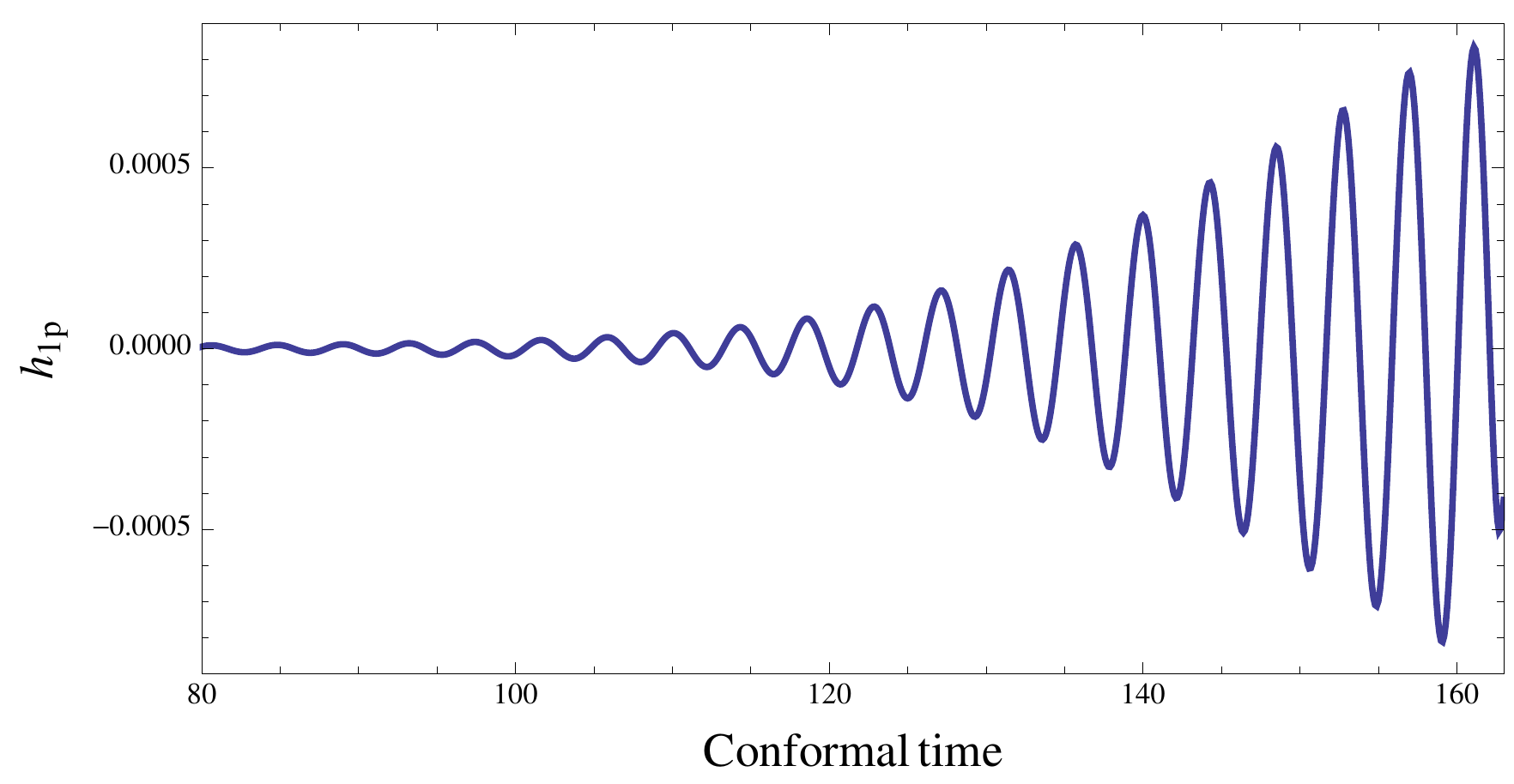}
 \end{subfigure}
 \begin{subfigure}    
\centering
\includegraphics[scale=0.42]{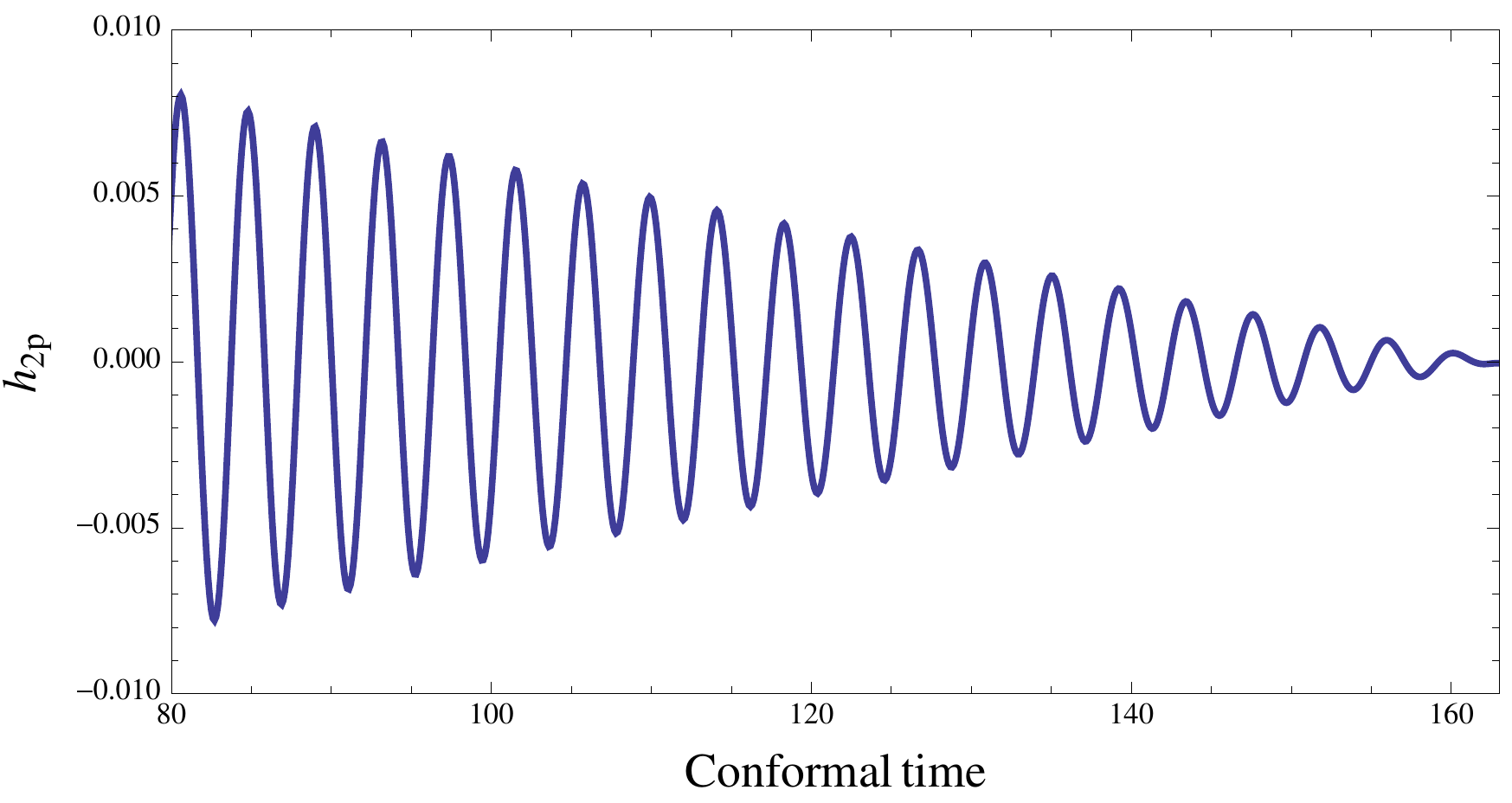}
 \end{subfigure}
\end{center}
\caption{Evolution of tensor perturbations as a function of the conformal time during the de-Sitter phase at late times in the matter-dominated era.}
\label{FigTensorDS}
\end{figure}

Fig.~\ref{FigTensorDS} shows numerical solutions for tensor perturbations as a function of $\tau$ (in arbitrary units) at late times for a given sub-horizon scale. In this particular case we set $m^2\beta_1=10^{-2}$ and all the other $\beta$s vanishing, and arbitrary initial conditions of the same order for both fields. In this case we observe that since $h_{1p}$ starts in the de-Sitter phase being at least two orders of magnitude smaller than $h_{2p}$, the previously described intermediate phase occurs, where $h_{1p}$ grows while $h_{2p}$ decays as expected for a sub-horizon scale. Generically, for different initial conditions, we would see a phase where $h_{1p}$ first decays and then it grows. 

\subsection{Bouncing branch}
As before, we only focus on early times as the evolution at late times will be the same as in the expanding branch. We study the radiation-dominated era and matter-dominated era. At early times, we consider only the leading order terms in $N$ in all the coefficients in eq.~(\ref{EqTensorh2})-(\ref{EqTensorh1}), as $N\gg 1$ at early times in this branch:
\begin{align}
& h_{2p}''+2\mathcal{H}h_{2p}'+x^2\mathcal{H}^2h_{1p}+m^2a^2N\beta_1(h_{2p}-h_{1p})=0,\label{Eqh2Bounce}\\
&h_{1p}''-(1+3w)\mathcal{H}h_{1p}'+\left(\frac{1+3w}{2}\right)^2x^2\mathcal{H}^2 h_{1p}-\frac{(1+3w)}{2}\frac{m^2a^2\beta_1}{N}(h_{1p}-h_{2p})=0.\label{Eqh1Bounce}
\end{align}
\subsubsection{Early times radiation-dominated era}
Let us consider $w=1/3$ in eq.~(\ref{Eqh2Bounce})-(\ref{Eqh1Bounce}), and find their solutions for super-horizon and sub-horizon scales.
\begin{description}
\item[Super-horizon scales:] the evolution equations are now
\begin{align}
& h_{2p}''+2\mathcal{H}h_{2p}'+m^2a^2N\beta_1(h_{2p}-h_{1p})=0,\\
&h_{1p}''-2\mathcal{H}h_{1p}'+\mathcal{O}(N^{-2})(h_{1p}-h_{2p})=0,
\end{align}
and are solved by 
\begin{align}
& h_{2p}=c_\pm \frac{e^{\pm iK\tau}}{\tau}+c_3+c_4\left[\tau^3-12\frac{\tau}{K^2}+\frac{24}{(K^4\tau)}\right], \nonumber\\
& h_{1p}=c_3+c_4\tau^3,
\end{align} 
where $c_\pm$, $c_3$ and $c_4$ are integration constants, and $K^2=m^2a^2N\beta_1=const$. Therefore, $h_{1p}$ and $h_{2p}$ grow as a power of $\tau$.

Notice that if $\beta_1$ were negative, the solution for $h_{2p}$ would include $e^{\pm |K|\tau}$ instead of oscillating functions, which would correspond to an exponential instability.

\item[Sub-horizon scales:] the evolution equations are now
\begin{align}
& h_{2p}''+2\mathcal{H}h'_{2p}+x^2\mathcal{H}^2h_{2p}+m^2a^2N\beta_1(h_{2p}-h_{1p})=0,\\
&h_{1p}''-2\mathcal{H}h_{1p}'+x^2\mathcal{H}^2 h_{1p}-\frac{m^2a^2\beta_1}{N}(h_{1p}-h_{2p})=0,\label{Eqh1BounceRad}
\end{align}
and when considering highest orders in $N$ only, the solutions are 
\begin{align}\label{TensorSubRad}
& h_{1p}\propto (1\mp ik\tau) e^{\pm ik\tau},\nonumber \\
& h_{2p}=\left(\frac{c_{1\pm}}{\tau}+c_{2\pm}+c_{3\pm}\tau\right) e^{\pm ik\tau}+\frac{c_{4\pm}}{\tau}e^{\pm i\omega\tau},
\end{align}
where $\omega^2=k^2+m^2\beta_1a^2N$, and where the coefficients $c_{1\pm}$, $c_{2\pm}$, $c_{3\pm}$ and $c_{4\pm}$ are integration constants, related to those of $h_{1p}$. Note that $\omega = {\rm constant}$ as during the radiation-dominated era at early times $\tilde{\rho}\approx \beta_4N^2\propto a^{-4}$, and therefore $a^2N$ is constant. Unlike GR, here we observe that $h_{1p}$ grows linearly with time, while $h_{2p}$ also includes a growing modes as a consequence of the interactions with $h_{1p}$. The growing mode in $h_{1p}$ is a consequence of the fact that the metric $f_{\mu\nu}$ is bouncing, and therefore at early times the term with $h_{1p}'$ in eq.~(\ref{Eqh1BounceRad}) has a negative sign. 

Notice that if $\beta_1$ were negative, $\omega^2$ would be negative for some values of $k$, and for those cases there would be an exponential instability in the solution for $h_{2p}$.

\end{description}

\begin{figure}[H]
\begin{center}
\begin{subfigure}  
\centering
\includegraphics[scale=0.416]{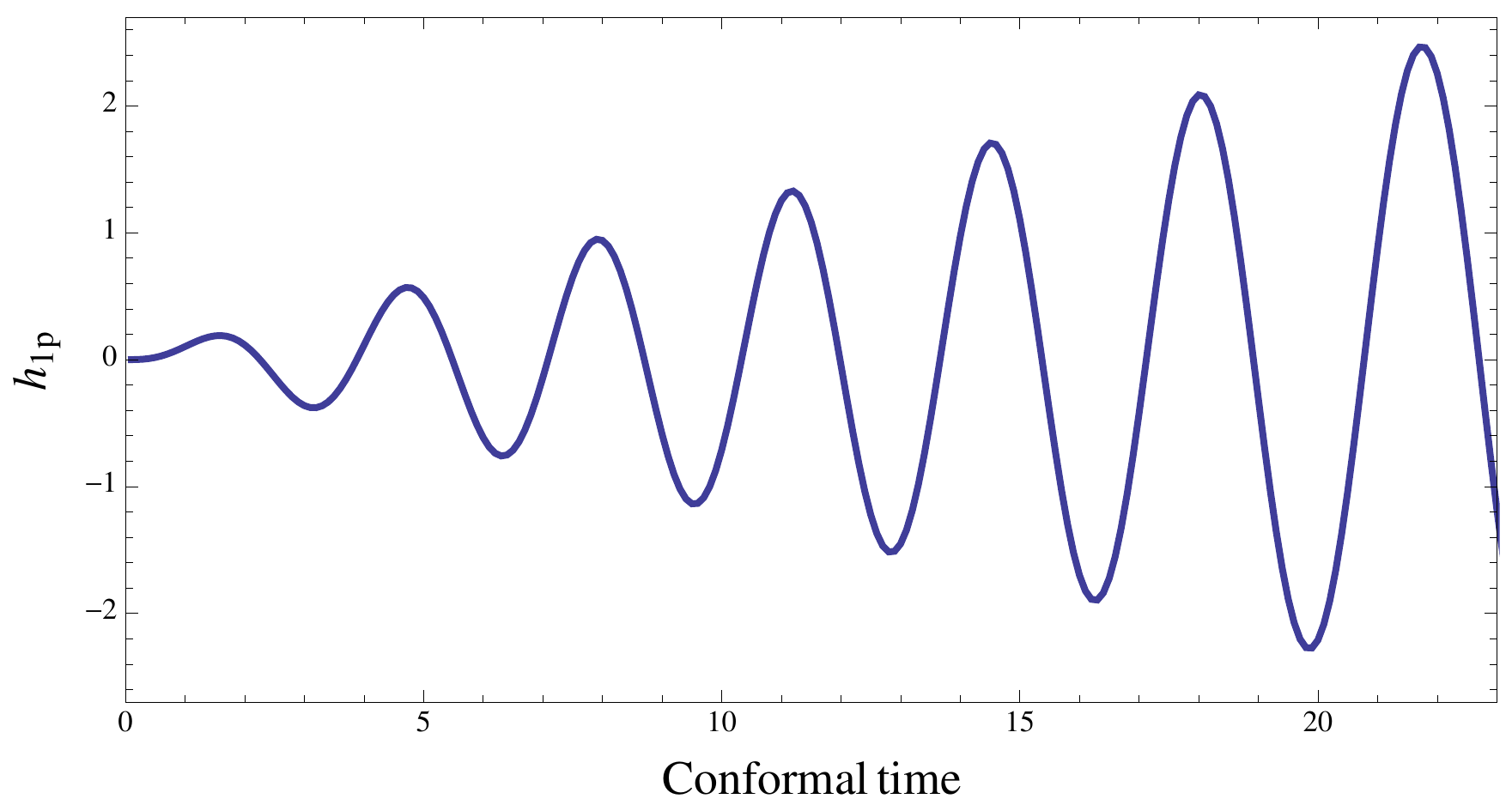}
 \end{subfigure}
 \begin{subfigure}    
\centering
\includegraphics[scale=0.434]{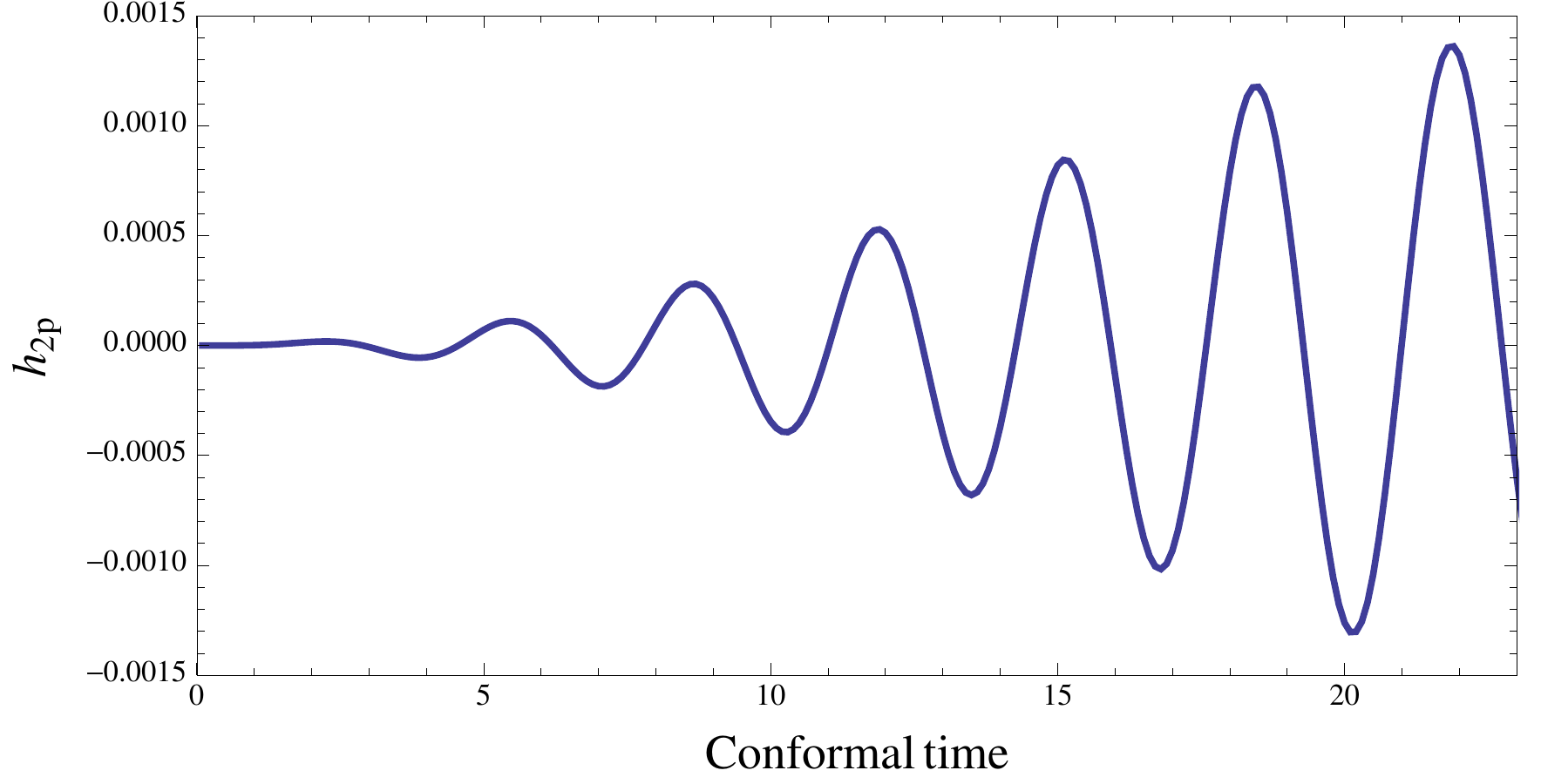}
 \end{subfigure}
\end{center}
\caption{Evolution of tensor perturbations as a function of the conformal time, during early times in the radiation-dominated era for a given sub-horizon scale.}
\label{FigTensorInfiniteRad}
\end{figure}

Fig.~\ref{FigTensorInfiniteRad} shows numerical solutions for the evolution of both tensor perturbations as a function of $\tau$, at early times during the radiation-dominated era for a given sub-horizon scale. In this case we set $m^2\beta_1=m^2\beta_4=10^{-2}$, and arbitrary initial conditions of order one for both fields. As expected due to the analytical solutions, we see a growth in both fields in this stage.

\subsubsection{Early times matter-dominated era}
Now, let us consider $w=0$ in eq.~(\ref{Eqh2Bounce})-(\ref{Eqh1Bounce}), and find their solutions for super-horizon and sub-horizon scales. Note that during the matter-dominated era at early times $\tilde{\rho}\approx \beta_4N^2\propto a^{-3}$, and then $a^2N\propto N^{-1/3}$ and $a^2/N\propto N^{-7/3}$. Therefore, mixing terms can be ignored in the equations of motion as $N\gg 1$ at early times in this branch.
\begin{description}
\item[Super-horizon scales:] the evolution equations are now
\begin{align}
& h_{2p}''+2\mathcal{H}h_{2p}'=0,\\
&h_{1p}''-\mathcal{H}h_{1p}'=0,
\end{align}
and are solved by $h_{2p}=c_1+c_2/\tau^3$; $h_{1p}=c_3+c_4\tau^3$, where $c_1$, $c_2$, $c_3$ and $c_4$ are integration constants. We find then that $h_{1p}$ grows as a power of $\tau$ and $h_{2p}$ decays, in a similar way to the radiation-dominated era solutions.
\item[sub-horizon scales:] the evolution equations simplify to
\begin{align}
& h_{2p}''+2\mathcal{H}h_{2p}'+x^2\mathcal{H}^2h_{2p}+\mathcal{O}(N^{-1/3})(h_{2p}-h_{1p})=0,\\
&h_{1p}''-\mathcal{H}h_{1p}'+\frac{x^2\mathcal{H}^2}{4} h_{1p}+\mathcal{O}(N^{-7/3})(h_{1p}-h_{2p})=0,
\end{align}
and are solved by $h_{1p}\propto (1\mp ik\tau/2) e^{\pm ik\tau/2}$ and $h_{2p}\propto \frac{(1\mp ik\tau)}{\tau^3}e^{\pm ik\tau}$.
\end{description}

\begin{figure}[H]
\begin{center}
\begin{subfigure}  
\centering
\includegraphics[scale=0.42]{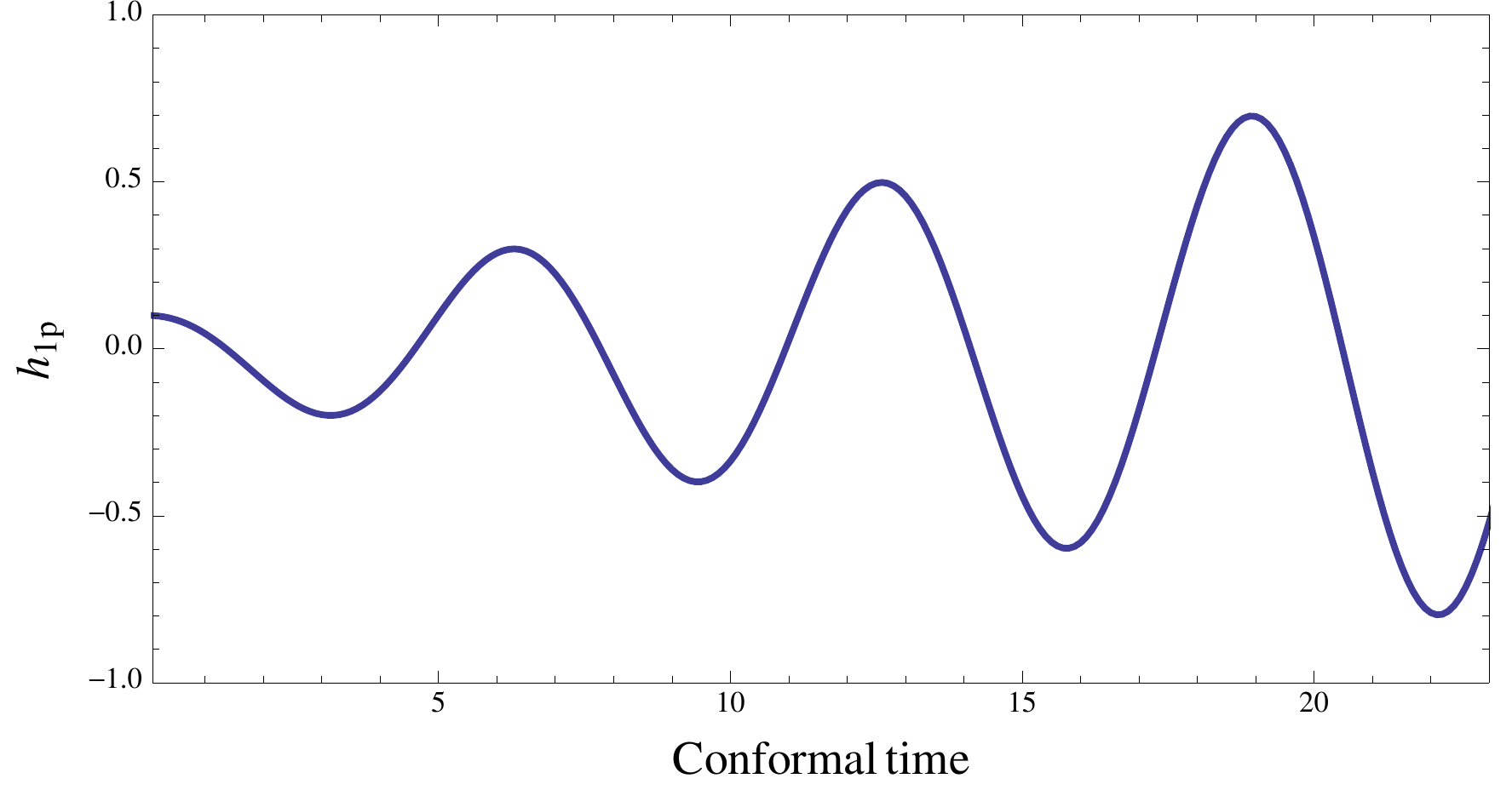}
 \end{subfigure}
\begin{subfigure}    
\centering
\includegraphics[scale=0.428]{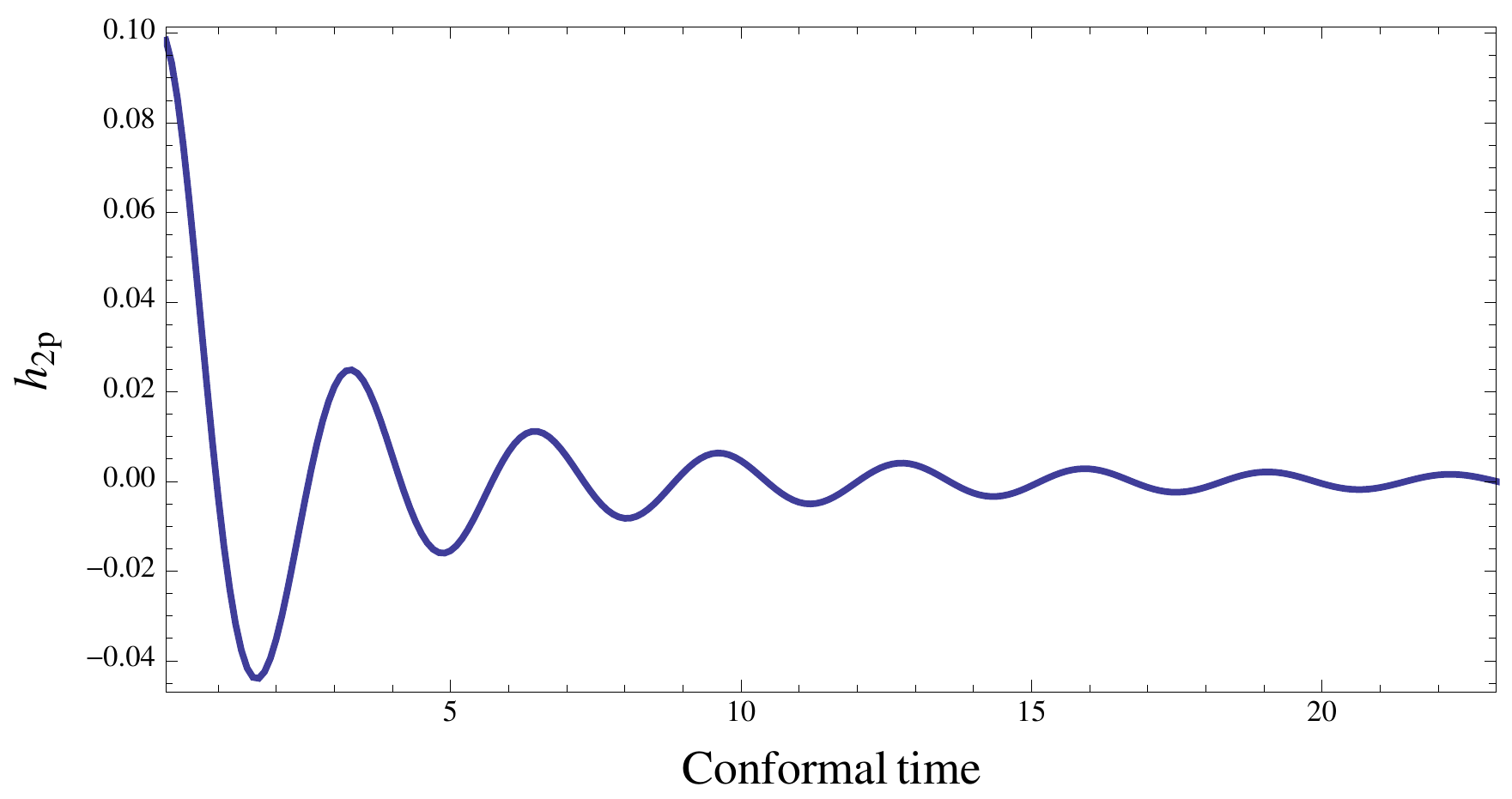}
 \end{subfigure}
\end{center}
\caption{Evolution of tensor perturbations during early times in the matter-dominated era.}
\label{FigTensorInfiniteMat}
\end{figure}

Fig.~(\ref{FigTensorInfiniteMat}) shows numerical solutions for the evolution of tensor perturbations as a function of $\tau$ (in arbitrary units), at early times in the matter-dominated era for a given sub-horizon scale. Again, in this case we set $m^2\beta_1=m^2\beta_4=10^{-2}$, and arbitrary initial conditions of the same order for both fields. Unlike during the radiation-dominated era, in this case $h_{1p}$ grows linearly with time, but $h_{2p}$ decays.

We have found that $h_{1p}$ grows as a power law at early times for super-horizon and sub-horizon scales. At late times, this could mean that $h_{1p}$ could start in de-Sitter phase being some orders of magnitude higher than $h_{2p}$. This would produce the same effect described previously for late times solutions in the expanding branch, but in this case $h_{2p}$ would grow at late times due to $h_{1p}$.

%--------------------------------------------------------------------------------------------------------------------------------------------------
\section{Discussion}
\label{Discussion}
In this paper we have undertaken a comprehensive analysis of the evolution of cosmological linear perturbations in massive bigravity and have found approximate analytical solutions in a wide range of regimes. We have confirmed the main results of previous works on linear perturbations, but also extended their analysis to vector and tensor modes. In doing so we have found that massive bigravity has a number of instabilities which manifest themselves as growing solutions. In particular, we have found that most choices of parameters generate exponential instabilities in the scalar, vector or tensor modes. A subset of model space does not have exponential instabilities: when $\beta_3=\beta_2=0$ with $\beta_1$ and $\beta_4$ being positive, which corresponds to a particular case of the bouncing branch. However, even for this subset of models we have found growing power-law solutions in the vector and tensor modes, contrary to GR, and a violation of the Higuchi bound, which would bring instabilities when studying the model beyond the classical linear regime. For vector and tensor fields, this growth is a consequence of a bounce in $f_{\mu\nu}$ along with effects from the interaction terms between both metrics. Analogously to scenarios with exponential instabilities, these growing modes could be a source of concern as the perturbation theory could break down at some early time. However, this latter case is not as bad because, as we will show later, we can prevent modes from growing too large by considering particular initial conditions. This resulting fine-tuning is much less restrictive than that required for the exponential solutions.

As mentioned in the introduction, such growing modes may be a hint that all is not well and that the initial values problem may not be well-posed. If indeed this is the case, it would not be surprising as extra degrees of freedom may lead to such behaviour. For example there have been efforts in trying to determine whether scalar-tensor theories have a well-posed initial value problem, while a study of Einstein-Aether theories has shown that caustics will generically arise there \cite{Contaldi:2008iw}. We believe a detailed analysis of the initial value problem in massive bigravity is essential to place it on a firm footing.

An alternative view could be to take the solutions we have found and speculate on their cosmological consequences. To do this accurately, one would have to explore the correct set of initial conditions which would arise in such a theory due to (for example) inflation. One would then have to incorporate our equations into a complete and realistic model of the universe that incorporates the various components, the correct thermal history and the Boltzmann equation for the relativistic degrees of freedom. We leave that for future work. Nevertheless, we can attempt to estimate the effect of the new solutions we have found by focusing on a few observables. 

In what follows we will focus solely on tensor modes; we found a growing mode for vectors but we do not address its effect for now. Recall from the previous section that for super-horizon scales during the radiation-dominated era, $h_{2p}$ grows as $\tau^3$ due to the interaction terms with $h_{1p}$. Therefore, from the end of the inflationary era until the recombination era, $h_{2p}$ might deviate substantially from its value in GR. As a result we might expect a larger effect from gravitational waves in the Cosmic Microwave Background (CMB). An estimate of how much $h_{2p}$ could grow in this stage (on super-horizon scales) gives us:
\begin{equation}\label{Eqh2cmb}
h_{2{\rm rec}}\approx h_{2i}+\frac{(K\tau_{\rm eq})^2}{6}\left[h_{1i}-h_{2i}+\frac{\tau_r^3}{15}\tau_i h_{1i}'\right]+\frac{(K\tau_{\rm eq})^2}{9}\left[h_{1i}-h_{2i}+\frac{\tau_r^3}{6}\tau_i h_{1i}'\right]\left(1-\left(\frac{\tau_{\rm eq}}{\tau_{\rm rec}}\right)^3\right),
\end{equation}
where $h_{2{\rm rec}}$ is the value of the tensor perturbation $h_{2p}$ at recombination given an initial value of $h_{2i}$ at some initial time $\tau_i$. The subindex ${\rm eq}$ corresponds to a value at the matter-equality time, and we have defined $\tau_r=\tau_{\rm eq}/\tau_i=a_{\rm eq}/a_i$. Here, we have also used that $K^2=m^2a^2N\beta_1\propto m$, and $K\tau_{\rm eq}\ll 1$ (which would happen for a sufficiently small $m$), and calculated the first order corrections in $K^2$. 

Note that in GR the value at the recombination era would be $h_{2i}$ for a super-horizon scale, given that $h_{2i}'=0$ and, therefore, the second and third terms in eq.~(\ref{Eqh2cmb}) correspond to the modifications introduced by massive gravity to this tensor perturbation, which are proportional to $m$. Even though $K\tau_{eq}\ll 1$, the modification is not necessarily small as it depends also on the initial conditions for $h_{1p}$. 

If we choose $\tau_i$ to be the end of the inflationary era (for example where $a_i\sim 10^{-28}$), we have that $\tau_r^3\tau_i\sim 10^{107}$. Therefore, we would need $h_{1p}$ to be effectively zero at the end of the inflationary era, and $h_{1p}$ would then be constant for super-horizon scales. Otherwise, $h_{1p}$, and as a consequence $h_{2p}$, could grow large and break the validity of perturbation theory. Assuming $h_{1i}'=0$ and some preferred values found in \cite{Konnig:2014xva} when constraining scalar perturbations with observational data, the largest modification introduced by massive gravity in $h_{2p}$ at the epoch of the recombination, according to eq.~(\ref{Eqh2cmb}), would be: 
\begin{equation}\label{h2recSuper}
\Delta h_{2{\rm rec}}=h_{2{\rm rec}}-h_{2i}=10^{-6}\left(h_{1i}-h_{2i}\right).
\end{equation} 
Further research at early times is needed in order to give exact numbers as we would need to know the initial condition for both tensor perturbations. 

In a similar way, we can study the evolution for sub-horizon perturbations. For a scale that crosses the horizon during the radiation-dominated era, there will be a modification in the evolution of $h_{2p}$, with respect to GR, coming from the interaction with $h_{1p}$, as we can see in eq.~(\ref{TensorSubRad}). From the horizon crossing time $\tau_c$ until the recombination era $\tau_{\rm rec}$, the modification to $h_{2p}$ is given by:
\begin{equation}\label{Eqh2cmbsub}
\Delta h_{2{\rm rec}}=\left(\frac{\tau_{\rm eq}}{\tau_{\rm rec}}\right)^2\left(\frac{K^2}{k^2}\right)\left[c_1h_{2c}+x_{\rm eq}^2\left(c_2h_{1c}+c_3\frac{h_{1c}'}{k}\right)\right],
\end{equation}
%\begin{equation}\label{Eqh2cmbsub}
%\Delta h_{2{\rm rec}}=\left(\frac{\tau_{eq}}{\tau_{\rm rec}}\right)^2\left(\frac{K^2}{k^2}\right)\left[c_1h_{2c}+c_2\frac{h_{2c}'}{k}+x_{eq}^2\left(c_3h_{1c}+c_4\frac{h_{1c}'}{k}\right)\right],
%\end{equation}
where $x_{\rm eq}=k\tau_{\rm eq}$, and the subindex $c$ indicates that the quantity is evaluated at the horizon-crossing time. Here, again, we have considered only first order corrections in $K^2$, and the coefficients $c_1$, $c_2$, and $c_3$ are functions of $\sin(x_{\rm rec})$ and $\cos(x_{\rm rec})$, so they all roughly have the same order of magnitude. 

Note that, since in eq.~(\ref{TensorSubRad}) $h_{2p}$ has a linear growing mode, one could have expected to have larger modifications for larger $k$, as larger $k$ enter the horizon before and consequently spend more time growing. However, as we observe in eq.~(\ref{Eqh2cmbsub}), for larger $k$ the modification is smaller. This happens because the coefficients $c_{1\pm}$, $c_{2\pm}$ and $c_{3\pm}$ in eq.~(\ref{TensorSubRad}) are related to those of $h_{1p}$. In particular, $c_{3\pm}\sim kh_{1c}$, $c_{2\pm}\sim k^2h_{1c}/K^2$ and $c_{1\pm}\sim k^3h_{1c}/K^4$. Therefore, for sufficiently small $m$, the dominant term will be $c_{1\pm}$ and therefore the growing mode will be suppressed compared to the decaying mode, which is what actually happens for observable scales with the preferred values found in \cite{Konnig:2014xva}. 

In addition, note in eq.~(\ref{Eqh2cmbsub}) that, since $x_{\rm eq}\gg 1$, the contribution from $h_{1c}$ to $\Delta h_{2{\rm rec}}$ is much larger that the contribution from $h_{2c}$. A numerical estimate at a scale of order $1Mpc$ gives us 
\begin{equation}
\Delta h_{2{\rm rec}}\sim 10^{-24}h_{2c}+10^{-5}h_{1c}+10^{45}h_{1c}',
\end{equation}
where, again, we see that some kind of mechanism is needed to get $h_{1c}'=0$ at early times, in order to avoid large modifications to GR. In addition, since the value of $h_{2{\rm rec}}$ in GR is estimated to be $h_{2{\rm rec}}^{(GR)}\sim 10^{-10}h_{2c}$, the initial condition $h_{1c}\sim h_{2c}$ will not lead to a small modification to GR. In fact, it will lead to a correction $10^5$ times larger than the GR value, contrary to what we found on super-horizon scales according to eq.~(\ref{h2recSuper}). 

It is clear that, without an appropriate set of initial conditions for cosmological perturbations, we are unable to make definitive statements about the observational viability of these models. They do, however, give us an indication as to what we might expect and it seems that there might be problems with both branches of massive bigravity. The full equations presented in this paper are what is required to modify existing software packages for precise calculations of the growth of large scale structure and the evolution of the cosmic microwave background. With such machinery in hand it should be possible to explore what initial conditions are observationally viable and can be used to place stringent constraints on any theory of the early universe in massive bigravity. 

Finally, it is important to remark that even if the initial value problem is not solved for the model studied in this paper, it does not mean that massive gravity should be left out as a cosmological model, as there are simple modifications to the simplest paradigm that could be explored. One simple and interesting modification can arise in eq.~(\ref{sq}), as asymmetries could be introduced in $f_{\mu\nu}$, while maintaining isotropy in $g_{\mu\nu}$. Other changes could also be considered in the type of coupling with matter (some variations have already been studied in \cite{Enander:2014xga,Akrami:2013ffa, Comelli:2014bqa, Akrami:2014lja,Aoki:2013joa,Hassan:2012wr,Hassan:2014gta,deRham:2014naa,Yamashita:2014fga,Noller:2014sta,Gumrukcuoglu:2014xba,deRham:2014fha,DeFelice:2014nja,Schmidt-May:2014xla,Bamba:2013hza,Aoki:2014cla}), or in the choices of the square root matrix $\sqrt{g^{-1}f}$.

%--------------------------------------------------------------------------------------------------------------------------------------------------
%\section{Conclusions}
%\label{Conclusions}

%--------------------------------------------------------------------------------------------------------------------------------------------------

\begin{acknowledgments}

We are grateful to James Bonifacio, James Scargill, Hans Winther, Luca Amendola, Pedro Alvarez, Tessa Baker, Mariele Mota, Johannes Noller and Adam Solomon for useful discussions and comments. We are particular grateful to Marco Crisostomi for encouraging us to check the properties of the perturbations at the bounce. ML was funded by Becas Chile. PGF acknowledges support from Leverhulme, STFC, BIPAC and the Oxford Martin School.

\end{acknowledgments}

%--------------------------------------------------------------------------------------------------------------------------------------------------------
\appendix
\section{Scalar perturbations equations}
In this section we present the relevant analysis and equations related to scalar perturbations.

\subsection{Auxiliary variables}
\label{AppScalarsAux}
As explained in Section \ref{SectionCosmologicalPerturbations}, the fields $B_1$, $B_2$, $\phi_1$ and $\phi_2$ appear as auxiliary variables in the equations of motion, and therefore they can be worked out in terms of the remaining fields $E_1$, $E_2$ and $\psi_2$, by using their own equations of motion-namely eq.~(\ref{G00}), (\ref{G0i}), (\ref{F00}) and (\ref{F0i}). The explicit expressions for the four auxiliary variables are:
\begin{align}
B_2&=\frac{1}{D_a}\left[   k^2\left(\frac{3}{2}Za^2m^2X + k^2N(1+X)\right)\mathcal{H} E_2' + \frac{3}{2}\mathcal{H}k^2N^2a^2E_1'm^2Z\right.  \nonumber\\
& + \frac{3}{4}\rho_*m^2XZ(1+w)(3\psi_2+k^2 E_2)a^4+\frac{1}{2}\left( m^2Z(1+X)(k^2E_2-k^2 E_1+3\psi_2)N^2\right. \nonumber\\
&\left.\left. +\rho_*(1+X)(1+w)\left( 3\psi_2+k^2 E_2\right) N + 3m^2XZ\psi_2 \right)k^2a^2+\psi_2k^4 N(1+X)\right],
\end{align} 

\begin{align}
B_1&=\frac{1}{4XD_a}\left[ 4Nk^2\mathcal{H}E_1'\left( \frac{3}{2}a^2\left( \rho_*X(1+w) + Nm^2Z + \rho_*(1+w) \right) + k^2(1+X) \right)\right. \nonumber\\
& -2Za^2Xm^2 \left(-3k^2\mathcal{H}E_2'+ \frac{3}{2} \left( (k^2E_2-k^2E_1+3\psi_2)X - k^2E_1\right)(1+w)\rho_*a^2 \right. \nonumber\\
& \left. \left.+k^2\left( (k^2 E_2-k^2E_1+3\psi_2)X+k^2(E_2-E_1) \right) \right) \right],
\end{align}

\begin{align}
\phi_2&=\frac{-1}{8\mathcal{H}D_a}\left[  8\mathcal{H}\psi_2' \left(\frac{9}{4}\rho_*m^2XZ(1+w)a^4+\frac{3}{2}k^2a^2\left(\rho_*(1+X)(1+w)N+m^2Z(X+N^2)\right) \right.\right. \nonumber\\
& \left.  + k^4N(1+X)\frac{}{}\right) + 2a^2 \left( 2k^2\mathcal{H}\left( \frac{3}{2}\rho_*m^2XZ(1+w)a^2+N(\rho_*(1+X)(1+w))k^2\right. \right. \nonumber\\
& \left. +Nm^2Z\frac{}{}\right)E_2' -2m^2N^2Z\mathcal{H}k^4E_1' + \frac{3}{2} ZX(1+w)\rho_*m^2a^4 \left(\rho_*(1+w)(3\psi_2+k^2 E_2) \right. \nonumber\\
&  \left. +m^2Z(k^2E_2-k^2E_1+3\psi_2)N\right) + k^2a^2\left( N(1+w)^2(1+X)(3\psi_2+k^2E_2)\rho_*^2+ \right. \nonumber\\
&     Z(1+w)m^2\rho_*\left( 3X\psi_2 + N^2( X(k^2 E_2-k^2E_1+3\psi_2)+2k^2E_2-k^2E_1+6\psi_2) \right)  \nonumber\\
& \left.\left. \left.    + Nm^4Z^2(X+N^2)(k^2E_2-k^2E_1+3\psi_2) \right) +2N\left( Nm^2Z+\rho_*(1+X)(1+w) \right)\psi_2k^4\frac{}{}\right)\right],
\end{align}

\begin{align}
\phi_1&=\frac{Za^2m^2}{4\mathcal{H}ND_a}\left[  -Nk^2\mathcal{H}E_1' ( 3\rho_*(1+w)a^2+2k^2)+\frac{3}{2}\rho_*m^2XZ(1+w)(k^2 E_2-k^2 E_1+3\psi_2)a^4\right. \nonumber\\
&+k^2a^2 \left( m^2 Z(k^2 E_2-k^2 E_1+3\psi_2)N^2+\rho_*(1+w)(3\psi_2+k^2E_2)N\right.\nonumber\\
& \left.\left.  +m^2XZ(k^2E_2-k^2E_1+3\psi_2) \right) + 2\mathcal{H} k^4 E_2' N  +2\psi_2k^4N\right],
\end{align}

where $D_a$ is given by:
\begin{align}
D_a &= \mathcal{H}\left[ \frac{3}{2}k^2a^2  \left(m^2N^2Z+\rho_*(1+X)(1+w)N + m^2XZ\right) + k^4N(1+X)\right. \nonumber\\
&\left.  + \frac{9}{4}\rho_*m^2 XZ(1+w)a^4\frac{}{}\right]
\end{align}

At a first glance, one might expect that the original system of equations (\ref{G00})-(\ref{Fij}), with seven scalar fields, has three dynamical degrees of freedom, as the equations of motion for $E_1$, $E_2$ and $\psi_2$ are independent and contain second derivatives. However, when eliminating the four auxiliary variables, and replacing them in the three remaining equations, we get that, in eq.~(\ref{Gii}) all first and second derivatives of $\psi_2$ cancel out, so that $\psi_2$ becomes an explicit auxiliary variable. Therefore, it can be written in terms of $E_1$ and $E_2$. Next, we show the expression for $\psi_2$ when worked out from eq.~(\ref{Gii}): 
\begin{align}
\psi_2&=\frac{k^2}{2D_p}\left[-2\mathcal{H}k^2N^2E_1'\left(\frac{3}{2}m^2Za^2(X-1)N^2+(-3\mathcal{H}^2X+3\mathcal{H}^2+k^2)N+\frac{3}{2}a^2m^2Z(X-1)\right)\right. \nonumber\\
& +2k^4\mathcal{H}N^3 E_2'+\frac{3}{2}k^2X^2Z^2a^4m^4(E_1-E_2 X)N^5+(E_1-E_2X)m^2a^2\left(\frac{9}{4}m^4Xa^4(X-1)Z^2\right.\nonumber\\
&\left.   +k^2(k^2X-6X^2\mathcal{H}^2+3\mathcal{H}^2)\frac{}{}\right)ZN^4+\left(\frac{3}{2}m^4a^4\left(-2 X^3E_2(k^2-3\mathcal{H}^2)\right.\right.\nonumber\\
& +\left. \left(k^2E_1 -6\mathcal{H}^2(E_2+E_1)\right)X^2+2E_1(3\mathcal{H}^2+k^2)X-k^2E_1\right)Z^2-2\mathcal{H}^2k^2\left(3X^3E_2\mathcal{H}^2\right.\nonumber\\
& \left.  \left. -(k^2E_2+3\mathcal{H}^2 E_1)X^2+(k^2E_1-3E_2\mathcal{H}^2)X+3\mathcal{H}^2E_1+k^2(-E_2+E_1)\right)\frac{}{}\right)N^3\nonumber\\
& +\left(\frac{9}{2}m^4Xa^4(X-1)(E_1-E_2 X)Z^2+(-9E_2\mathcal{H}^4+6k^2\mathcal{H}^2 E_2)X^3 \right.\nonumber\\
&  -(k^2-3\mathcal{H}^2)\left(3\mathcal{H}^2( E_1+E_2)+k^2 E_2\right)X^2+\left(-9\mathcal{H}^4 E_1-6k^2\mathcal{H}^2\left(E_1+\frac{1}{2}E_2\right)+k^4E_2\right)X\nonumber\\
&  \left. +\left(6\mathcal{H}^2E_1+k^2(-E_2+E_1)\right)k^2\frac{}{}\right)m^2a^2ZN^2+3m^4a^4\left(-\frac{1}{2}E_2(k^2-6\mathcal{H}^2)X^3 -\frac{1}{2}k^2E_1\right.\nonumber\\
& \left.\left. -3\mathcal{H}^2(E_1+E_2)X^2 +E_1\left(3\mathcal{H}^2+k^2\right)X \right)Z^2N+\frac{9}{4}m^6XZ^3a^6(X-1)(E_1-E_2X)\frac{}{}\right],
\end{align}
where $D_p$ is given by 
\begin{align}
D_p&=X\left[  3m^2a^2\left( \frac{9}{8}m^4Xa^4(X-1)Z^2+\left(k^2X+\frac{3}{2}\mathcal{H}^2-3X^2\mathcal{H}^2\right)k^2\right)ZN^4   \right.\nonumber\\
& + \frac{9}{4}X^2m^4Z^2a^4k^2N^5 +\left(\frac{9}{2}m^4a^4\left( (k^2-3\mathcal{H}^2)X^2+\left(\frac{1}{2}k^2+3\mathcal{H}^2\right)X-\frac{1}{2}k^2\right)Z^2\right. \nonumber\\
&\left.  -6k^4X\mathcal{H}^2-9k^2\mathcal{H}^4+9X^2k^2\mathcal{H}^4+k^6\frac{}{}\right)N^3  + 3m^2a^2ZN^2\left(\frac{9}{4}m^4Xa^4(X-1)Z^2 \right. \nonumber\\
& \left. +\left(\frac{9}{2}\mathcal{H}^4-3\mathcal{H}^2k^2\right)X^2+\left(k^4-\frac{9}{2}\mathcal{H}^4-\frac{3}{2}\mathcal{H}^2k^2\right)X+3\mathcal{H}^2k^2\right)\nonumber\\
&\left. +\frac{9}{4}m^4a^4 \left( \left(k^2-6\mathcal{H}^2\right)X^2+\left(k^2+6\mathcal{H}^2\right)X-k^2\right)Z^2N+\frac{27}{8}m^6XZ^3a^6(X-1)\right].
\end{align}
Therefore, as expected, only two degrees of freedoms are remain: $E_1$ and $E_2$.

\subsection{Complete equations of motion}
\label{AppScalarEq}
In this subsection we present the full equations of motion for the two propagating, physical scalar degrees of freedom: $E_1$ and $E_2$. The equation for $E_2$ takes the following form:
\begin{align}
&E_2^{''}-\frac{27}{4D_2}\left(w+\frac{1}{3}\right)  k^2 N^2 a^4 \rho_* m^2 \mathcal{H} Z (1+w) E_1'-\frac{3\mathcal{H}}{D_2}\left[ -\frac{3}{2}(1+w)^2 a^4 \left(\frac{3}{2} a^2 X m^2 Z\right. \right. \nonumber\\
&\left. \left. +k^2 N(1+X)\frac{}{}\right)\rho_*^2+ \frac{1}{2}(1+w)\left(\frac{3}{2}m^2a^2 Z\left(\left(3w-1\right) X-2 N^2\right) + k^2 N\left(3(w-1) X+ 3w \right.\right. \right. \nonumber\\
&\left. \left. \left.  +1\right)\frac{}{}\right) a^2 k^2 \rho_*+ (X-1) k^6 N \left(w-\frac{1}{3}\right) \right]E_2'+\frac{k^2}{D_2} \left[  k^6 N w (X-1)   \right. \nonumber\\
&   -\frac{3}{4}(1+w)^2 a^4 \left(\frac{3}{2} a^2 X m^2 Z+k^2 N (1+X)\right) 	\rho_*^2 +\frac{1}{2} a^2 (1+w)k^2\rho_* \left\{\frac{3}{2}Z m^2 a^2 \left(3X w  \right.\right.  \nonumber\\
&\left.\left. \left. + N^2\left( \left(3w+1\right) X-1\right)\right) +N k^2 \left(\left(3w-1\right) X+3w+1\right)\right\} \right]E_2\nonumber\\
&+\frac{m^2}{D_2} a^2 N\left[\frac{9}{4} k^2 \tilde{Z} \rho_* m^2 a^4 Z(1+w) N^2+k^2 \left\{3(1+w)\left(\tilde{Z}-(1+3w)\frac{Z}{4}\right)k^2X\rho_*a^2\right.\right.\nonumber\\
&\left. + \frac{9}{4} \tilde{Z}\rho_*^2 (1+w)^2 (1+X) a^4+\left(\tilde{Z}-\frac{3}{2}\left(w+\frac{1}{3}\right)Z\right)k^4 (X-1)\right\}N\nonumber\\
& \left. +\frac{9}{4}(1+w)\tilde{Z} ZX\rho_* m^2 a^4 \left(\frac{3}{2}\rho_*(1+w) a^2+k^2\right)\right](E_2-E_1)=0,
\end{align}
and the equation for $E_1$ takes the following form:
\begin{align*}
&E_1^{''}+\frac{2\mathcal{H} }{D_1}\left[ \frac{9}{4}a^4\left\{\frac{3}{4}a^2m^2\left(4X^2+\left(3w-1\right)X-\left(1+3w\right)\right)Z^2+\frac{1}{2}\left(3X^2+\left(3w+1\right)X\right.\right. \right. \nonumber\\
& \left. \left. -3w \right)k^2NZ +N\tilde{Z}k^2(1+X)\right\}(1+w)^3N\rho_*^3  +3a^2 \left\{ -\frac{9}{16}a^4m^4(1+X)(X-1)^2(N^2+1)Z^3   \right. \nonumber\\
& -\frac{9}{8}a^2m^2Z^2\left(\frac{1}{3}k^2N \left( N^2X(X^2-X-1)+X^3+2+3w(1-X+N^2)-5X^2\right)\right. \nonumber\\
& \left. -m^2\tilde{Z}(X-1)^2(N^2+1)a^2\frac{}{}\right) +\frac{3}{4}k^2\left(m^2\tilde{Z}(X-1)^2(N^2+1)a^2\right. \nonumber\\
& \left.\left.  +\left(2X^2+\left(w-\frac{1}{3}\right)X-\left(w+\frac{1}{3}\right)\right)k^2N \right)NZ+\tilde{Z}k^4N^2X\right\} (1+w)^2\rho_*^2  \nonumber\\
&  +k^2 (X-1)\left\{ -\frac{9}{8}a^4m^4(X-1)(1+X)(N^2+1)Z^3-\frac{3}{2}a^2(N^2+1)(X-1)m^2\left(\frac{}{}Nk^2(1+X)\right.\right. \nonumber\\
& \left. \left. -\frac{3}{2}m^2a^2\tilde{Z}\right)Z^2 +\frac{1}{2}\left(6m^2\tilde{Z}(N^2+1)(X-1)a^2+\left(1+3X\right)k^2N\right)k^2NZ+k^4\tilde{Z}N^2\right\}(1+w)\rho_*\nonumber\\
& \left. +(N^2+1)k^6(X-1)^2 Zm^2N \left(\tilde{Z} -\frac{1}{2}(X+1)Z\right)\right]E_1'\nonumber\\
&-2\frac{\mathcal{H}k^2}{D_1}\left[ -\frac{9}{8}a^4\rho_*m^4(1+X)(X-1)^2(N^2+1)(1+w)Z^3- \frac{1}{2} m^2Z^2 \left\{\frac{}{} k^4N(X-1)^2(1  \right. \right. \nonumber\\
& +X)(N^2+1) -\frac{27}{2}\left(\frac{2}{3}X^2+\left(w-\frac{1}{3}\right)X-\frac{1}{2}w-\frac{1}{6}\right)a^4(1+w)^2N\rho_*^2  + \frac{3}{2}\left( Nk^2(1+X)  \right. \nonumber\\
& +\mbox{continues in next page}
\end{align*}
\begin{align}
& \left.\left.-3m^2a^2\tilde{Z} \right) a^2(1+w)(N^2+1)(X-1)^2\rho_*  \right\} + k^2NZ\left( \frac{9}{2}a^2(1+w)^2\left(Xw+\frac{1}{2}X^2-\frac{1}{6}\right)N\rho_*^2  \right.\nonumber\\
& + \frac{3}{2}(1+w)(X-1)\left(m^2a^2 \tilde{Z}(X-1)N^2 +k^2(X+w)N+m^2a^2 \tilde{Z}(X-1)\right)\rho_* \nonumber\\
&  \left. \left.   +k^2\tilde{Z}m^2(X-1)^2(N^2+1) \frac{}{}\right)  +\left(\frac{3}{2}a^2(1+X)(1+w)\rho_*+k^2(X-1)\right)\rho_*\tilde{Z}(1+w) k^2 N^2 \right]E_2'\nonumber\\
&+\frac{\rho_*k^2(1+w)}{D_1}\left[ -\frac{3}{4}m^4 Z^2 \left( (1+X)Z-2\tilde{Z}\right)a^4k^2(X-1)^2N^4-\frac{1}{2}\left\{-3\rho_*\left(\left(-\frac{1}{2}X^3+2X^2 \right. \right. \right.\right. \nonumber\\
& \left. +\left(3w-\frac{1}{2}\right)X-\frac{3}{2}w-1\right)Z \left.  +\tilde{Z}(X-1)^2\right)(1+w)a^2  +k^2(X-1)^2((1+X)Z\nonumber\\
&\left.   -2\tilde{Z}) \frac{}{}\right\} m^2Za^2k^2N^3 + \left\{ \left(-\frac{3}{2}m^4(1+X)(X-1)^2Z^3  +3m^4\tilde{Z}(X-1)^2 Z^2  + \frac{3}{4}\left(X^2-3w\right. \right. \right. \nonumber\\
& \left. \left. +\left(3w-2\right)X-1\right) \rho_*^2(1+w)^2 Z-\frac{3}{2}\rho_*^2 X\tilde{Z}(1+w)^2(1+X)\right)a^4 + \frac{1}{2}\rho_*a^2k^2(1+w)\left(\frac{}{}\left(-1 \right.\right.\nonumber\\
& \left. \left. +\left(2+3w\right)X^2+\left(3w-1\right)X\right)Z-4\tilde{Z}\left(\frac{1}{2}+X\right)(X-1)\right) +\frac{1}{3}\left(\left(-1+\left(1+3w\right)X\right)Z\right. \nonumber\\
& \left.\left. -2\tilde{Z}(X-1)\right)k^4(X-1)\right\} k^2N^2-\frac{1}{2}m^2Za^2\left\{\frac{9}{4}ZX^3\rho_*^2(1+w)^2a^4 -3k^2(1+w)\left( \tilde{Z}(X-1)^2\right. \right. \nonumber\\
&\left.\left.  +Z\left(-X^3+\frac{5}{2}X^2+\left(-\frac{1}{2}+3w\right)X-\frac{3}{2}w-1\right)\right)\rho_*a^2 +((1+X)Z-2\tilde{Z})k^4(X-1)^2\right\} N\nonumber\\
&\left. -\frac{3}{4}m^4Z^2\left((1+X)Z-2\tilde{Z}\right)a^4k^2(X-1)^2 \right]E_1 +\frac{1}{D_1}\left[  -a^2(X-1)^2m^4Z^2k^6N^4 \left(\tilde{Z} \right.\right.\nonumber\\
&\left. \left. -(X+1)\frac{Z}{2}\right) - \frac{2}{3}m^2Zk^4 \left\{-\frac{1}{2}(X-1)k^2Z\left(-\frac{3}{2}\rho_*\left(1+3X-X^2+3w\right)(1+w)a^2+(X^2 \right. \right.\right.  \nonumber\\
& \left. \left. -1)k^2\frac{}{}\right)  + \left(\frac{9}{4}\rho_*^2(1+w)^2(1+X)a^4 +\frac{3}{2}\rho_*k^2X(X -1)(1+w)a^2 +k^4(X-1)^2\right)\tilde{Z}\right\} N^3   \nonumber\\
&   - \frac{2}{3}k^2 \left\{ -\frac{3}{2}k^4 m^4a^2(1+X)(X-1)^2 Z^3  + 3a^2m^4 \left(\frac{9}{8}\rho_*^2 X(1+w)^2a^4  + k^4(X-1)^2\right)\tilde{Z} Z^2 \right. \nonumber\\
&  -\frac{3}{2}\rho_*\left(\frac{3}{2}\rho_*\left( X^2w+\frac{2}{3}X+w\right)(1+w)a^2+(X-1)k^2\left(Xw-\frac{1}{3}\right)\right) k^4 (1+w)Z \nonumber\\
& \left. +\rho_* \left(\frac{3}{2}a^2(1+X)(1+w)\rho_*+k^2(X-1)\right)\tilde{Z}k^2\left( \frac{3}{2}a^2(1+X)(1+w)\rho_*+k^2X\right)(1+w)\right\}N^2\nonumber\\
& - \frac{2}{3}m^2Zk^2\left\{-\frac{1}{2}\left(\frac{9}{4}\rho_*^2X\left(X^2-1+\left(1+3w\right)X\right)(1+w)^2a^4+\frac{9}{2}\rho_*(X-1)\left(\frac{2}{3}X^2 -\frac{2}{3}-w\right. \right.\right. \nonumber\\
& \left. +\left(w-\frac{2}{3}\right)X\right)k^2(1+w)a^2 \left.  +k^4(1+X)(X-1)^2\frac{}{}\right)k^2Z +\tilde{Z}\left(\frac{3}{2}\rho_*(1+w)a^2+k^2\right) \nonumber\\
& \left. \cdot \left(\frac{9}{2}\rho_*^2X(1+w)^2(1+X)a^4+3\rho_*k^2X(X-1)(1+w)a^2 +k^4(X-1)^2 \frac{}{}\right)\right\}N \nonumber\\
& -a^2 \left(\tilde{Z} \left(\frac{27}{8}X^2\rho_*^3(1+w)^3a^6+\frac{9}{4}k^2\rho_*^2X^2(1+w)^2a^4+k^6(X-1)^2\right)\right. \nonumber\\
& \left. \left. -\frac{k^6}{2}(1+X)(X-1)^2 Z \right) m^4 Z^2   \right](E_2-E_1)=0,
\end{align}
where $D_1$ and $D_2$ are given by:
\begin{align}
D_1=&Z\rho_*(1+w) N\left[\frac{9}{4}\rho_* m^2\left( N^2 k^2+\left(\frac{3}{2}\rho_* (1+w)a^2+ k^2\right) X\right) a^4 (1+w) Z\right. \nonumber\\
&\left. +\left(\frac{3}{2}\rho_*(1+X)(1+w)a^2+k^2(X-1)\right) \left(\frac{3}{2}\rho_*(1+w)a^2+ k^2\right)Nk^2\right] \\
 D_2=& \frac{27}{8} m^2 ZX\rho_*^2 (1+w)^2 a^6+\frac{9}{4} k^2\rho_* \left(m^2  (X+N^2)Z+\rho_*(1+X) (1+w) N\right)(1+w)a^4\nonumber \\
& +3 k^4\rho_* NX(1+w)a^2+k^6 N(X-1).
\end{align}

\subsection{Exponential instabilities}
\label{AppInstability}
As mentioned previously, in the expanding branch scalar perturbations have an exponential instability at early times for sub-horizon scales; the instability is independent of the particular values of the parameters $\beta$s. However, during the bouncing branch, different solutions can be found for different parameters. For this reason, we distinguish the following cases: (a) $\beta_3\not=0$; (b) $\beta_3=0$ and $(\beta_4-3\beta_2)\not=0$; (c)$\beta_3=0$ and $(\beta_4-3\beta_2)=0$; (d) $\beta_3=\beta_2=0$. 

In what follows we will see that in cases (a), (b) and (c), $E_1$ develop an exponential instability at early times. For simplicity, let us study the equations of motion during the radiation-dominated era for sub-horizon scales. Generically, the equations of motion can be written as
\begin{equation}
E_a^{''}+f_{ab}(x,N)E_b'+g_{ab}(x,N)E_b=0; \; x=k\mathcal{H}^{-1},
\end{equation}
but when approximated at early times in the bouncing branch ($N\gg 1$) and for sub-horizon scales ($x \gg 1$), these coefficients become:

\begin{description}
\item[Case (a):] 
\begin{align}
&f_{11}=\frac{16}{3}\frac{\beta_4}{\beta_3}\frac{\mathcal{H}}{N}x^2 , \; f_{12}=\frac{8}{9}\frac{\beta_4^2}{\beta_3^2}\frac{\mathcal{H}}{N^2}x^2, \; f_{22}=2\mathcal{H} , \; f_{21}=-18\mathcal{H}, \\
&g_{11}=-\frac{1}{9}x^2\mathcal{H}^2 , \; g_{12}=\frac{4}{27}x^2\mathcal{H}^2, \; g_{22}=-\frac{\beta_3}{\beta_4}N\mathcal{H}^2 , \; g_{21}=\frac{\beta_3}{\beta_4}N\mathcal{H}^2 , 
\end{align}
\item[Case (b):] 
\begin{align}
&f_{11}=-2\mathcal{H} , \; f_{12}=2\mathcal{H}, \; f_{22}=\frac{12}{x^2}\frac{(\beta_4-3\beta_2)}{\beta_4}\mathcal{H} , \; f_{21}=\frac{54}{x^4}\frac{(\beta_4-3\beta_2)\beta_2}{\beta_4^2}\mathcal{H}, \\
&g_{11}=-\frac{1}{3}x^2\mathcal{H}^2 , \; g_{12}=-\frac{2}{3}x^2\mathcal{H}^2 , \; g_{22}=\frac{1}{3}x^2\mathcal{H}^2 , \; g_{21}=6\frac{\beta_2}{\beta_4}\mathcal{H}^2, 
\end{align}
\item[Case (c):]
\begin{align}
&f_{11}=-6\mathcal{H} , \; f_{12}=6\mathcal{H}, \; f_{22}=-\frac{36}{Nx^2}\frac{\beta_1}{\beta_4} \mathcal{H}, \; f_{21}=-\frac{27}{Nx^4}\frac{\beta_1}{\beta_4}\mathcal{H} , \\
&g_{11}=-\frac{1}{3}x^2\mathcal{H}^2 , \; g_{12}=-\frac{14}{3}x^2\mathcal{H}^2, \; g_{22}=\frac{1}{3}x^2\mathcal{H}^2 , \; g_{21}=4\mathcal{H}^2.
\end{align}
\end{description}
As we can see in all cases, the coefficient $g_{11}$ has a negative sign, which will induce an exponential instability in the solutions for $E_1$.

\subsection{Density contrast}
\label{AppDensityContrast}

The explicit form of the density contrast $\delta_{GIk}$ as a function of $E_i$ is:

\begin{align}
\delta_{GIk} & = \frac{2(1+w)}{D_d}\left[ -27k^2a^4\rho_* m^2 \mathcal{H} N^2 Z(1+w)E_1' + 9\rho_* k^4m^2ZX(1+w)(N^2 E_1+E_2)a^4   \nonumber\right. \\
& +\frac{9}{2}\rho_* \mathcal{H} (1+w)a^2E_2'\left( 9\rho_* m^2 XZ(1+w)a^4 + 6 k^2Na^2\left( (1+X)(1+w)\rho_*+Zm^2 N \right) \right.\nonumber \\
& \left.+ 4k^4N(X-1) \right)  + 6k^6N \left( E_2(1+X)(1+w)\rho_* - m^2NZ(X-1)(E_2-E_1) \right)a^2 \nonumber \\
& \left. + 4k^8E_2 N(X-1)  \right],
\end{align}
where $D_d$ is given by:

\begin{align}
D_d&=27m^2 ZX\rho_*^2(1+w)^2a^6+18k^2\rho_*(1+w)a^4\left( \rho_*(1+X)(1+w)N+m^2Z(X+N^2)\right)\nonumber \\
&+ 24k^4\rho_*NX(1+w)a^2+8k^6N(X-1).
\end{align}

\section{Ghost-like instabilities}\label{AppHiguchi}
As it was shown in \cite{Hassan:2011ea}, bimetric massive gravity given by eq.~(\ref{MGaction}) is said to be ghost-free in the sense that it propagates the right number of degrees of freedom: five for a massive graviton and two for a massless graviton, and avoids an extra ghost-like scalar field (with negative sign in its kinetic term). However, as realised for the first time by Higuchi in \cite{Higuchi:1986py}, the helicity-0 mode of the massive graviton might behave as a ghost for some values of the parameters of the theory in de-Sitter space-time, leading to instabilities on the solutions beyond the classical linear regime. The condition to have positive kinetic terms only in the action is known as the Higuchi bound. In addition, the helicity-1 vector field could also propagate as a ghost for some parameters, while the tensor fields are always safe from becoming ghosts (see \cite{Deser:2001wx}).

In the case of FRW backgrounds, described by eq.~(\ref{sq})-(\ref{sg}), a Higuchi bound for scalar and vector fields was found in \cite{Fasiello:2013woa} for the bimetric massive gravity model addressed in this paper, by analysing the quadratic action for linear perturbations. In this section, we analyse the satisfiability of these Higuchi bounds for the relevant cases considered in this paper.

\subsection{Scalar fields}
According to \cite{Fasiello:2013woa}, the Higuchi bound for the helicity-0 mode in the second branch of background solutions, satisfying $X\mathcal{H}=h$, is:
\begin{equation}\label{HiguchiScalar}
\tilde{m}^2\left( 1+ \frac{1}{N^2}\right) - 2H^2 \ge 0,
\end{equation}
where $H$ is the Hubble parameter and $\tilde{m}$ is given by: 
\begin{equation}
\tilde{m}^2=m^2NZ,
\end{equation}
 where $Z$ was defined previously as $Z=\beta_1+2\beta_2N+\beta_3N^2$.
%This bound is valid for arbitrary matter content and arbitrary FRW-type solutions.

In what follows, we consider the expanding and bouncing branches, and analyse the Higuchi bound in two relevant limit cases: early and late times.

\begin{description}
\item[Expanding branch:] In this branch we have $\beta_1> 0$. Using the Friedmann equation given by eq.~(\ref{EqFried}), the bound (\ref{HiguchiScalar}) becomes:
\begin{equation}\label{ExpandHig1}
 m^2N(\beta_1+2\beta_2N+\beta_3N^2)\left( 1+ \frac{1}{N^2}\right) -  \frac{2}{3}\left[\rho_0+ m^2\left(\beta_0+3N\beta_1+3\beta_2N^2+\beta_3N^3\right) \right]\ge 0,
\end{equation}
or equivalently, using the constraint (\ref{Density}),
\begin{equation}\label{ExpandHig2}
N(\beta_1+2\beta_2N+\beta_3N^2)\left( 1+ \frac{1}{N^2}\right) -  \frac{2}{3}\left(\frac{\beta_1}{N}+3\beta_2+3\beta_3N+\beta_4N^2\right) \ge 0.
\end{equation}

\begin{description}

\item[1. Early times:] At early times, $N \ll 1$. Considering only the leading terms in $1/N$, the bound (\ref{ExpandHig2}) becomes:
\begin{equation}
\frac{\beta_1}{3N} \ge 0,
\end{equation}
which is satisfied for the cases considered in this paper, as it was assumed that $\beta_1>0$ and $N>0$. 

\item[2. Late times:] At late times we approach a de-Sitter space-time where $\rho_0\rightarrow 0$ and $N\rightarrow \bar{N}$, where $\bar{N}$ satisfies eq.~(\ref{Density}) with $\rho_0=0$. In this regime the bound (\ref{ExpandHig1}) becomes:
\begin{equation}
 \bar{N}(\beta_1+2\beta_2\bar{N}+\beta_3\bar{N}^2)\left( 1+ \frac{1}{\bar{N}^2}\right) -  \frac{2}{3}\left(\beta_0+3\bar{N}\beta_1+3\beta_2\bar{N}^2+\beta_3\bar{N}^3\right) \ge 0.
\end{equation}
This bound can be satisfied for different values of the parameters. One interesting case is when $\beta_1$ is the only non-zero parameter. In this case the bound becomes:
\begin{equation}
 \left( \frac{1}{\bar{N}^2}-1\right)  \ge 0 \quad \Rightarrow \quad \bar{N}<1,
\end{equation}
which is actually satisfied, as in this $\beta_1$-only model, $\bar{N}=1/\sqrt{3}$.

\end{description}

Finally, we have found that the Higuchi bound can be satisfied in the expanding branch for appropriate values of the parameters at early times and late times\footnote{A more careful analysis is needed to check that the Higuchi bound is satisfied at all times, which will be left as future work.}. However, this does not guarantee instability-free solutions, as we could have tachyonic instabilities, which is what happens in this branch as described in section \ref{SecScalarPert}, where growing exponential solutions were found.

\item[Bouncing branch:] In this branch we have $\beta_3=\beta_2=0$ and $\beta_4\not= 0$ with $\beta_1\not= 0$.
Here, $Z=\beta_1$. Using the Friedmann equation given by eq.~(\ref{EqFried}), the bound (\ref{HiguchiScalar}) becomes:

\begin{equation}\label{BounceHig1}
 m^2N\beta_1\left( 1+ \frac{1}{N^2}\right) -  \frac{2}{3}\left[\rho_0+ m^2\left(\beta_0+3N\beta_1\right) \right]\ge 0,
\end{equation}
or equivalently, using the constraint (\ref{Density}),
\begin{equation}\label{BounceHig2}
N\beta_1\left( 1+ \frac{1}{N^2}\right) -  \frac{2}{3}\left(\frac{\beta_1}{N}+\beta_4N^2\right) \ge 0.
\end{equation}

\begin{description}

\item[1. Early times:] At early times, $N\gg 1$. Using eq.~(\ref{BounceHig2}) and considering leading terms in $N$, the bound becomes: 
\begin{align}
& m^2N\beta_1 -  \frac{2}{3}m^2\beta_4N^2 \ge 0 \\
\Rightarrow   \quad &  \approx -  \frac{2}{3}m^2 \beta_4N^2 \ge 0, 
\end{align}
which can only be satisfied if $\beta_4<0$, which is not viable as we would have negative energy density (see eq.~(\ref{BounceCondB4})).

\item[1. Late times:] At late times we approach a de-Sitter space-time where $\rho_0\rightarrow 0$ and $N\rightarrow \bar{N}$, where $\bar{N}$ satisfies eq.~(\ref{Density}) with $\rho_0=0$. Using eq.~(\ref{BounceHig1}), the bound becomes:
\begin{align}
& m^2\bar{N}\beta_1\left( 1+ \frac{1}{\bar{N}^2}\right) -  \frac{2}{3}m^2(\beta_0+3\beta_1\bar{N}) \ge 0\\
 \Rightarrow \quad & \frac{\beta_1}{\bar{N}}\left(1-\bar{N}^2\right)-\frac{2}{3}\beta_0 \ge 0.
\end{align}
For the interesting case of self-acceleration, where $\beta_0=0$, this bound is generically not satisfied as $\bar{N}\ge 1$ (see \cite{Konnig:2014xva}). It can only be satisfied if $\beta_4=2\beta_1$, where $\bar{N}=1$. %For the case with positive cosmological constant $\beta_0>0$, this bound will never be satisfied.

\end{description}

Finally, we have found that the Higuchi bound is not satisfied in the bouncing branch. This means that in the quadratic action for perturbations, the helicity-0 mode has a negative kinetic term, becoming a ghost-like degree of freedom. In this case, this does not translate into instabilities in the solutions as we found well-behaved solutions in section \ref{SecScalarPert}. However, instabilities might appear when studying higher order perturbations.

\end{description}

\subsection{Vector fields}

According to \cite{Fasiello:2013woa}, the Higuchi bound for the vector modes in the second branch of background solutions, satisfying $X\mathcal{H}=h$, is:
\begin{equation}\label{HiguchiVector}
\tilde{m}^2 > 0.
\end{equation}
For the relevant cases considered in this paper $m^2>0$ and $N>0$, so this condition becomes:
\begin{equation}\label{HiguchiVector2}
Z=\beta_1+2\beta_2N+\beta_3N^2 > 0.
\end{equation}

Analogously to the scalar modes, we now consider the expanding and bouncing branches, and analyse the Higuchi bound in two relevant limit cases: early and late times.

\begin{description}
\item[Expanding branch:] In this branch $\beta_1>0$. 

\begin{description}
\item[Early times:] Early times are characterised by $N\ll 1$. Then, in this regime eq.~(\ref{HiguchiVector2}) becomes simply $Z\approx \beta_1$, which is satisfied.

\item[Late times:] At late times we approach a de-Sitter space-time where $\rho_0\rightarrow 0$ and $N\rightarrow \bar{N}$, where $\bar{N}$ satisfies eq.~(\ref{Density}) with $\rho_0=0$. Condition (\ref{HiguchiVector2}) becomes:
\begin{equation}
\beta_1+2\beta_2\bar{N}+\beta_3\bar{N}^2 > 0,
\end{equation}
which can be satisfied for appropriate values for $\beta$s. In particular, for the $\beta_1$-only model, this condition will be satisfied.

\end{description}

\item[Bouncing branch:] This branch is characterised for $\beta_2=\beta_3=0$, and therefore $Z=\beta_1$. This meas that at all times, the condition (\ref{HiguchiVector}) is satisfied if $\beta_1>0$, which corresponds to the case considered in subsection \ref{SubSecBounceVector}, as there it was shown that $\beta_1<0$ introduced exponential instabilities in scalar, vector and tensor modes, and therefore that case was ruled out.

\end{description}

Finally, we have found that the Higuchi bound for vector modes can be satisfied at early and late times for appropriate values of parameters in the expanding branch, while it is always satisfied in the bouncing branch.

%--------------------------------------------------------------------------------------------------------------------------------------------------------
 
\bibliographystyle{apsrev4-1}
\bibliography{RefMassiveGravity}

\end{document}